\documentclass[aps,pra,twocolumn,10pt,amsmath,amssymb,showpacs,superscriptaddress,notitlepage,longbibliography]{revtex4-1}
\usepackage[colorlinks=true,linkcolor=blue,anchorcolor=red,citecolor=blue, urlcolor=blue]{hyperref}
\usepackage{bm}
\usepackage{graphicx}
\usepackage{xcolor}
\usepackage{wasysym}
\usepackage{makecell}
\usepackage{multirow}
\usepackage{comment}

\begin{document}
\title{
Can out-of-equilibrium linear response reveal anyonic statistics?\\
}

\author{Gu Zhang}
\email{zhanggu217@nju.edu.cn}
\affiliation{National Laboratory of Solid State Microstructures, School of Physics, Jiangsu Physical Science Research Center and Collaborative Innovation Center of Advanced Microstructures, Nanjing University, Nanjing 210093, China}
\affiliation{Beijing Academy of Quantum Information Sciences, Beijing 100193, China}

\author{Igor Gornyi}
\email{igor.gornyi@kit.edu}
\affiliation{Institute for Quantum Materials and Technologies and Institut f\"ur Theorie der Kondensierten Materie, Karlsruhe Institute of Technology, 76131 Karlsruhe, Germany}

\author{Yuval Gefen}
\email{yuval.gefen@weizmann.ac.il}
\affiliation{Department of Condensed Matter Physics, Weizmann Institute of Science, Rehovot 761001, Israel}

\begin{abstract}

Determination of the anyonic statistical braiding phase has relied on either Aharonov-Bohm interference experiments or cross-correlations measured in collider platforms. Here we consider collider setups, supporting stationary far-from-equilibrium anyonic beams, characterized by an effective chemical potential and an effective temperature, which carry universal information about the elementary charge and the statistical braiding phase.
We develop an \textit{out-of-equilibrium} linear response theory designed to describe charge and thermal quantum transport under deviations from this effective equilibrium (unlike linear response designed to describe transport near genuine equilibrium).
Our effective linear-response transport coefficients directly reflect the fractional charge and statistics of the anyons involved, avoiding the need to measure higher-order current correlations.
Moreover, the emergence of finite thermoelectric (Peltier and Seebeck) coefficients signifies the presence of real anyon collisions (as opposed to virtual braiding in the time domain), intimately associated with the breaking of a characteristic particle-hole symmetry specific to anyonic gases.

\end{abstract}

\maketitle

\section{Introduction}
\label{sec:intro}

Being a fundamental paradigmatic example of topological quantum platforms and possibly heralding future quantum information technologies, braiding of anyonic
operators~\cite{LaughlinPRL83, Arovas1984, Kitaev2001UFN, MongPRX14,Feldman2021}
has been a focal point over more than two decades
~\cite{HorodeckiRevModPhys09, NielsenChuangBook, WildeBook, GuhneTothReview, StreltsovAdessoPlenioRevModPhys17}.
Adding to the crucial appeal of this topic is its potential relevance to fault-tolerant topological quantum manipulations. 
Recent encouraging experiments have demonstrated braiding of Laughlin particles in Hong-Ou-Mandel (HOM)~\cite{BartolomeiScience20, PierrePRX23, LeeNature23, RuellePRX23,DeNC25}, Fabry-Perot~\cite{NakamuraNatPhys19, NakamuraNatPhys20, NakamuraNC22, Nakamura2023}, as well as Mach-Zehnder interferometers~\cite{HeiblumNP2023}, following a series of theoretical proposals~\cite{ThoulessGefenPRL91, SafiDevilardPRL01,  KaneFisherPRB03, VishveshwaraPRL03, KimPRL05, LawPRB06, FeldmanKitaevPRL06, FeldmanPRB07, PonomarenkoAverinPRL07, RosenowHalperinPRL07, ViolaPRL12, CampagnanoPRL12, CampagnanoPRB13, RosenowLevkivskyiHalperinPRL16, CampagnanoPRB16, SimNC16, LeePRL19, RosenowSternPRL20} and inspiring further advances of theory~\cite{MartinDeltaT20, SafiFDPRB20, MorelPRB22, KyryloPRB22, GuPRB22, LeeSimNC22, schillerPRL23, JonckheerePRL23, JonckheerePRB23, IyerX2023, AndreevNC25,OneHalfPRB24, ThammBerndPRL24, KivelsonMurthyPRL25, LandscapePRL25, MatteoSciPhys25, MatteoSciPhys25, KarmakarX25}. Important for our present study are anyonic colliders, representing an efficacious platform to manifest and manipulate anyonic statistics and the induced quantum-statistics-based entanglement. This topic is especially appealing, considering the fact that in transport systems, the proposal to generate and detect entanglement (in terms of, e.g., Bell and CHSH inequalities) is still limited to integer systems~\cite{BeenakkerPRL03, BüttikerPhysE03, SamuelssonPRL03, SamuelssonPRL04, SamuelssonPRB05, RoychowdhuryPRB16, LuoPRL22, GuNC24}.
One notes that collider measurements require sensitive fluctuation detection and careful separation of statistical effects from quasiparticle charge, scaling dimension, pulse shape, temperature, and edge dynamics.

Among these approaches, stationary anyon-collider experiments~\cite{BartolomeiScience20, PierrePRX23, LeeNature23, RuellePRX23,DeNC25} employ dilute, noisy injected beams generated by upstream quantum point contacts (QPCs), placing the incident channels intrinsically far from equilibrium.
The outgoing currents reflect quantum partitioning at this junction.
In the strongly diluted, point-like collider limit, the leading statistics-sensitive process is time-domain braiding: a quasiparticle–quasihole pair created and annihilated at the collider acquires a phase by braiding in time with non-equilibrium anyons passing the junction~\cite{RosenowLevkivskyiHalperinPRL16, SimNC16, SafiFDPRB20, LeeSimNC22, schillerPRL23}.
Away from this limit, direct tunneling of beam anyons and real anyonic collisions also contribute.
Conventionally, these processes are diagnosed through current-noise auto correlations and cross-correlations between separate drains, often combined into a generalized Fano factor whose predicted negative value constitutes an anyonic signature~\cite{RosenowLevkivskyiHalperinPRL16, BartolomeiScience20, LeeSimNC22, PierrePRX23, LeeNature23, RuellePRX23}.
Such measurements and their interpretation are demanding: the statistics-dependent contribution must be separated from injection and partition noise and from effects of fractional charge, scaling dimension, pulse shape, temperature, edge dynamics, and QPC transmission.
Cross-correlations are particularly delicate, requiring simultaneous low-noise detection in distinct drains and careful background subtraction. These challenges motivate observables that preserve sensitivity to anyonic statistics while relying only on average charge and heat currents.

Given the difficulty of measuring current cross-correlations in collider geometries, together with the sensitivity of such observables to non-universal details, a natural question arises: can purely d.c. charge and heat transport measurements be used to extract the universal anyonic braiding phase?
This challenge is particularly acute because the injected anyonic beam is stochastic, stationary, intrinsically non-thermal and far from equilibrium.
To pursue this direction, we note that such stationary non-equilibrium states can be characterized in terms of effective equilibrium parameters: {\it effective} chemical potential, $e V_\text{eff}$, and an {\it effective}  temperature, $T_\text{eff}$.
These quantities are operationally defined by weakly and locally coupling the non-equilibrium system to an equilibrium reference probe with well- defined chemical potential $eV$ and temperature $T$~\cite{EngquistAndersonPRB81}.
This calibration procedure has previously been implemented for edge channels~\cite{NosigliaParkRosenowGefenPRB18, LandscapePRL25}, including interacting composite edges in the incoherent regime~\cite{ProtopopovGefenMirlinAoP17, SpanslattPRB21}.

Once the possibility of characterizing a stationary far-from-equilibrium system by the effective equilibrium parameters $V_\text{eff}$ and $T_\text{eff}$ has been established, the next question is whether one can formulate a consistent linear-response theory about that state, and whether its transport coefficients retain universal information on the anyonic braiding phase.
By \textit{effective linear response} we mean the following operational construction.
One first tunes the coupled channels to effective equilibrium, defined by the simultaneous vanishing of the d.c. charge and heat currents, and then introduces infinitesimal mismatches in their effective or physical voltages and temperatures.
The induced currents define a $2\times 2$ matrix [see matrix elements in Eq.~\eqref{eq:coefficients_definition}] comprising charge and heat conductances and thermoelectric cross-coefficients.
The reference state nevertheless remains microscopically far from equilibrium: at least one incoming channel carries a stochastic non-equilibrium anyonic beam (thus being non-thermal) upstream from the central junction. Effective equilibrium thus identifies a local no-current operating point rather than restoring a thermal distribution throughout the system.
This differs fundamentally from generalized linear-response theories~\cite{BenaSafiPBR07,SpeckSeifertPRE09, SeifertEPL2010,SafiJoyezPRB11,KonopikLutzPRR19, ManaparambilWeymannPRB23, BlairPRR24}, in which incoming channels and reservoirs are assumed to be at equilibrium, as in a conventional Landauer–Büttiker setup. Here we develop this construction for non-thermal, non-equilibrium anyonic edges and ask whether the resulting d.c. coefficients can disclose fractional statistics without requiring higher-order current correlations.

The answer depends crucially on which processes are retained at the collider. In the strongly diluted limit, the transmission probability through the diluter, $\mathcal{T}_1$ [see Eq.~\eqref{eq:d1_t1} and above], becomes much smaller than unity, such that the leading process is time-domain braiding (TDB)~\cite{RosenowLevkivskyiHalperinPRL16, SimNC16, SafiFDPRB20, LeeSimNC22, schillerPRL23}, which can also be formulated in terms of an anyonic topological vacuum bubble~\cite{SimNC16}.
In a TDB event, a virtual quasiparticle–quasihole pair is created and annihilated at the collider, while its world lines braid in time with non-equilibrium anyons passing the junction. The injected anyons neither tunnel through the central collider nor undergo real collisions there.
We refer to the regime in which TDB dominates and direct tunneling is negligible as the collision-free limit; this restriction is essential for the universality of the effective linear-response theory.
An equilibrium channel possesses particle–hole symmetry about its chemical potential; remarkably, a non-equilibrium channel governed exclusively by TDB retains an analogous symmetry about its effective chemical potential (see the solid curves in Fig.~\ref{fig:distributions}).
At effective equilibrium, where $V_\text{1,eff} = V_\text{2,eq}$, the symmetry centers of the two channels coincide. The off-diagonal response coefficients, and hence the effective Seebeck and Peltier coefficients, then vanish.
In the zero-ambient-temperature limit, TDB also leaves a single relevant energy scale, providing the basis for the universal structure of the remaining coefficients.
The charge and heat conductances, together with the Lorenz number, assume universal forms [see Eqs.~\eqref{eq:universal_coefficients}-\eqref{lorenz-numbers-eq-eq}, Eq.~\eqref{eq:g_f_eq_eq} and Fig.~\ref{fig:thermoelectric_constants_Lorenz}] controlled by the anyonic data.
In particular, the Lorenz number is independent of microscopic QPC details and retains sensitivity to the braiding phase [see Eq.~\eqref{eq:universal_coefficients} and Appendix~\ref{sec:gamma_chi_functions}]; in the ideal Laughlin case, the conductances depend only on the filling fraction [through the scaling dimension, the effective charge, and the braiding phase, see Eq.~\eqref{eq:universal_coefficients}].

When the dilution is not extreme, TDB ceases to provide a complete description. Anyons emitted by the diluters may tunnel through the central collider and/or undergo real anyonic collisions.
These processes introduce additional energy scales, including the source-bias scale, and make the response sensitive to microscopic parameters such as the diluter and collider transmissions.
They also destroy the effective particle–hole symmetry of the collision-free regime [see Eq.~\eqref{eq:full_neq_distributions}, and the dashed curves in Fig.~\ref{fig:distributions}].
Consequently, the charge and heat conductances, as well as the Lorenz number, are no longer universal functions solely of the effective charge, scaling dimension, and braiding phase, but acquire a non-universal dependence on device details.
Simultaneously, the off-diagonal response coefficients become finite: a temperature mismatch generates a charge current, while a voltage mismatch generates a heat current, yielding nonzero Seebeck and Peltier coefficients (see Fig.~\ref{fig:coefficients_collision}).
Their emergence therefore provides a smoking-gun signature of particle–hole-symmetry breaking, the breakdown of the TDB-only or collision-free paradigm, and the accompanying loss of universality.
We evaluate both regimes for non-equilibrium Laughlin anyons in the collider geometry of Fig. 1(a) [Eqs.~\ref{eq:coefficients_definition} and \eqref{eq:transport_parameters_definition}]: first retaining TDB alone, and then incorporating real tunneling and collisions. 
The definition of effective linear response, the universal Lorenz number in the TDB regime, and the thermoelectric diagnosis of departures from it thus extend the scope of previous studies of anyonic braiding~\cite{RosenowLevkivskyiHalperinPRL16, SimNC16, SafiFDPRB20, LeeSimNC22, schillerPRL23}.
We note that the present approach allows one to avoid non-universal factors that should be accounted for when employing anyonic Aharonov-Bohm setups  or anyonic collider platforms.

The outline of the paper is the following.
We begin, in Sec.~\ref{sec:setup}, with the introduction of the setups, as well as observables under current consideration.
This is followed by Sec.~\ref{sec:experimental_proposal}, where we present the concrete structures that can realize these setups and enable the detection of the proposed observables in practical experiments.
Having established the the connection between the theoretical proposals and their experimental implementations, we proceed to Sec.~\ref{sec:quests}, to discuss the key anyonic signatures and the protocols for their extraction under various experimental conditions.
These protocols are among the central results of the current work.
Starting with Sec.~\ref{sec:model},
we turn to the microscopic theoretical description by introducing the explicit system Hamiltonian used to model the proposed setups.
This is followed by Sec.~\ref{sec:correlation_functions}, where we provide the correlation functions, and establish their connection to observables.

Secs.~\ref{sec:collision_free_results} and \ref{sec:phase_three_games} contain the central message of our work.
Firstly, Sec.~\ref{sec:collision_free_results} provides the effective linear response coefficients (charge conductance, heat conductance, Peltier, Seebeck coefficients, and the Lorenz number) in the collision-free limit.
Crucially, due to the particle-hole symmetry in the collision-free limit, these coefficients depend universally on anynoic signatures, including the fractional charge, braiding phase and scaling dimension.
With these universal observables, we continue in Sec.~\ref{sec:phase_three_games}, to illustrate the extraction of anyonic braiding phase in the collision-free limit, under different circumstances.
The results presented in the preceding sections are affected by direct tunneling of non-equilibrium anyons, which leads to particle-hole symmetry breaking. The resulting effects, including modifications of effective parameters, the emergence of Seebeck and Peltier coefficients, and, most importantly, the consequences for the experimental extraction of anyonic signatures, are discussed in detail in Sec.~\ref{sec:beyond_tdb}.

Further supporting information is provided in the Appendixes.
They include: The charge and heat tunneling currents beyond the collision-free limit in Appendix~\ref{sec:currents_beyond}; Explicit expressions for the universal functions in Eq.~\eqref{eq:universal_coefficients} in Appendix~\ref{sec:gamma_chi_functions}; Integral forms of the effective linear-response coefficients in Appendix~\ref{app:eff_res}; The proof of the Onsager relation under extended conditions in Appendix~\ref{app:onsager_proof}; Effective parameters after including direct tunneling of non-equilibrium anyons in Appendix~\ref{app:eff_para_direct}; and, Channel-resolved conductances in the collision-free limit in Appendix~\ref{app:chi_gamma}. The last Appendix also discusses the different conductance contributions from equilibrium and non-equilibrium anyonic channels, with the latter affected by anyonic braiding.

\section{The setup and observables}
\label{sec:setup}

%%%%%%%%%%%%%%%%%%%%%%%%%
\begin{figure}[!ht]
  \includegraphics[width= 0.75 \linewidth]{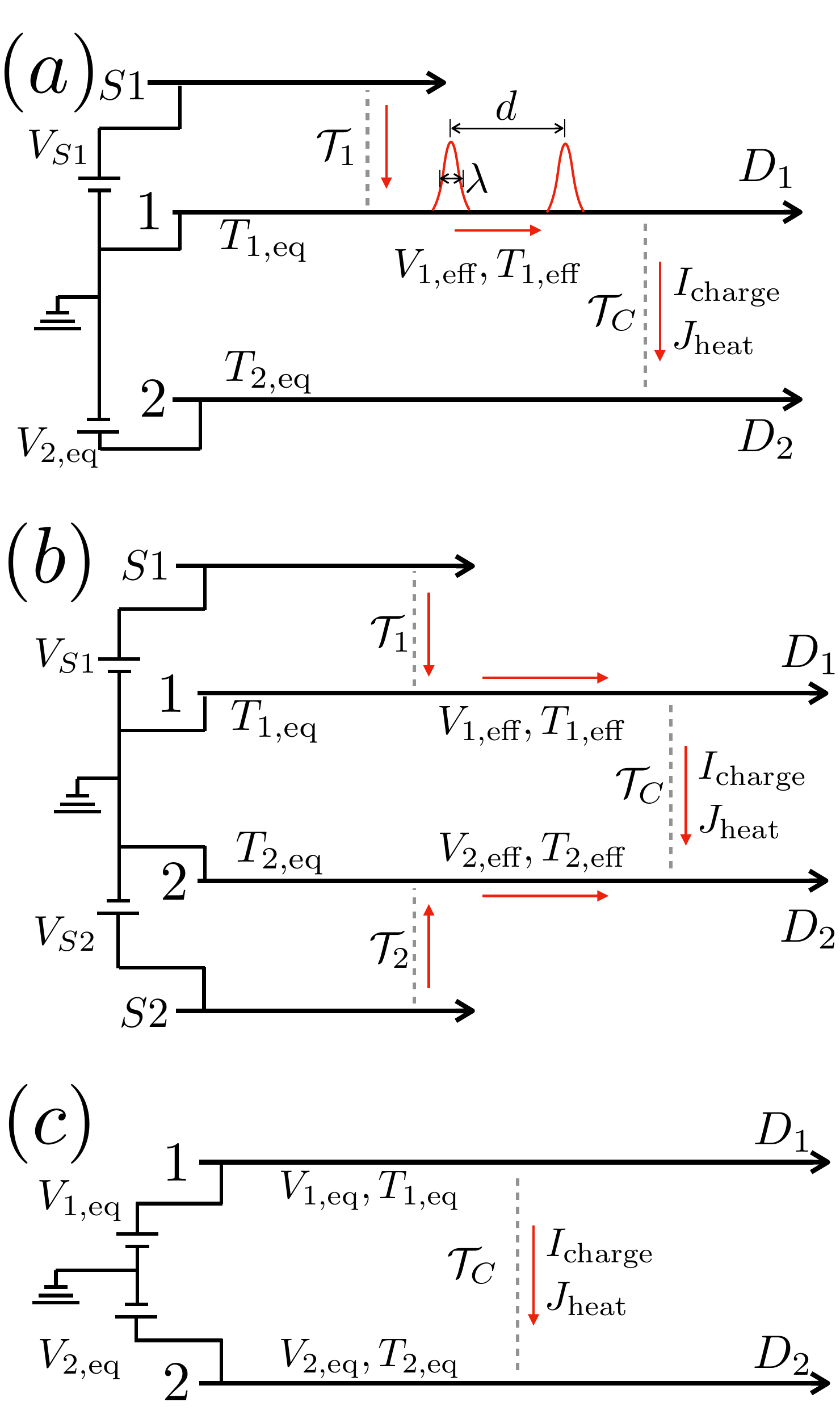}
  \caption{\textbf{The theoretical schematics of the setups considered by this work.}
  Tunnelings between different channels are marked out by the red arrows.
  (a) The mostly focused setup of our work, where channel 1 contains non-equilibrium anyons (the red pulses) from the source channel, $S1$, through the diluter that is described by the tunneling probability, $\mathcal{T}_1 \equiv 2\pi I_1/(q e^2 V_{S1})$. 
  Importantly, channel 1 becomes strongly out-of-equilibrium, when the typical half-width ($\lambda$) of non-equilibrium anyons is much smaller than the typical distance ($d$) between two neighboring anyons.
  Channel 1 communicates with an equilibrium channel, 2, through the central collider, with the transmission probability $\mathcal{T}_C$ [see Eq.~\eqref{eq:tc_relations}]. (b) and (c) are the two extra setups, where both channels 1 and 2 are out-of-equilibrium and in-equilibrium, respectively. 
  }
  \label{fig:three_setups}
\end{figure}
%%%%%%%%%%%%%%%%%%%%%%%%%

\subsection{Schematic setups}
\label{sec:schematics}

Within this work, we explore the transport properties of the anyonic HOM setups shown in Fig.~\ref{fig:three_setups}.
Especially, we focus on the setup of Fig.~\ref{fig:three_setups}(a), which consists of three edge-state channels: the source channel $S1$, and two functionality channels, 1 and 2, that communicate between them through the central collider.
These chiral edge-state channels, carrying charge-$e^*$ Laughlin quasiparticles, are located at the boundary of a Laughlin-state quantum Hall sample that has the filling fraction $\nu$ in the bulk. 

Channel 2 is at equilibrium, characterized by the corresponding chemical potential $V_\text{2,eq}$ and temperature $T_\text{2,eq}$. Channel 1
is injected with the non-equilibrium current $I_1$ (and the corresponding noise $S_1$) from the source channel $S1$ (with its chemical potential $V_{S1}$ and temperature $T_{S1}$) through the diluter located at $x=0$.
This diluter is characterized by the transmission probability $\mathcal{T}_1 \equiv 2\pi I_1/(q e^2 V_{S1})$,  with $q\equiv e^*/e$, and it is addressed as strongly diluted, when $\mathcal{T}_1 \ll 1$.
The non-equilibrium anyonic beam of channel 1 (upstream from the collider) is described by two characteristic length scales [see Fig.~\ref{fig:three_setups}(a)]: the typical width of non-equilibrium anyonic pulses [the red curves of Fig.~\ref{fig:three_setups}(a)], $\lambda \sim v/e^* V_{S1}$, and the typical distance, $d \sim v e^*/I_1$, between two neighboring ones.
When the latter scale becomes much larger than the former one, $d \gg \lambda$, channel 1 contains spatially well-separated non-equilibrium anyonic pulses, and is thus referred to as highly non-equilibrium.
Clearly, as $\lambda/d \sim I_1/e^*2 V_{S1} = \mathcal{T}_1/(2\pi q)$, arriving at the highly non-equilibrium regime requires a strong dilution at the diluter.
These two channels, 1 and 2, communicate through the central collider (located at $x = L > 0$) that is characterized by its transmission probability, $\mathcal{T}_C$ [see Eq.~\eqref{eq:tc_relations}].
Through out this work, we assume $\mathcal{T}_C \ll 1$, to limit our investigation to the leading order of the tunneling through the central collider.
For clarify, we stress that $\mathcal{T}_1$ and $\mathcal{T}_C$ are observables that can be directly measured experimentally.
They therefore differ from $\mathcal{T}_1^{(0)}$ and $\mathcal{T}_C^{(0)}$, which are defined as the squared magnitudes of the tunneling amplitudes in the tunneling Hamiltonian at the diluter, Eq.~\eqref{eq:hdiluter}, and that at the central collider, Eq.~\eqref{eq:hcollider}, respectively.
The corresponding tunneling charge current and heat current through the central QPC, as shown by Fig.~\ref{fig:three_setups}(a), are denoted as $I_\text{charge}$ and $J_\text{heat}$, respectively.

In addition to the setup of Fig.~\ref{fig:three_setups}(a), here we consider two extra setups, shown by Figs.~\ref{fig:three_setups}(b) and \ref{fig:three_setups}(c).
In contrast to that of Fig.~\ref{fig:three_setups}(a), here both edges of the latter two setups are either in-equilibrium [Fig.~\ref{fig:three_setups}(c)] or strongly out-of-equilibrium [Fig.~\ref{fig:three_setups}(b)].
The study of the quantum transport of these two setups can provide us extra information on the fractional features of quasiparticles hosted by the system.
Actually, the final setup [i.e., Fig.~\ref{fig:three_setups}(c)] contains two in-equilibrium channels, where non-equilibrium anyons, necessary for time-domain braiding, are absent.
The comparison between its corresponding quantum transport coefficients and that of the former two setups can thus more directly disclose the fractional braiding phase, $\theta$, of transport quasiparticles.

\subsection{Effective equilibrium of strongly non-equilibrium systems}
\label{sec:eff_equi}

In the schematic setups, channel 1 of Fig.~\ref{fig:three_setups}(a) and both channels of Fig.~\ref{fig:three_setups}(b) are strongly out-of-equilibrium, such that strictly neither of them can be simply characterized by two parameters, i.e., the temperature and the chemical potential.
Fortunately, following Ref.~\cite{EngquistAndersonPRB81} for fermionic and Ref.~\cite{LandscapePRL25} for anyonic systems, a channel that is strongly out-of-equilibrium can nevertheless be characterized by effective parameters within the effective equilibrium framework.
More specifically, for a non-equilibrium channel [say, channel 1 of Fig.~\ref{fig:three_setups}(a)], its equilibrium parameters can be calibrated by attaching it (at another QPC) to an in-equilibrium one [say, channel 2 of Fig.~\ref{fig:three_setups}(a)], described by the voltage $V_\text{2,eq}$ and $T_\text{2,eq}$.
This system containing channels 1 and 2 at the colliding QPC is defined to arrive at the effective equilibrium, when the tunneling charge current and the heat current both vanish, $I_\text{charge} = J_\text{heat} = 0 $.
Following this definition, the effective parameters, $V_\text{1,eff}$ and $T_\text{1,eff}$, of a non-equilibrium anyonic channel can be defined to equal the real parameters of an in-equilibrium channel ($V_\text{1,eff} = V_\text{2,eq}$ and $T_\text{1,eff} = T_\text{2,eq}$), when the system has arrived at the effective equilibrium.

Following definitions above, channel 1 of Fig.~\ref{fig:three_setups}(a) and both middle channels (i.e., 1 and 2) of Fig.~\ref{fig:three_setups}(b) can be described by the corresponding effective parameters.
Importantly, as long as the central collider permits weak transport of a \textit{determined} type of quasiparticles through the same tunneling process used to define the effective parameters, these parameters exhibit ``transitivity'': namely, connecting two channels with the same potential and temperature, either real or effective, yields vanishing charge and heat currents between them.

\subsection{Introduce transport coefficients}

We begin by defining quantum transport parameters of the focused setup, i.e., that of Fig.~\ref{fig:three_setups}(a), where channels 1 and 2 are strongly out-of-equilibrium, and in-equilibrium, respectively.
As outlined in Sec.~\ref{sec:eff_equi}, the identification of effective parameters of channel 1 requires the knowledge of both the tunneling charge current $I_\text{charge}$ and heat current $J_\text{heat}$, through the central QPC.
These two quantities can be evaluated, to leading order of the tunneling amplitude square, $\mathcal{T}_C^{(0)}$
[which 
is proportional to the measurable tunneling probability, $  \mathcal{T}_C^{(0)} \propto \mathcal{T}_C$, see Eqs.~\eqref{eq:tc_relations} and \eqref{eq:hcollider}], as follows~\cite{LandscapePRL25,MatteoSciPhys25} (from now on, we set $\hbar=k_B=1$):
\begin{equation}
\begin{aligned}
    I_\text{charge} & \equiv e^* \,\mathcal{T}_C^{(0)} \int_{-\infty}^\infty dt \left[\big\langle \psi^\dagger_1 (t) \psi_1(0) \big\rangle \big\langle \psi_2 (t) \psi_2^\dagger (0) \big\rangle\right.\\
    & \left. - \big\langle \psi_1 (t) \psi^\dagger_1(0) \big\rangle \big\langle \psi^\dagger_2 (t) \psi_2 (0) \big\rangle \right],\\
    J_\text{heat} & \equiv 2 i \mathcal{T}_C^{(0)} \int_{-\infty}^\infty dt \big\langle \psi^\dagger_1 (t) \psi_1(0) \big\rangle \partial_t \big\langle \psi_2 (t) \psi_2^\dagger (0) \big\rangle,
\end{aligned}
\label{eq:currents_correlations}
\end{equation}
in terms of the anyonic creation operators $\psi_1^\dagger$ and $\psi_2^\dagger$, and the annihilation ones $\psi_1$ and $\psi_2$, respectively (cf. Sec.~\ref{sec:model} for more details).
Notably, the heat current $J_\text{heat}$, defined by the second line of Eq.~\eqref{eq:currents_correlations} is referenced to the chemical potential of channel 2. A shift in this reference changes $J_\text{heat}$ by a term proportional to $I_\text{charge}$, which vanishes at effective equilibrium.
In Eq. (1) and the Hamiltonian in Eq. (13), the tunneling amplitude $\xi_C$ and its squared magnitude $\mathcal{T}_C^{(0)}$ are taken to be constants.
This simplification however does not undermine the applicability of our results to realistic experiments.
Indeed, a smooth energy dependence of the bare tunneling amplitude is renormalized into the experimentally accessible infrared tunneling coefficient, with the RG flow cut off at the relevant thermal/transport scale [e.g., the factor $\mathcal{T}_C^{(0)} (\pi \bar{T} \tau_0)^{4\delta - 2}$ entering the observables in Eq.~\eqref{eq:universal_coefficients}, with $\delta$ the scaling dimension].
This common renormalized factor cancels from the Lorenz ratios and other combinations used in the hard-game inversion (cf. Sec.~\ref{sec:quests}).

In addition to the chemical potential and temperature, systems in equilibrium are conventionally characterized by extra quantum transport parameters obtained from linear response, which can disclose e.g., quantized charge and heat conductances of quantum Hall systems, and exotic features of quantum criticality.
As among our central conclusions, the linear response technique, applicable initially to close-to-equilibrium systems, can also be extended to braiding-involved strongly out-of-equilibrium systems, to be denoted as the \textit{non-equilibrium} effective linear response within this work.
More specifically, the same as linear response, the effective linear response framework involves four parameters, $L_\text{IV}^s$, $L_\text{JT}^s$, $L_\text{JV}^s$ and $L_\text{IT}^s$ to characterize the system that is slightly deviated from the effective equilibrium.
Here $s$ marks out the status of two involved channels. It equals neq-eq, neq-neq and eq-eq, for the setups of Figs.~\ref{fig:three_setups}(a), \ref{fig:three_setups}(b) and \ref{fig:three_setups}(c), respectively.
For instance, for the mostly-focused setup of Fig.~\ref{fig:three_setups}(a) (where $s \ =\  $neq-eq), these transport coefficients are defined by
\begin{equation}
\begin{aligned}
    L_\text{IV}^\text{neq-eq} & \equiv \partial_{V_\text{1,eff} - V_\text{2,eq}}\, I_\text{charge}\, \Big|_\text{eff. equil.}, \\
    L_\text{JT}^\text{neq-eq} & \equiv \partial_{T_\text{1,eff} - T_\text{2,eq}} J_\text{heat}\, \Big|_\text{eff. equil.},\\
    L_\text{JV}^\text{neq-eq} & \equiv \partial_{V_\text{1,eff} - V_\text{2,eq}} J_\text{heat}\, \Big|_\text{eff. equil.}, \\  L_\text{IT}^\text{neq-eq}  & \equiv \partial_{T_\text{1,eff} - T_\text{2,eq}}\, I_\text{charge}\, \Big|_\text{eff. equil.},
\end{aligned}
\label{eq:coefficients_definition}
\end{equation}
characterizing the currents through the collider tunneling bridge
(subscripts $I$ and $J$ correspond to the tunneling charge $I_\text{charge}$ and heat current $J_\text{heat}$, respectively) induced in response to \textit{infinitesimal} biases in (effective or real) voltage and temperature (subscripts $V$ and $T$) between the channels in the stationary state.
The notation ``eff. equil.'' stands for $ V_\text{1,eff} = V_\text{2,eq}\equiv \bar{V},\,T_\text{1,eff} = T_\text{2,eq} \equiv \bar{T}$, where as the reminder, $I_\text{charge} = J_\text{heat} = 0$.

In practical setups, these effective linear response coefficients can be obtained in basically two steps.
As the first step, one needs identify the effective parameters of the effective equilibrium, after tuning both $I_\text{charge}$ and $J_\text{heat}$ to zero.
Then, chemical potential and temperature of either the equilibrium channel or the non-equilibrium one are slightly shifted, and the resulting charge and heat currents are calculated (or measured experimentally). After taking the corresponding derivatives with respect to the potential or temperature differences, the coefficients are evaluated under the condition of effective equilibrium.
In fact, the quantities defined by Eq.~\eqref{eq:coefficients_definition} are not all independent.
This is a manifestation of the fact that even though we are dealing with a far-from-equilibrium scenario, the Onsager relations are satisfied near our effective equilibrium point.
Specifically we find that
$L^\text{neq-eq}_\text{JV} = L^\text{neq-eq}_\text{IT} \bar{T}$
(see details in Sec.~\ref{sec:onsager_relation} and an extended discussion in Appendix~\ref{app:onsager_proof}).

Similarly, coefficients defined in Eq.~\eqref{eq:coefficients_definition} apply to also the setups of Figs.~\ref{fig:three_setups}(b) and \ref{fig:three_setups}(c).
As the difference, in the setup of Fig.~\ref{fig:three_setups}(b), $V_\text{2,eq}$ of Eq.~\eqref{eq:coefficients_definition} should be replaced by $V_\text{2,eff}$, due to its non-equilibrium feature;
In that of Fig.~\ref{fig:three_setups}(c), instead $V_\text{1,eff}$ of Eq.~\eqref{eq:coefficients_definition} should be replaced by $V_\text{1,eq}$, and also ``eff. equil.'' for effective equilibrium should be replaced by ``equil.'', for real equilibrium.
The kinetic coefficients defined by Eq.~\eqref{eq:coefficients_definition} for the setup of Fig.~\ref{fig:three_setups}(a), and similarly for that of Figs.~\ref{fig:three_setups}(b) and \ref{fig:three_setups}(c), are further related to charge conductance $G^s$, heat conductance $\kappa^s$, Seebeck ($\mathcal{S}^s$) and Peltier ($\mathcal{P}^s)$ coefficients (cf. Refs.~\cite{Callen48,BenentiPhysRep17}):
\begin{equation}
\begin{aligned}
    G^s & = L_\text{IV}^s,\quad 
    \kappa^s  = L_\text{JT}^s - \frac{L_\text{JV}^s L_\text{IT}^s}{L_\text{IV}^s},\\
    \mathcal{S}^s & = \frac{L^s_\text{IT}}{L^s_\text{IV}},\quad 
    \mathcal{P}^s  = \frac{L^s_\text{JV}}{L^s_\text{IV}}.
\end{aligned}
\label{eq:transport_parameters_definition}
\end{equation}
Notice that in Eq.~\eqref{eq:transport_parameters_definition}, the Seebeck and Peltier coefficients contain the conductance $G^s$ on the denominator. For convenience, these coefficients can be referred to as the ``off-diagonal'' (i.e., response of charge and heat transport to, reversely, a bias in temperature and electric voltage, respectively) quantum transport coefficients, that are normalized by the ``diagonal'' element, $L^s_\text{IV}$.

\section{``Experimental'': how to realize the setup of Fig.~\ref{fig:three_setups}(a) and what to measure}
\label{sec:experimental_proposal}

%%%%%%%%%%%%%%%%%%%%%%%%%
\begin{figure}[!ht]
  \includegraphics[width= 1.0 \linewidth]{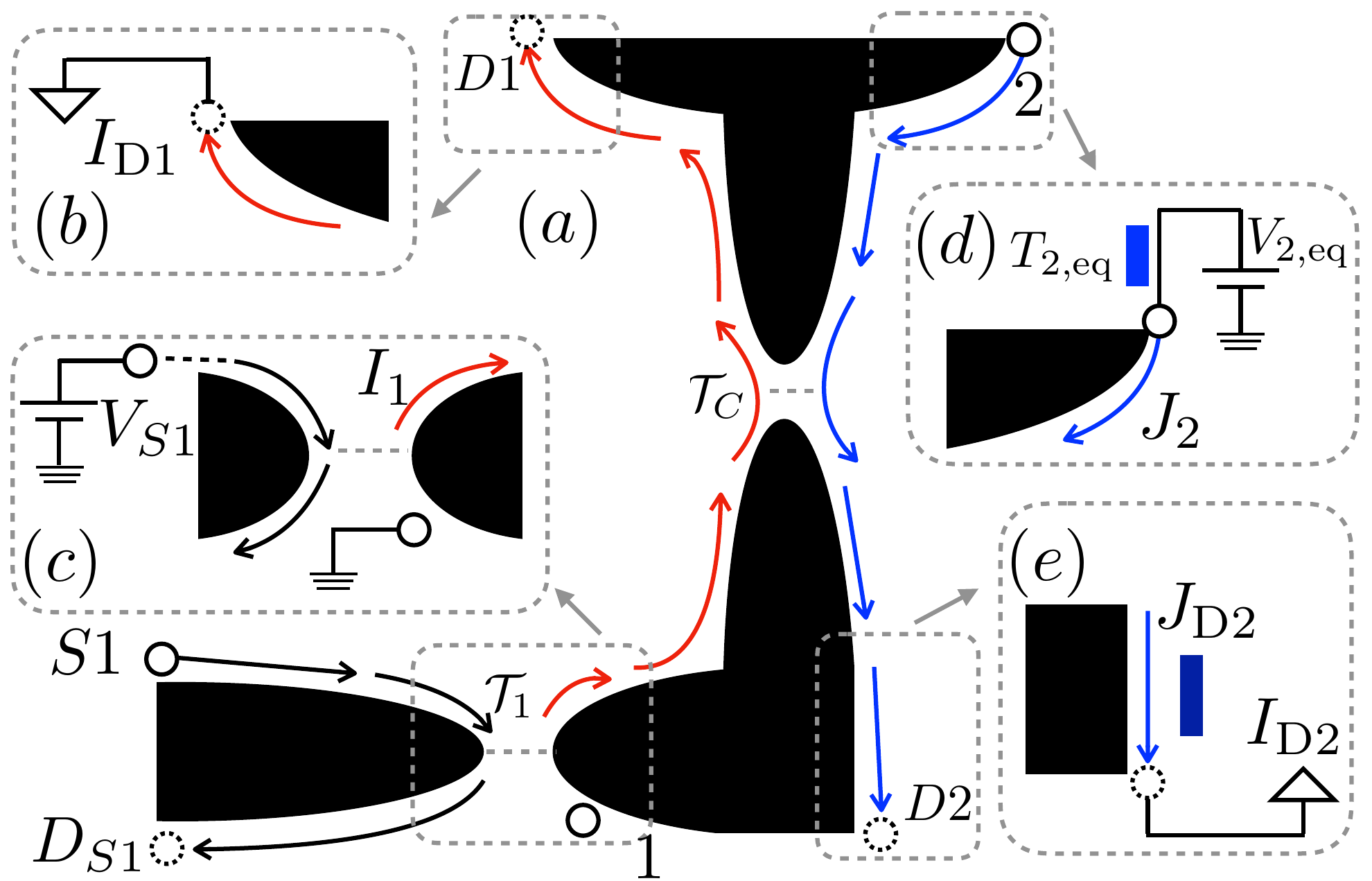}
  \caption{\textbf{The experimental realization of the setup in Fig.~\ref{fig:three_setups}(a).}  
  (a) The full picture of the setup. Here the black areas refer to that influenced by the depleting metallic gates.
  Chiral channels, $S1$, 1, and 2, highlighted respectively by the black, red and blue curves, are thus developed at the boundary of these depleted areas. Here the sources and drains of these edge states are marked out by the solid and dashed circles, respectively.
  The setup contains two QPCs (highlighted by the dashed gray lines): (i) the diluter that connects channels $S1$ and 1, and (ii) the central collider that connects channels 1 and 2.
  Both QPCs have the filling fraction that equals to that of the bulk, thus allow the tunneling of anyons, with weak transmission probabilities $\mathcal{T}_1$ and $\mathcal{T}_C$, respectively.
  These two experimentally accessible quantities, defined by $\mathcal{T}_1 \equiv 2\pi I_1/(q e^2 V_{S1})$ and Eq.~\eqref{eq:tc_relations}, respectively, differ from $\mathcal{T}^{(0)}_1$ and $\mathcal{T}^{(0)}_C$, which denote the squared magnitudes of the tunneling amplitudes, $|\xi_1|^2$ in $H_\text{diluter}$ [Eq.~\eqref{eq:hdiluter}], and $|\xi_C|^2$ in $H_\text{collider}$ [Eq.~\eqref{eq:hcollider}], respectively.
  (b) The current $I_{D1}$ is measured at the drain $D1$ of channel 1.
  (c) The zoom-in of the diluter, where channel 1 receives the current $I_1$ from the source channel $S1$ (biased by the voltage $V_{S1}$) after the diluter.
  (d) The zoom-in of the source of channel 2. Here the blue box refers to the Ohmic contact which tunes the temperature $T_\text{2,eq}$ and the corresponding heat current $J_2$ in channel 2. (e) The zoom-in of the source $D2$ of channel 2. Here the darker-blue box refers to the floating contact that measures the heat current $J_{D2}$ that arrives at the diluter.
  At the drain $D2$, the arriving current $I_{D2}$ is also detected.
  }
  \label{fig:neq-eq_setup}
\end{figure}
%%%%%%%%%%%%%%%%%%%%%%%%%

\subsection{The setup for experimental observations}

Here we illustrate how to realize the mostly focused setup of this work, i.e., that of Fig.~\ref{fig:three_setups}(a), containing an equilibrium channel and a non-equilibrium one, in real experiments. The other two setups, shown by Figs.~\ref{fig:three_setups}(b) and \ref{fig:three_setups}(c), can be realized similarly, by adding one or removing one source channel, respectively.
To begin with, the basic structure is shown by Fig.~\ref{fig:neq-eq_setup}(a). Here the areas colored black refer to that where quasiparticles are depleted by the metallic gates. The white area instead refers to the topologically non-trivial area, containing a constant filling fraction $\nu$.
The boundary of the depleted areas thus supports edge states highlighted by the arrows.
Here the black, red and blue arrows refer to the source channels $S1$, and these middle two channels, 1 and 2, respectively, in the schematic picture Fig.~\ref{fig:three_setups}(a).
Figs.~\ref{fig:neq-eq_setup}(b) to \ref{fig:neq-eq_setup}(e) are the zoom-in plots of the structure near the source, drains, or the diluter.
Here we would like to especially highlight the blue area in Fig.~\ref{fig:neq-eq_setup}(d) and the darker blue area in Fig.~\ref{fig:neq-eq_setup}(e).
They refer to the heater that controls the particle temperature in channel 2, and the floating contact that measures the heat that arrives at the drain, respectively.

Verifying our theoretical predictions requires the knowledge of two sets of measurements: (i) The identification of the effective equilibrium, i.e., the measurement of the effective chemical potential and effective temperature; and (ii) Close to the effective equilibrium, the measurement on the effective linear response coefficients defined by Eq.~\eqref{eq:transport_parameters_definition}.
When all ambient temperatures are vanishingly small, the determination of these quantities requires only the measurement on the charge currents [$I_1$ of Fig.~\ref{fig:neq-eq_setup}(c) and $I_{D1}$ of Fig.~\ref{fig:neq-eq_setup}(b), respectively] and thermal currents [$J_2$ of Fig.~\ref{fig:neq-eq_setup}(d) and $J_{D2}$, respectively].
Consequently, these quantities can in principle be extracted solely from average charge and heat transport measurements, without collider noise or drain-current cross-correlations used in previous experiments~\cite{BartolomeiScience20,PierrePRX23,LeeNature23}.

\subsection{Identifying effective equilibrium parameters in the time-domain braiding limit}
\label{sec:free_parameters}

Now we begin to move to identify the effective parameters.
This task is comparatively simpler in the collision-free limit, where time-domain braiding dominates.
In this limit, due to the particle-hole symmetry [see solid curves of Fig.~\ref{fig:distributions}, and Secs.~\ref{sec:ph_symmetric} for more discussions], the charge and heat transport are uncorrelated, such that the effective chemical potential and effective temperature can be obtained by \textit{independently} tuning the potential and temperature of channel 2.
However, when away from this limit, the direct tunneling of non-equilibrium anyons results in the breaking of particle-hole symmetry, and thus the interplay between charge and heat transport (see dashed curves of Fig.~\ref{fig:distributions}, and Sec.~\ref{sec:ph_symm_breaking} for details).
The identification of the effective equilibrium point thus requires the simultaneous tuning of $V_\text{2,eq}$ and $T_\text{2,eq}$.
Below we illustrate the experimental protocol to identify the effective equilibrium point, for systems within and beyond the collision-free limit, separately.

Identification of \textit{the effective parameters} in the collision-free limit, when ambient temperatures of channels $S1$ and 1 equal zero:\\
1. Pinch-off the central QPC (i.e., set $\mathcal{T}_C = 0$), and measure the current that arrives at $D1$, to obtain $I_1$ right after the diluter;\\
2. Keep $\mathcal{T}_C$ small but finite and measure the current $I_{D1}$ that arrives at $D1$. Generically, $I_{D1} \neq I_1$, given a finite charge current $I_\text{charge}$ that tunnels through the central collider;\\
3. Tune the bias $V_\text{2,eq}$ to change the value of $I_{D1}$.
Assuming that $I_{D1}$ coincides with $I_1$ when $V_\text{2,eq} = V_\text{2,eq}^0$.
The \textit{effective chemical potential} of channel 1 then equals $V_\text{1,eff}^0 = V_\text{2,eq}^0$;\\
4. Fix $V_\text{2,eq} = V_\text{2,eq}^0$ and pinch off the central collider.
For a given temperature of the channel 2, $T_\text{2,eq}$ [tuned by the Ohmic contact shown by the blue box of Fig.~\ref{fig:neq-eq_setup}(d)], measure the heat current that arrives at the floating contact [the darker-blue box of Fig.~\ref{fig:neq-eq_setup}(e)]. This heat current equals $J_2$ when $\mathcal{T}_C = 0$;\\
5. Tune $\mathcal{T}_C$ to be small but finite, and measure the heat current, $J_{D2}$, that arrives at the drain $D2$.
Assuming that $J_{D2}$ coincides with $J_2$ when $T_\text{2,eq} = T_\text{2,eq}^0$.
The \textit{effective temperature} of channel 1 then equals $T_\text{1,eff}^0 = T_\text{2,eq}^0$.

Notice that in the protocol above, the superscript ``0'' of effective parameters, $V_\text{1,eff}^0$ and $T_\text{1,eff}^0$, is added to highlight the fact that these parameters are obtained in the collision-free limit: to distinguish from these beyond the collision-free limit, cf. $V_\text{1,eff}$ and $T_\text{1,eff}$ in Sec.~\ref{sec:parameters_beyond}.
As disclosed by Ref.~\cite{LandscapePRL25}, in the time-domain braiding limit, we have
\begin{equation}
    V_\text{1,eff}^0 = I_1 \sin (\theta)/(e^*)^2,
    \label{eq:veff0}
\end{equation}
for the effective chemical potential.
The corresponding effective temperature equals
\begin{equation}
    T_\text{1,eff}^0 \equiv S_1/[(e^*)^2 s_c(\theta,\delta)],
    \label{eq:teff0}
\end{equation}
with $s_c$ universally [see Eq.~\eqref{eq:eff_temp_definition} of Sec.~\ref{sec:universal_collision_free}] depending on the braiding phase $\theta$ and the scaling dimension $\delta$ (see Sec.~\ref{sec:quests}).
As the reminder, here $I_1$ and $S_1$ refer to the non-equilibrium current and the corresponding noise, respectively, of channel 1.

\subsection{Identifying effective parameters beyond the time-domain braiding limit}
\label{sec:parameters_beyond}

When the transmission through the diluter, $\mathcal{T}_1$, becomes large enough, the direct tunneling of non-equilibrium anyons has to be taken into consideration.
As will be discussed in Sec.~\ref{sec:ph_symm_breaking}, this direct tunneling breaks the symmetry between particle and hole sectors, thus bringing in the interplay between charge and heat transport.
As the consequence, of this case, the identification of the effective equilibrium requires a simultaneous tuning of both $V_\text{2,eq}$ and $T_\text{2,eq}$, following the steps below:\\
1. Tune the potential and temperature of channel 2 to a given set of values, $V_\text{2,eq}$ and $T_\text{2,eq}$;\\
2. Pinch off the central QPC, and measure two quantities: (i) the current $I_1$ that arrives at the drain $D1$, and the heat current $J_2$ that arrives at the drain $D2$;\\
3. Turn on the central QPC such that $\mathcal{T}_C$ is small but finite. Measure two quantities: (i) the current $I_{D1}$ that arrives at the drain $D1$, and the heat current $J_{D2}$ that arrives at the drain $D2$. With measurement outcomes in step 2, we obtain the charge current $I_\text{charge} = I_1 - I_{D1}$ and the heat current $J_\text{heat} = J_{D2} - J_2$;\\
4. Changes the values of $V_\text{2,eq}$ and $T_\text{2,eq}$;\\
5. Based on the measurements, make the plot of $\text{sgn}( I_\text{charge} J_\text{heat} )$ as the function of $V_\text{2,eq}$ and $T_\text{2,eq}$ [see Fig.~\ref{fig:coefficients_collision}(b)].
The effective equilibrium point is thus the ``quadripoint'' of different areas [i.e., the point highlighted by the white star of Fig.~\ref{fig:coefficients_collision}(b)].

Notice that within our consideration, the energy-dependence of the transmission probabilities is not important on the evaluation of effective parameters.
Indeed, within the focus setup, i.e., Fig.~\ref{fig:three_setups}(a), the equilibrium channel serves as the calibrator of the effective parameters of the non-equilibrium channel.
The value of $\mathcal{T}_C$ is thus unimportant, as long as it is small enough such that the QPCs remain within the weak tunneling limit, where the renormalization of the tunneling amplitude is cut off at a sufficiently large infrared scale set by the non-equilibrium current, voltage bias, or system temperature.

\subsection{Protocol to obtain the effective linear response coefficients}
\label{sec:protocol_effective_linear_response}

As among our central achievements, within this work we have developed the framework of effective linear response [see Eqs.~\eqref{eq:coefficients_definition} and \eqref{eq:transport_parameters_definition}], to describe the response of the charge and heat transport ($I_\text{charge}$ and $J_\text{heat}$), to a slight deviation from the effective equilibrium.
The major theoretical results will be provided in Secs.~\ref{sec:collision_free_results} and \ref{sec:beyond_tdb}, for situations within and beyond the collision-free limit.
Here we illustrate in-advance the protocol to experimentally measure these effective linear response coefficients defined in Eq.~\eqref{eq:coefficients_definition}.

Obtaining the effective linear response coefficients, after identifying the effective equilibrium, takes the steps below [cf. the setup of Fig.~\ref{fig:three_setups}(a), and its potential experimental realization in Fig.~\ref{fig:neq-eq_setup}]:\\
1. Tune the parameters of channel 2 to $V_\text{2,eq} = V_\text{1,eff}$ and $T_\text{2,eq} = T_\text{1,eff}$, such that the system arrives at the effective equilibrium;\\
2. Fix the voltage $V_\text{2,eq} = V_\text{1,eff}$, and slightly detune the temperature of channel 2, into $T_\text{2,eq} \to T_\text{1,eff} + \delta T_2$. Then perform the measurements below:\\
2.1 When $\mathcal{T}_C = 0$, measure the charge current $I_1$ and the heat current $J_2$, that arrive at the drain $D1$ and $D2$, respectively;\\
2.2 When $\mathcal{T}_C$ is finite, measure the charge current $I_{D1}$ and the heat current $J_{D2}$, that arrive at the drain $D1$ and $D2$, respectively.
We then obtain the tunneling charge current $I_\text{charge} = I_1 - I_{D1}$ and the tunneling heat current $J_\text{heat} = J_{D2} - J_2$;\\
2.3 We then obtain the effective linear response coefficients, $L_\text{IT}^\text{neq-eq}$ and $L_\text{JT}^\text{neq-eq}$,
\begin{equation}
    L_\text{IT}^\text{neq-eq} = -\lim_{\delta T_2 \to 0} \frac{I_\text{charge}}{\delta T_2},\ L_\text{JT}^\text{neq-eq} = -\lim_{\delta T_2 \to 0} \frac{J_\text{heat}}{\delta T_2};
    \label{eq:lit_ljt}
\end{equation}
3. Fix the temperature $T_\text{2,eq} = T_\text{1,eff}$, and slightly detune the voltage of channel 2, into $V_\text{2,eq} \to V_\text{1,eff} + \delta V_2$. Then perform the measurements below:\\
3.1 When $\mathcal{T}_C = 0$, measure the charge current $I_1$ and the heat current $J_2$, that arrive at the drain $D1$ and $D2$, respectively;\\
3.2 When $\mathcal{T}_C$ is finite, measure the charge current $I_{D1}$ and the heat current $J_{D2}$, that arrive at the drain $D1$ and $D2$, respectively.
We then obtain the tunneling charge current $I_\text{charge} = I_1 - I_{D1}$ and the tunneling heat current $J_\text{heat} = J_{D2} - J_2$;\\
3.3 We then obtain the effective linear response coefficients, $L_\text{IV}^\text{neq-eq}$ and $L_\text{JV}^\text{neq-eq}$,
\begin{equation}
    L_\text{IV}^\text{neq-eq} = -\lim_{\delta V_2 \to 0} \frac{I_\text{charge}}{\delta V_2},\ L_\text{JV}^\text{neq-eq} = -\lim_{\delta V_2 \to 0} \frac{J_\text{heat}}{\delta V_2}.
    \label{eq:liv_ljv}
\end{equation}
Eqs.~\eqref{eq:lit_ljt} and \eqref{eq:liv_ljv} then give us the values of effective linear response coefficients defined in Eq.~\eqref{eq:coefficients_definition}.
Following in-addition Eq.~\eqref{eq:transport_parameters_definition}, we obtain the charge conductance $G^\text{neq-eq}$, the heat conductance $\kappa^\text{neq-eq}$, as well as the Seebeck and Peltier coefficients, $\mathcal{S}^\text{neq-eq}$ and $\mathcal{P}^\text{neq-eq}$, of the setup presented by Fig.~\ref{fig:three_setups}(a).
Actually, one can further define the Lorenz number $\mathcal{L}^\text{neq-eq}$, following Eq.~\eqref{eq:Lorenz_number}, to describe the relation between heat and charge quantum transport.
Notice that protocols above, developed for the setup of Fig.~\ref{fig:three_setups}(b), can be naturally extended to the other two setups of Fig.\ref{fig:three_setups}.

\section{Quest for extracting fractional statistics}
\label{sec:quests}

The development of the effective equilibrium and effective linear response extends the powerful close-to-equilibrium theory, especially the linear response theory to non-equilibrium systems, thus enriching the toolkit in the extraction of the non-trivial information of exotic systems.
As an important example, here we illustrate how to extract fractional braiding phase of chiral Laughlin anyons based on our theoretical results.

\subsection{The quantities on the agenda}

Laughlin anyons become extremely attractive, due to several unique anyonic signatures, including the fractional charge $e^* = q e$, fractional braiding statistics $\theta$, as well as the scaling power factors $\delta$ of a single anyonic operator.
Ideally, these quantities are all in close connection to the filling fraction $\nu$, i.e., $q = \theta/2\pi=2 \delta= \nu$, such that the detection of any of these quantities can immediately disclose the fractional statistical information.
However, practically these quantities do not necessarily coincide, so that
it is impossible to obtain them from a single measurement.
For the convenience of the reader, we outline in Appendix~\ref{app:fractional_signatures} the microscopic origins of these anyonic signatures and how they enter the transport theory.

\subsection{Knowledge of parameters, and ``games'' to exact the anyonic braiding phase}
\label{sec:games}

As outlined above (cf. more discussions in Appendix~\ref{app:fractional_signatures}), the equality $q =\theta/2\pi = 2 \delta = \nu$ should be viewed as a special property of the ideal Laughlin state, not as a definition of $q$, $\theta$, or $\delta$.
Thus, practically the difficulty of obtaining the fractional braiding phase depends on the number of known anyonic signatures, and the relation among their.
Below, depending on this ``difficulty'', we consider four different ``games'': one ``easy'' game, two ``intermediate'' games, and one ``hard'' game.
Within this section, we provide only the conclusion.
Derivation details and discussions are provided from Secs.~\ref{sec:easy} to \ref{sec:hard}.

To begin with, we consider the easy game, conditional on the idea Laughlin relations, $q = \theta/2\pi = 2 \delta = \nu$.
Of this case, one can obtain the fractional braiding phase by simply measuring either the charge conductance $G^s$ or heat conductance $\kappa^s$, of any setup $s$ of Fig.~\ref{fig:three_setups} [see Eq.~\eqref{eq:universal_coefficients}, with the explicit function expressions in Sec.~\ref{sec:gamma_chi_functions}], when the system is in equilibrium, or at the effective equilibrium.
The easy game tests whether different transport observables indeed consistently reproduce the same underlying parameter, whereas in the hard game (see below), $q$, $\theta$ and $\delta$ do not need to satisfy the ideal Laughlin relations.

Now we consider two intermediate games, where it is assumed to have the knowledge of either the fractional charge, $e^*$, or the scaling dimension, $\delta$.
Of the former case, the braiding statistics can be obtained by detecting the effective chemical potential in the collision-free limit [see Eq.~\eqref{eq:intermediate_1}], as in the collision-free limit, the effective chemical potential depends only on the braiding phase $\theta$ and the fractional charge $e^*$. The latter is assumed as known for the first intermediate game.
An alternative intermediate game instead assumes that $\delta$ is known in advance.
Of this case, the braiding phase can be obtained by measuring, in addition to the effective parameters (the effective chemical potential and effective temperature), the Lorenz number $\mathcal{L}^\text{neq-neq}$ [see Eq.~\eqref{eq:intermediate_2}] of the setup of Fig.~\ref{fig:three_setups}(b).

Finally, we move to the hard game, where none of these anyonic signatures is known apriori.
Here the transport theory establishes a mapping,
\begin{equation}
 (q, \theta, \delta) \longrightarrow
\left\{V_\text{1,eff},T_\text{1,eff},G^s,\kappa^s,\mathcal{L}^s,\ldots\right\},   
\end{equation}
from the quasiparticle charge, statistical angle, and scaling exponent to experimentally accessible transport quantities, including the effective parameters ($V_\text{1,eff} $ and $ T_\text{1,eff}$), conductances [$G^s$ and $\kappa^s$ of Eq.~\eqref{eq:transport_parameters_definition}], and the Lorenz number [$\mathcal{L}^s$ of Eq.~\eqref{eq:Lorenz_number}].
The experimental protocol is designed to invert this mapping without imposing the ideal Laughlin relation of the easy game.
Of this case, one can extract the braiding phase by an extra measurement, in addition to that of the second intermediate game, on the Lorenz number $\mathcal{L}^\text{eq-eq}$ of the setup of Fig.~\ref{fig:three_setups}(c) (see Sec.~\ref{sec:hard}).
Indeed, the measurement on $\mathcal{L}^\text{eq-eq}$ discloses us the scaling dimension $\delta$, after which the hard game reduces to the second intermediate game outlined above.

We stress, however, that solutions above to different games are not the only options.
Indeed, even for the hard game, there are only three unknown anyonic signatures, $q$, $\theta$ and $\delta$,
so that in principle they can be extracted by measuring a suitable set of three functionally independent transport observables (say, $O_1, O_2$ and $O_3$), i.e.,
\begin{equation}
    \text{det}\left[ \frac{\partial (O_1,O_2,O_3)}{\partial (q,\theta,\delta)} \right] \neq 0,
\end{equation}
as long as these observables depend universally on these anyonic signatures.
Here, ``universal'' means that, apart from contact normalization and dimensional scales [$\mathcal{T}_C^{(0)} (\pi \bar{T} \tau_0)^{4\delta - 2}$ of e.g., $G^s$ and $\kappa^s$ of Eq.~\eqref{eq:universal_coefficients}], the quasiparticle dependence is fixed solely by $q$, $\theta$, and $\delta$.

\subsection{Limit of the validity of the above
}

From Sec.~\ref{sec:games}, we have illustrated how to obtain the fractional anyonic braiding phase $\theta$ in games with different difficulty.
Crucially, the analysis above relies strongly on the universal features (i.e., the dependence on only the anyonic signatures) of the effective equilibrium parameters, and corresponding effective linear response coefficients.
As will be explained in Sec.~\ref{sec:ph_symmetric} with more details, their requested universal features are direct consequence of the symmetry between particle and hole sectors, and the latter further requires the system to stay in the collision-free limit, where only time-domain braiding processes are involved in the tunneling through the central collider.

Practically, however, central collider of a real experimental setup allows also the direct tunneling of non-equilibrium anyons (see Fig.~\ref{fig:time-domain_and_real}).
This process, which becomes increasingly important upon the enhancement of $\mathcal{T}_1$, will sabotage the particle-hole symmetry (see Fig.~\ref{fig:distributions}), and ensue introducing extra non-universal contributions to these effective linear response coefficients.
As the consequence, it is a practically demanding and meaningful task, to evaluate the importance of the direct tunneling of non-equilibrium anyons, especially their influence on the experimental measurement of anyonic braiding signature.

As will be discussed shortly in Sec.~\ref{sec:collision_seebeck}, in addition to the protocol to identify the fractional anyonic statistics, within this work we suggest further a criteria to identify the importance of the direct tunneling of non-equilibrium anyons, by experimentally checking the amplitude of thermoelectric coefficients (i.e., the Seebeck and Peltier coefficients).
This criteria is chosen, as finite thermoelectric coefficients is a direct consequence of the particle-hole symmetry breaking that is also induced by the direct tunneling of non-equilibrium anyons.
These coefficients are thus more sensitive than the tunneling probability through the diluter ($\mathcal{T}_1$), to manifest the importance of the direct tunneling of non-equilibrium anyons.
This fact is actually shown in Figs.~\ref{fig:coefficients_collision}(c) and \ref{fig:coefficients_collision}(d), where a small tunneling probability at the diluter, $\mathcal{T}_1 \approx 0.1$, suffices to produce a significant enough Seebeck coefficient $e\mathcal{S}^\text{neq-eq} \approx 1$.

\section{Model}
\label{sec:model}

Now we begin to introduce how we model these setups. Here we focus on the setup of Fig.~\ref{fig:three_setups}(a), which consists of one highly non-equilibrium channel (channel 1) and an in-equilibrium one (channel 2).
The modelings and correlation functions of these two channels can be easily extended to the highly non-equilibrium channels of the setup of Figs.~\ref{fig:three_setups}(b), and the equilibrium ones of \ref{fig:three_setups}(c), respectively.

As the starting point, the setup of Fig.~\ref{fig:three_setups}(a) consists of three chiral channel, $S1$, 1 and 2.
In the absence of QPCs, they are described by the Hamiltonian of free chiral Luttinger liquid channels, $H_\text{edge}$,
\begin{equation}
    H_\text{edge} = \frac{v}{4\pi} \sum_\alpha\int dx\, (\partial_x \phi_\alpha)^2,
\end{equation} 
where $v$ is the quasiparticle velocity and $\phi_\alpha$ is the bosonic mode of chiral channel $\alpha = S1,\, 1,\, 2$, obeying the commutation relation $[\phi_\alpha (x),\phi_{\alpha'} (x')] =i\pi\, \delta_{\alpha,\alpha'}\, \text{sgn}(x-x')$.

With these bosonic operators, the Hamiltonian of the diluter at $x=0$ can be written as
\begin{align}
    H_\text{diluter} =  \frac{\xi_1}{2\pi\tau_0} 
        e^{ -i\sqrt{\nu}\, \left[\phi_{S1} (0) - \phi_1 (0) \right]}    + \text{H.c.},
        \label{eq:hdiluter}
\end{align}
with $\sqrt{\nu}$ the prefactor of the bosonic field in the vertex operator, following the theory of the Laughlin edge~\cite{WenPRB90}.
Alternatively, the tunneling through the diluter can be written in terms of the anyonic operators $\psi^\dagger_{S1}$ and $\psi_1$ of the Laughlin channels with filling fraction $\nu$ as
$H_\text{diluter} = v \xi_1 \,\psi^\dagger_{S1} (0) \psi_1 (0)  + \text{H.c.}$, with $\xi_1$ the tunneling amplitude at the diluter.
In Eq.~\eqref{eq:hdiluter}, $\tau_0$ refers to the short-time cutoff.
Notice that different from the non-interacting fermionic situation, in anyonic systems the tunneling amplitude square, i.e., $\mathcal{T}_1^{(0)} \equiv |\xi_1|^2$,  generically differs from the transmission probability that is actually obtained in real experiments (cf. Sec.~\ref{eq:dilution} for more discussions).
This is due to the celebrated zero-bias anomaly that occurs widely in interacting systems.
Indeed, the non-equilibrium current $I_1$, generated due to the transmission through the diluter, equals
\begin{equation}
    I_1 =  (e^*)^2 V_{S1} \mathcal{T}_1^{(0)} \frac{\sin(\theta) \Gamma (1 - 4\delta)}{2\pi^2} (e^* V_{S1} \tau_0)^{4\delta-2},
    \label{eq:i1}
\end{equation}
where the scaling factor, i.e., $(e^* V_{S1} \tau_0)^{4\delta-2}$, refers to the chiral Luttinger liquid anomaly, and $\Gamma$ the gamma function.
This factor equals simply one for non-interacting fermions (where $\nu =1$).

At the central collider, the corresponding tunneling Hamiltonian is given by
\begin{align}
    H_\text{collider} &= \frac{\xi_C}{2\pi\tau_0} 
    e^{ -i\sqrt{\nu} \left[\phi_{1} (L) - \phi_{2} (L) \right] } + \text{H.c.}\notag \\
    &=  v \xi_C\, \psi^\dagger_1 (L) \psi_2 (L)  + \text{H.c.},
    \label{eq:hcollider}
\end{align}
where, as in Eq.~\eqref{eq:hdiluter}, the vertex operator has $\sqrt{\nu}$ as the prefactor of the bosonic field. 
In the ideal Laughlin case, this implies $q = 2 \delta = \theta/2 \pi = \nu$.
However, we distinguish among these quantities, as they involve, in general, different matrices (cf. Appendix~\ref{app:fractional_signatures}).
In Eq.~\eqref{eq:hcollider}, $\xi_C$ is the tunneling amplitude at the collider, 
which determines the corresponding bare transmission probability, $\mathcal{T}_C^{(0)}$ [cf. the prefactor of Eq.~\eqref{eq:currents_correlations} for the charge and heat current], via $\mathcal{T}_C^{(0)} = |\xi_C|^2$.
The experimentally accessible transmission probability through the central collider, $\mathcal{T}_C$, is actually proportional to the bare one $\mathcal{T}_C^{(0)}$ [defined after Eq.~\eqref{eq:hcollider}], i.e.,
\begin{equation}
    \mathcal{T}_C \equiv \frac{2\pi q}{(e^*)^2} \frac{\partial I_\text{charge}}{\partial \delta V} \propto \mathcal{T}_C^{(0)} ,
\label{eq:tc_relations}
\end{equation}
where $\delta V  = V_\text{1,eff} - V_\text{2,eq}$ refers to the difference in potentials (of channels 1 and 2), of the setup of Fig.~\ref{fig:three_setups}(a).
Notice that this definition of $\mathcal{T}_C$ can be easily generalized into the setups of Figs.~\ref{fig:three_setups}(b) and \ref{fig:three_setups}(c), where $\delta V$ instead equals $V_\text{1,eff} - V_\text{2,eff}$ and $V_\text{1,eq} - V_\text{2,eq}$, respectively.

Finally, following Fig.~\ref{fig:neq-eq_setup}, channels $S1$ and 2 contain a voltage bias, $V_{S1}$ and $V_\text{2,eq}$, respectively, with respect to channel 1, before the diluter.
Here we follow the technique introduced by e.g., Ref.~\cite{KaneFisherPRB92}, to include these two voltage biases by two phase factors that enter the correlation functions of channel 1 [see Eq.~\eqref{eq:neq_correlations}] and 2 [see Eq.~\eqref{eq:equilibrium_correlation}], respectively.

\section{Correlation functions}
\label{sec:correlation_functions}

Following Eq.~\eqref{eq:currents_correlations}, the evaluation of both the charge ($I_\text{charge}$) and heat ($J_\text{heat}$) currents requires the knowledge of the correlation functions at the central collider.
Again we focus on the setup of Fig.~\ref{fig:three_setups}(a),
where the composing channels 1 and 2 are strongly non-equilibrium and in-equilibrium, respectively.
Their correlation functions thus apply to both channels in the setups of Figs.~\ref{fig:three_setups}(b) and \ref{fig:three_setups}(c), respectively.

To begin with, the equilibrium channel 2 has a real chemical potential $V_{2,\text{eq}}$ and a real temperature $T_{2,\text{eq}}$ that both enter the correlation functions,
\begin{equation}
\begin{aligned}
    &\langle \psi^\dagger_2 (L,t) \psi_2 (L,0) \rangle_\text{eq} \!=\! \frac{ e^{i e^* V_{2,\text{eq}} t}}{2\pi v \tau_0} \, \frac{(\pi T_{2,\text{eq}} \tau_0)^{2\delta}}{\sin^{2\delta} \left[\pi T_{2,\text{eq}} (t-i\tau_0)\right]}\,,\\
    &\langle \psi_2 (L,t) \psi^\dagger_2 (L,0) \rangle_\text{eq} \!= \!\frac{e^{-i e^* V_{2,\text{eq}} t}}{2\pi v \tau_0}\,  \frac{(\pi T_{2,\text{eq}}\tau_0)^{2\delta}}{\sin^{2\delta} \left[\pi T_{2,\text{eq}} (t-i\tau_0)\right]} ,
\end{aligned}
    \label{eq:equilibrium_correlation}
\end{equation}
where $\tau_0$ is the short-time cutoff, as the reminder.
For the non-equilibrium channel 1, correlation functions are given by (see Supplementary Secs.~S1 to S3 for the derivation details)
\begin{equation}
\begin{aligned}
    & \langle \psi^\dagger_1 (L,t) \psi_1 (L,0) \rangle_\text{neq} 
     =\mathcal{D}_1 \, \frac{e^{ie^* V_{S1} t}}{2\pi v \tau_0} \,\frac{(\pi T_{S1} \tau_0)^{2\delta} }{\sin^{2\delta} [\pi T_{S1} (\tau_0 \!+\! i t)]} \\
   &\ \ + \frac{g(q,\theta, t)}{2\pi v \tau_0} \frac{\tau_0^{2\delta}}{(\tau_0 + i t)^{2\delta}} \left[ 1 + i\,\mathcal{D}_1 C  \,\sin\left(\frac{\theta}{2}\right) \,\text{sgn}(t) \right],
    \\
    & \langle \psi_1 (L,t) \psi^\dagger_1 (L,0) \rangle_\text{neq}  
     =\mathcal{D}_1 \, \frac{e^{-i e^* V_{S1} t}}{2\pi v \tau_0} \,\frac{(\pi T_{S1} \tau_0)^{2\delta} }{\sin^{2\delta} [\pi T_{S1} (\tau_0 \!+\! i t)]} \\
   &\ \ + \frac{g(q,\theta, -t)}{2\pi v \tau_0} \frac{\tau_0^{2\delta}}{(\tau_0 + i t)^{2\delta}} \left[ 1 - i\,\mathcal{D}_1 C  \,\sin \left(\frac{\theta}{2}\right) \,\text{sgn}(t) \right],\\
   & C \equiv \frac{4\pi\delta \  \Gamma (2-4\delta)}{\sin\left(\frac{\theta}{2}\right) \Gamma^2 ( 1 - 2\delta) },
\end{aligned}
\label{eq:neq_correlations}
\end{equation}
where $\mathcal{D}_1$, with its expression
\begin{equation*}
    \mathcal{D}_1  
    = \mathcal{T}_1^{(0)}  \frac{\Gamma^2 (1-2\delta) \sin^2 \left(\frac{\theta}{2}\right)}{\pi^2} (e^* V_{S1} \tau_0)^{4\delta-2},
    \label{eq:dilution}
\end{equation*}
is related to the experimentally accessible transmission probability, $\mathcal{T}_1 \equiv 2\pi I_1/(q e^2 V_{S1})$, via
\begin{equation}
    \mathcal{T}_1 = \mathcal{D}_1 \frac{2\pi q \Gamma (1 - 4\delta)}{\tan \left(\frac{\theta}{2}\right) \Gamma^2 (1 - 2\delta)}.
\label{eq:d1_t1}
\end{equation}
Eq.~\eqref{eq:dilution} also contains the same scaling factor $(e^* V_{S1} \tau_0)^{4\delta-2}$, as that of the non-equilibrium current $I_1$ [cf. Eq.~\eqref{eq:i1}].

In Eq.~\eqref{eq:neq_correlations}, 
The braiding processes at the collider are encoded in correlators (\ref{eq:neq_correlations}) through the function ($e^* = q e$, as the reminder)
\begin{align}
    g(q, \theta, t) \!\equiv\! \exp\left\{\! 
    -\frac{ [1 \!-\! \cos(\theta)] S_1|t|}{(e^*)^2}\! + \!\frac{ i\sin(\theta)  I_1  t }{e^*}\! \right\},
    \label{g-braid}
\end{align}
where the imaginary and real parts of the exponential factor depend on the tunneling current $I_1$ at the diluter and the corresponding noise $S_1$, respectively.
Note that the non-equilibrium correlation functions [Eq.~\eqref{eq:neq_correlations}] are \textit{not} simply given by introducing $T_\text{1,eff}$ into the expression for the equilibrium case [Eq.~\eqref{eq:equilibrium_correlation}].
Notably, following Eq.~\eqref{g-braid}, the factor $I_1 \sin (\theta)/e^*$, being proportional to $I_1$, controls the point of particle-hole symmetry. 
The other factor, $2S_1 \sin^2 (\theta/2)/(e^*)^2$, on the other hand determines the width of correlation functions.
With zero ambient temperatures, $S_1 = e^* I_1$, and these two quantities become strongly correlated.

The time-domain braiding contribution to the correlators [the terms with unity in the curly brackets in Eqs.~\eqref{eq:neq_correlations}] becomes dominant in the strongly dilute limit $\mathcal{T}_1 \ll 1$, where corrections from anyon collisions are comparatively small~\cite{LeeSimNC22,schillerPRL23,LeeNature23}.
As introduced in Sec.~\ref{sec:intro}, within this work we address this limit as ``collision-free'', to highlight the exclusion [setting $\mathcal{T}_1 =0$ in Eq.~\eqref{eq:neq_correlations}] of both the tunneling of anyons supplied by the diluter [first terms in Eq.~\eqref{eq:neq_correlations}] and collisions among them (second terms in curly brackets). The transmission coefficient of the diluter and the source bias voltage $V_{S1}$ enter the remaining terms,  
corresponding to time-domain braiding, only implicitly---through $S_1$ and $I_1$.

We can further define \textit{effective} particle and hole distribution functions as the Fourier transform of the correlation functions 
\begin{equation}
\begin{aligned}
    n_{\text{p},\alpha}(\epsilon) & \equiv v \int dt\, e^{-i\epsilon t} \langle \psi_\alpha^\dagger (t) \psi_\alpha (0)\rangle,
    \\
    n_{\text{h},\alpha}(\epsilon) & \equiv v \int dt\, e^{i\epsilon t} \langle \psi_\alpha (t) \psi_\alpha^\dagger (0)\rangle,
    \label{eq:distribution_definition}
\end{aligned}
\end{equation}
for anyonic channels 1 and 2,
with $v$ the quasiparticle transport velocity.
Formally, they are the Fourier transform of lesser and greater Green's functions, respectively.
We define these two effective distribution functions for two reasons.
Firstly, with these functions defined, the tunneling current (charge and heat), given by Eq.~\eqref{eq:currents_correlations}, can be alternatively written as
\begin{equation}
\begin{aligned}
    &I_\text{charge} \equiv e^*\,\mathcal{T}_C^{(0)} \!\int \! \frac{d\epsilon}{2\pi} \, [ n_\text{p,1} (\epsilon) n_\text{h,2} (\epsilon) - n_\text{p,2} (\epsilon) n_\text{h,1} (\epsilon)],\\
    &J_\text{heat} \! \equiv\!  \mathcal{T}_C^{(0)}\!\!\int \! \frac{d\epsilon}{2\pi}\! \, (\epsilon\! -\! e^* \bar{V}) [ n_\text{p,1} (\epsilon) n_\text{h,2} (\epsilon)\! -\! n_\text{p,2} (\epsilon) n_\text{h,1} (\epsilon)],
\end{aligned}
\label{eq:currents}
\end{equation}
where $\bar{V} = (V_\text{1,eff} + V_\text{2,eq})/2$ is the average of the chemical potentials in channels 1 and 2.
For the other two setups, $\bar{V}$ instead equals $(V_\text{1,eff} + V_\text{2,eff})/2$, and $(V_\text{1,eq} + V_\text{2,eq})/2$, respectively.

In comparison to that of Eq.~\eqref{eq:currents_correlations}, Eq.~\eqref{eq:currents} has two advantages. Firstly, it is physically more transparent, as it indicates that each tunneling event at the central collider effectively carries the charge $e^*$ and the energy of the quasiparticle, respectively.
In addition, the definition of Eq.~\eqref{eq:currents} enables us to define the particle-hole symmetry of the system.
More explicitly, a system is defined to display the particle-hole symmetry, if there exists an energy, with respect to which the effective particle and hole distribution functions of both involved channels (1 and 2) are symmetric.
Importantly, the presence of this particle-hole symmetry turns out to be the key factor that leads to both the absence of thermoelectric coefficients, and the universal feature of effective parameters, as well as these effective linear response coefficients.
More detailed discussions on the particle-hole symmetry of different cases will be provided in Secs.~\ref{sec:ph_symmetric} and \ref{sec:ph_symm_breaking}, concerning the situation within and beyond the collision-free limit, respectively.

\section{Effective linear response coefficients in the collision-free limit}
\label{sec:collision_free_results}

Now we begin to evaluate the effective linear response coefficients of the setup of Fig.~\ref{fig:three_setups}(a), starting with the analysis in the collision-free limit [see Eqs.~\eqref{eq:veff0} and \eqref{eq:teff0} featuring the corresponding effective equilibrium point].
As the reminder, within this limit $\mathcal{T}_1 \ll 1$, where non-equilibrium anyons are forbidden to directly tunnel through and collide at the central collider.
Their influence is later-on included in Sec.~\ref{sec:beyond_tdb}.

\subsection{Simplified correlation functions in the collision-free limit}

In the collision-free limit, the non-equilibrium and equilibrium correlators, Eqs.~\eqref{eq:equilibrium_correlation} and \eqref{eq:neq_correlations}, respectively, can be written into a uniform manner:
\begin{equation}
\begin{aligned}
& \langle  \psi_2^\dagger (t) \psi_2 (0)\rangle_\text{eq} \equiv f_2(t)\, e^{i e^* V_\text{2,eq} t } , 
\\
&\langle \psi_2 (t) \psi_2^\dagger (0) \rangle_\text{eq} \equiv f_2(t)\, e^{-i e^* V_\text{2,eq} t },
\end{aligned}
\label{eq:f2_correlations}
\end{equation}
and
\begin{equation}
\begin{aligned}
&\langle \psi_1^\dagger (t) \psi_1 (0)\rangle_\text{neq}\simeq \frac{g(q,\theta, t)}{2\pi v \tau_0} \frac{\tau_0^{2\delta}}{(\tau_0 + i t)^{2\delta}} \equiv f_\alpha(t) e^{ie^* V_\text{1,eff}^0 t } ,
\\
&\langle \psi_1 (t) \psi_1^\dagger (0)\rangle_\text{neq}\!\simeq\! \frac{g(q,\theta, -t)}{2\pi v \tau_0} \frac{\tau_0^{2\delta}}{(\tau_0 + i t)^{2\delta}}\!\!\equiv \!\! f_\alpha(t) e^{-i e^* V_\text{1,eff}^0 t },
\end{aligned}
\label{eq:f1_f2_correlations}
\end{equation}
where functions $f_\alpha(t)$ and $f_2(t)$ [see Eqs.~\eqref{eq:f1s} and \eqref{eq:f2}, respectively]
are independent of the potentials  $V_\text{1,eff}^0$ [see Eq.~\eqref{eq:veff0} for the effective chemical potential when time-domain braiding dominates] and $V_\text{2,eq}$. 
In this work, the non-equilibrium current $I_1$ and the associated tunneling noise $S_1$ are theoretically considered two explicitly independent variables, although experimentally they are directly correlated, especially when ambient temperatures of channels $S1$ and 1 equal zero. 
In the collision-free limit, they characterize the effective chemical potential and the effective temperature, respectively. 
To avoid possible complications related to the dependence of $I_1$ and $S_1$ on the same experimentally accessible knobs (e.g., $\mathcal{T}_1$ and $V_{S1}$), measuring effective linear response coefficients [i.e., Eq.~\eqref{eq:coefficients_definition}]
can be conveniently performed by measuring the response of currents $I_\text{charge}$ and $J_\text{heat}$ to changes in potential or temperature of the equilibrium channel 2.

\subsection{Simplified expressions of currents and effective linear response coefficients}
\label{sec:response_collision_free}

With correlation functions given by Eqs.~\eqref{eq:f2_correlations} and \eqref{eq:f1_f2_correlations}, we are ready to evaluate the currents and their corresponding effective linear response functions, close to the effective equilibrium.
Here we start with the charge current $I_\text{charge}$ and its corresponding response functions,
\begin{equation}
\begin{aligned}
    I_\text{charge} & = \!2 i e^* \mathcal{T}_C^{(0)}\! \int dt \,f_\alpha(t) f_2(t)\,  \sin [e^* (V_\text{1,eff}^0 - V_\text{2,eq}) t],\\
    L_\text{IV}^\text{neq-eq} & \!\! = \!2 i (e^*)^2 \mathcal{T}_C^{(0)}\!\!\! \!\int\!\! dt \ \! t \! \,f_\alpha(t) f_2(t) \cos [e^* (V_\text{1,eff}^0 - V_\text{2,eq}) t] ,\\
    L_\text{IT}^\text{neq-eq}     &\!\! = \!2 i e^* \mathcal{T}_C^{(0)}\! \!\! \!\int \!\! dt \sin [e^* (V_\text{1,eff}^0 \!-\!\! V_\text{2,eq}) t] \frac{\partial [f_\alpha (t) f_2(t)]}{\partial (T_\text{1,eff}^0 \!-\! T_\text{2,eq})}\! .
\end{aligned}
\label{eq:it_and_lit}
\end{equation}
Here, $L_\text{IV}^\text{neq-eq} = G^\text{neq-eq}$ is the tunneling conductance at the collider.
Note that $L_\text{IT}^\text{neq-eq}$ given by Eq.~\eqref{eq:it_and_lit} vanishes, since by definition $ V_\text{2,eq} = V_\text{1,eff}^0$ at effective equilibrium.
The Seebeck coefficient $\mathcal{S}^\text{neq-eq}$, which is proportional to $L_\text{IT}^\text{neq-eq}$ [cf. Eq.~\eqref{eq:transport_parameters_definition} for its definition] also equals zero.

The heat current and its corresponding response functions are instead expressed as
\begin{equation}
\begin{aligned}
    J_\text{heat}  & =  - i \mathcal{T}_C^{(0)} \int dt\,\left[ f_\alpha' (t) f_2 (t) - f_\alpha (t)  f_2' (t) \right] \\
    &\quad \times \cos\!\big[ e^* (V_\text{1,eff}^0 - V_\text{2,eq}) t\big], \\
    L_\text{JT}^\text{neq-eq}  & =  - i \mathcal{T}_C^{(0)} \int dt\,\frac{\partial \left[ f_\alpha' (t) f_2 (t) - f_\alpha (t)  f_2' (t) \right]}{\partial (T_\text{1,eff}^0 - T_\text{2,eq})} \\
    &\quad \times \cos\!\big[ e^* (V_\text{1,eff}^0 - V_\text{2,eq}) t\big], \\
   L_\text{JV}^\text{neq-eq} &= i e^* \mathcal{T}_C^{(0)} \int dt \, t \, \left[ f_\alpha' (t) f_2 (t) - f_\alpha (t)  f_2' (t) \right]\\
   & \quad \times \sin\!\big[ e^* (V_\text{1,eff}^0 - V_\text{2,eq}) t\big],
\end{aligned}
\label{eq:jq_and_lvj}
\end{equation}
where $f_\alpha'(t) \equiv \partial_t f_\alpha (t)$ and $f_2'(t) \equiv \partial_t f_2 (t)$.
Both $L_\text{JV}^\text{neq-eq}$ and the associated Peltier coefficient, $\mathcal{P}^\text{neq-eq}$, 
vanish in the collision-free limit.
Notice that the vanishing of both $\mathcal{S}^\text{neq-eq}$ and $\mathcal{P}^\text{neq-eq}$ is arrived for all functions $f_\alpha$ and $f_2$ that are independent of (real or effective) potentials.
As will be discussed in Sec.~\ref{sec:ph_symmetric}, this conclusion is the direct consequence of the particle-hole symmetry for a system within the collision-free limit.
Remarkably, this conclusion is not limited to the specific setup of Fig.~\ref{fig:three_setups}(a), but remains valid for the other setups [e.g., those of Figs.~\ref{fig:three_setups}(b) and \ref{fig:three_setups}(c)] where correlation functions indeed share the structure of Eq.~\eqref{eq:f1_f2_correlations}.
The differences among these setups are only implicitly contained in explicit expressions of the functions $f_\alpha$ and $f_2$.
This observation immediately leads to the conclusion that in all three setups of Fig.~\ref{fig:three_setups}, the thermoelectric coefficients become zero as long as the particle-hole symmetry is present.

Finally, notice that $L_\text{IV}^\text{neq-eq}$ and $L_\text{JT}^\text{neq-eq}$ of Eqs.~\eqref{eq:it_and_lit} and \eqref{eq:jq_and_lvj} are both finite, and correspond to the charge conductance $G^\text{neq-eq}$ and heat conductance $\kappa^\text{neq-eq}$ [cf. their definitions in Eq.~\eqref{eq:transport_parameters_definition}], respectively.
Again, these two integrals can be straightforwardly extended to the other two setups, i.e., these of Figs.~\ref{fig:three_setups}(b) and \ref{fig:three_setups}(c), after two replacement: (i) replacing the expressions of $f_\alpha$ and $f_2$ [i.e., Eqs.~\eqref{eq:f2_correlations} and \eqref{eq:f1_f2_correlations}] for the setup of Fig.~\ref{fig:three_setups}(a), by that of the other two setups; and (ii) replacing differences in potentials ($V_\text{1,eff}^0 - V_\text{2,eq}$) and temperatures ($T_\text{1,eff}^0 - T_\text{2,eq}$) by the setup-specific ones, i.e., by $V_\text{1,eff}^0 - V_\text{2,eff}$ and $T_\text{1,eff}^0 - T_\text{2,eff}$ for Fig.~\ref{fig:three_setups}(b), while by $V_\text{1,eq} - V_\text{2,eq}$ and $T_\text{1,eq} - T_\text{2,eq}$ for Fig.~\ref{fig:three_setups}(c), respectively.
Here we choose to present $G^s$ and $\kappa^s$ for all the setups (with the label $s$) in the next section, to discuss the universal feature of both of them, as well as the corresponding Lorenz number.

\subsection{Lorenz number in the collision-free limit}
\label{sec:universal_collision_free}

As stated above, the particle-hole symmetry is not a unique feature of the setup of Fig.~\ref{fig:three_setups}(a) in the collision-free limit.
Indeed, it is also featured in the setup of Fig.~\ref{fig:three_setups}(b), in the collision-free limit, and in that of Fig.~\ref{fig:three_setups}(c).
Unlike the setup of Fig.~\ref{fig:three_setups}(a), where channels 1 and 2 are non-equilibrium and equilibrium, respectively, for the other two setups, these channels are either both out of equilibrium or both at equilibrium, respectively.
It is thus a rewarding task to compare the transport features of all three setups in the presence of particle-hole symmetry.

To characterize all three setups within a uniform framework, we define the Lorenz number,
\begin{equation}
    \mathcal{L}^{s} \equiv \frac{(e^*)^2 \kappa^s}{\bar{T} G^s },
\label{eq:Lorenz_number}
\end{equation}
as the ratio between the heat and charge conductances (the definition remains valid beyond the collision-free limit).
Here, the superscript $s$ refers to neq-eq, neq-neq, and eq-eq setups of Figs.~\ref{fig:three_setups}(a), \ref{fig:three_setups}(b), and \ref{fig:three_setups}(c), respectively.
The factor $(e^*)^2$ in the numerator of Eq.~\eqref{eq:Lorenz_number} renders $\mathcal{L}$ dimensionless and compensates for the fractional charge of Laughlin quasiparticles.

Within the collision-free limit, for all three setups of Fig.~\ref{fig:three_setups}, the charge conductance, heat conductance, and the Lorenz number can be written into a universal form, as
\begin{equation}
\begin{aligned}
    G^s & = (e^*)^2\,  \gamma^s \left(\theta,\delta \right)\, \mathcal{T}_C^{(0)} \, (\pi \bar{T} \tau_0)^{4\delta-2},\\
     \kappa^s   & =  \chi^s\left(\theta,\delta \right)\, \mathcal{T}_C^{(0)} \, (\pi \bar{T}\tau_0)^{4\delta - 2}\,  \bar{T},\\
     \mathcal{L}^s & =  \frac{\chi^s \left(\theta,\delta \right)}{\gamma^s \left(\theta,\delta \right)},
\end{aligned}
\label{eq:universal_coefficients}
\end{equation}
where the differences among these setups are completely encoded by the two functions $\gamma^s$ and $\chi^s$ [see Eqs.~\eqref{eq:g_f_neq_eq} to  \eqref{eq:g_f_eq_eq} for their explicit expressions of all three setups].
Again, the factor $(\pi \bar{T} \tau_0)^{4\delta-2}$ refers to the Luttinger liquid anomaly that is the same for all three setups.

Following Eq.~\eqref{eq:universal_coefficients}, in all three cases, the Lorenz number is a function of only two (out of three) anyonic signatures, i.e., the fractional braiding phase $\theta$ and the scaling dimension $\delta$ of a single anyonic operator. In particular, for the neq-neq and eq-eq cases, the explicit expressions for it take the following compact forms:
\begin{align}
     \mathcal{L}^\text{neq-neq}(\theta, \delta)  & = 4 s_c^2(\theta,\delta) \sin^4\left(\frac{\theta}{2}\right)/(1 - 4\delta),
     \label{lorenz-number-neq-neq}\\
     \mathcal{L}^\text{eq-eq}(\delta) & = {4\pi^2 \delta^2}/(1 + 4\delta) ,\label{lorenz-numbers-eq-eq}
\end{align}
where $s_c(\theta,\delta) \equiv S_1/[(e^*)^2 T_\text{1,eff}^0]$ [cf. Eq.~\eqref{eq:teff0}] is the solution of the integral equation
\begin{equation}
    \int dt' \frac{e^{-\frac{s_c}{\pi}2 \sin^2 (\theta/2) |t'|} \cot (\tau_0' + i t')}{(\tau_0' + it')^{2\delta} \sin^{2 \delta} (\tau_0' + it')}   = 0,
    \label{eq:eff_temp_definition}
\end{equation}
which contains only $\delta$ and $\theta$, and the positive infinitesimal $\tau_0' \equiv \tau_0 \pi T_\text{1,eff}^0$ that is introduced to avoid singularities of the integral.
Note that, for the integer case, $\mathcal{L}^\text{eq-eq}(\nu = 1)$ correctly reduces to the value for non-interacting fermionic systems, $\mathcal{L}_\text{fermion} = \pi^2 /3$, as it should be.

A close look at the transport coefficients $G^s$, $\kappa^s$, and $\mathcal{L}^s$ [Eqs.~\eqref{eq:universal_coefficients} to \eqref{lorenz-numbers-eq-eq}], as well as the functions $\gamma^s(\theta,\delta)$ and $\chi^s(\theta,\delta)$ [Eq.~\eqref{eq:g_f_neq_eq} to \eqref{eq:g_f_eq_eq} in Sec.~\ref{sec:gamma_chi_functions}] reveals the following interesting observation: in the $s =$\,neq-neq and neq-eq setups, these coefficients depend  \textit{explicitly} on the effective charge ($e^*$ for $G^s$), scaling dimension $\delta$, and the braiding phase $\theta$. However, the dependence on both the effective charge and the renormalization factor [cf. $(\pi \bar{T} \tau_0)^{4\delta-2}$ of Eq.~\eqref{eq:universal_coefficients}, arising from the scaling dimension] cancel out in the Lorenz number defined by Eq.~\eqref{eq:Lorenz_number}.
By contrast, in the eq-eq scenario, the braiding phase does not show up. This last observation is a consequence of the fact that no time-domain braiding takes place for the eq-eq case (cf. Refs.~\cite{RosenowLevkivskyiHalperinPRL16,LeeSimNC22,schillerPRL23,LeeNature23,LandscapePRL25}, where non-equilibrium anyonic beams are the prerequisites of time-domain braiding). The remaining dependence of $\mathcal{L}^\text{eq-eq}$ on $\delta$ [Eq.~\eqref{lorenz-numbers-eq-eq}] is due to the residual (after the cancellation of scaling dimension) effect of anyonic correlations near the Laughlin surface.

%%%%%%%%%%%%%%%%%%%%%%%%%
\begin{figure}[!ht]
  \includegraphics[width= 1 \linewidth]{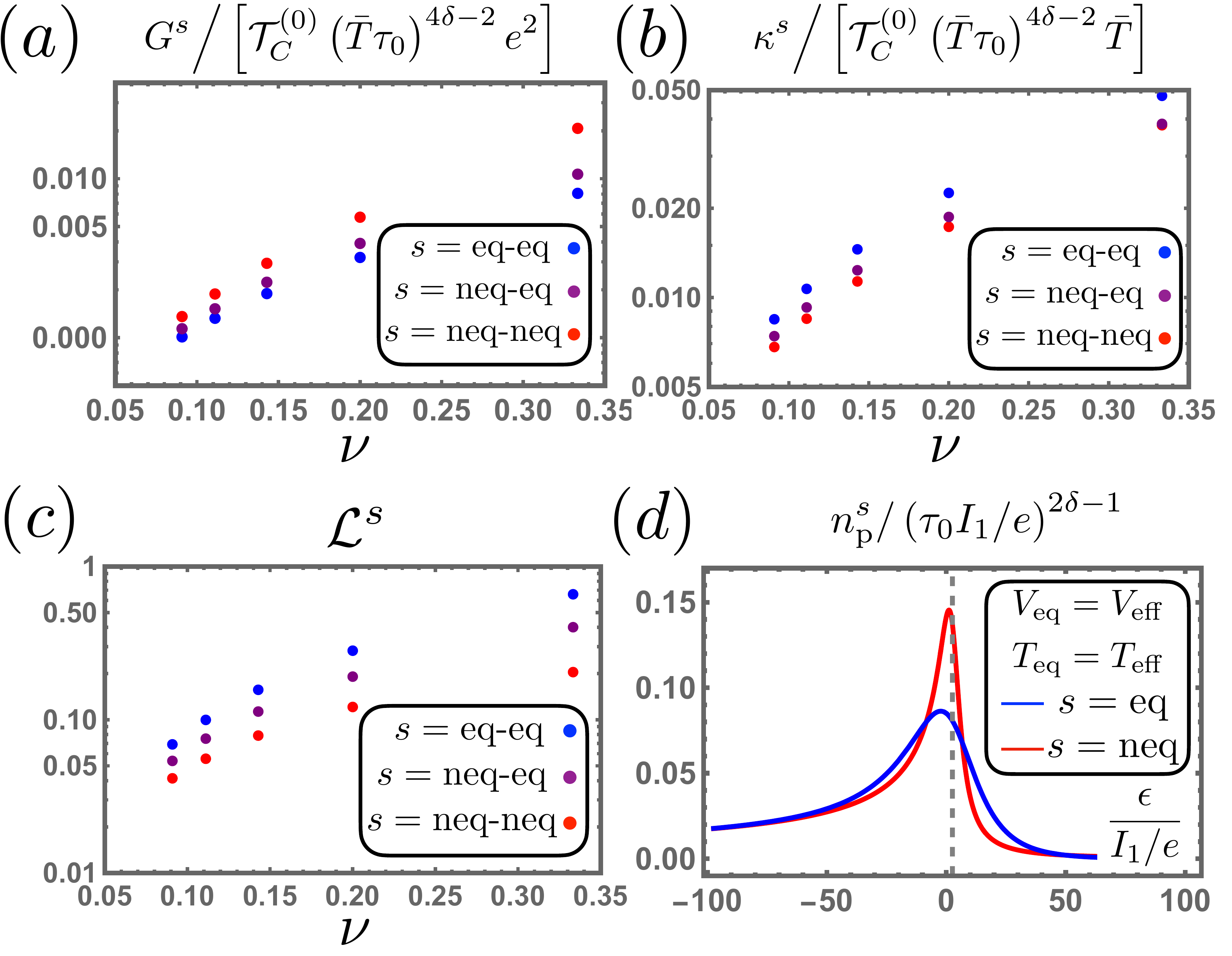}
  \caption{
  \textbf{Charge and heat conductances in the effective linear response regime, and collision-free limit.}
  In addition, we assume the ideal case where $q = \theta/2\pi=2\delta = \nu$.
  The magenta, red, and blue symbols correspond to setups shown in Figs.~\ref{fig:three_setups}(a), \ref{fig:three_setups}(b) and \ref{fig:three_setups}(c), respectively.
  For non-equilibrium channels, we exclude collisions of non-equilibrium anyons. The normalized charge conductance (a), normalized heat conductance (b), and the Lorenz number (c) are shown as functions of the filling fraction. Panel (d) shows the \textit{effective} particle distribution functions, $n_\text{p}^s$, for equilibrium ($s$=eq, blue curve) and non-equilibrium ($s$=neq, red curve) anyonic channels that have the same (effective or real) potential and temperature.
  }
  \label{fig:thermoelectric_constants_Lorenz}
\end{figure}
%%%%%%%%%%%%%%%%%%%%%%%%%

In Fig.~\ref{fig:thermoelectric_constants_Lorenz},
we present $G^s$, $\kappa^s$ [defined by Eq.~\eqref{eq:coefficients_definition}], and the corresponding Lorenz number $\mathcal{L}^s$ in the effective linear response regime, for all three setups, as functions of the filling fraction $\nu$, for the ideal case where $q=\theta/2\pi=2\delta=\nu$.
We observe that all three quantities ($G^s$, $\kappa^s$ and $\mathcal{L}^s$) increase with increasing filling fraction $\nu$.
The Lorenz number in all three setups differs from $\mathcal{L}_\text{fermion} \approx 3.29$ characteristic of non-interacting fermions.
This deviation originates from the crucial differences between the Fermi and Laughlin surfaces. 
This should be contrasted with the results of  Refs.~\cite{DurstAdamLeePRB00,KubalaPRL08,GargPRL09,LavasaniPRB19,GiulianoIOP2022,MajidiNanoLett2022}, where the deviation from $\mathcal{L}_\text{fermion}$ occurred from impurities or interactions.

Furthermore, the effect of time-domain braiding between anyons [for the setups of Figs.~\ref{fig:three_setups}(a) and \ref{fig:three_setups}(b)] is revealed by the difference between braiding-influenced Lorenz numbers ($\mathcal{L}^\text{neq-eq} $ and $\mathcal{L}^\text{neq-neq} $), and the braiding-irrelevant one, $\mathcal{L}^\text{eq-eq} $. 
Thus, the Lorenz number is capable of disclosing the braiding feature of Laughlin anyons. 
More specifically, all three quantities that correspond to the setup of Fig.~\ref{fig:three_setups}(a), involving the tunneling between an equilibrium channel and the non-equilibrium one, are between those of the other two setups [Figs.~\ref{fig:three_setups}(b) and \ref{fig:three_setups}(c)], where both channels are either in or out-of equilibrium.
This setup dependence of transport coefficients manifests the difference between effective and real parameters.

Actually, this setup dependence can be more intuitively understood by comparing the effective distribution functions of an equilibrium channel, and that of an out-of-equilibrium one.
As shown in Fig.~\ref{fig:thermoelectric_constants_Lorenz}(d), the effective particle distribution in a non-equilibrium channel (the red curve) is more heavily concentrated around the effective Laughlin surface (the gray dashed line).
A change of the effective chemical potential would thus induce a greater modification of the electric current, corresponding to a larger electric conductance.
However, since the heat current contains an extra factor $\epsilon - e^* \bar{V}$, which vanishes near the Laughlin surface (i.e., when $\epsilon \approx e^* \bar{V}$),
this concentration of particles near the Laughlin surface reduces the heat conductance of a non-equilibrium channel.

\subsection{The particle-hole symmetry in the time-domain braiding limit}
\label{sec:ph_symmetric}

Secs.~\ref{sec:response_collision_free} and \ref{sec:universal_collision_free} disclose two key features of a system in the collision-free limit, i.e., the vanishing of the thermoelectric coefficients, and the universal feature of the effective linear response coefficients, respectively.
Crucially, as the key message of this section, both features are the direct consequence of the particle-hole symmetry of the system in the collision-free limit.

We start with the reminder that a channel is defined to possess the particle-hole symmetry, if there exists an energy, with respect to which the effective particle and hole distribution functions, defined by Eq.~\eqref{eq:distribution_definition}, are symmetric.
Following this definition, this particle-hole symmetry is present for all equilibrium channels.
Indeed, following Eqs.~\eqref{eq:equilibrium_correlation} and \eqref{eq:distribution_definition}, the effective particle and hole distribution functions of an equilibrium channel, e.g., channel 2 of the setup of Fig.~\ref{fig:three_setups}(a), have the explicit finite-temperature expressions (with $\epsilon$ the energy of the corresponding quasiparticle state),
\begin{equation}
\begin{aligned}
    n_\text{p,2} & = \frac{1}{2\pi \Gamma (2\delta)} \left( \frac{1}{2\pi T_\text{2,eq} \tau_0} \right)^{1-2\delta} \exp \left(- \frac{\epsilon - e^* V_\text{2,eq}}{2 T_\text{2,eq}} \right)\\
    &\quad \times \Bigg| \Gamma \left( \delta + i \frac{\epsilon - e^* V_\text{2,eq}}{2 \pi T_\text{2,eq}} \right) \Bigg|^2,\\
    n_\text{h,2} & = \frac{1}{2\pi \Gamma (2\delta)} \left( \frac{1}{2\pi T_\text{2,eq} \tau_0} \right)^{1-2\delta} \exp \left( \frac{\epsilon - e^* V_\text{2,eq}}{2 T_\text{2,eq}} \right)\\
    &\quad \times \Bigg| \Gamma \left( \delta + i \frac{\epsilon - e^* V_\text{2,eq}}{2 \pi T_\text{2,eq}} \right) \Bigg|^2,
\end{aligned}
\label{eq:distributions_equi}
\end{equation}
which are clearly symmetric with respect to the energy of the ``Laughlin surface'', i.e., at $\epsilon_0 \equiv e^* V_\text{2,eq}$. Indeed, $n_\text{p,2}(\epsilon) = n_\text{h,2} (2e^* V_\text{2,eq} - \epsilon)$ following Eq.~\eqref{eq:distributions_equi}.

The situation however becomes more complicated for a channel that is strongly out-of-equilibrium, e.g., channel 1 of the setup of Fig.~\ref{fig:three_setups}(a).
Firstly, within the collision-free limit (allowing only time-domain braiding processes at the central collider), following Eqs.~\eqref{eq:distribution_definition} and \eqref{eq:f1_f2_correlations}, the corresponding effective particle and hole distribution functions have the explicit expressions
\begin{equation}
\begin{aligned}
    &n_\text{p,1} \big|_\text{TDB} = \frac{ \Gamma (1-2\delta)}{\pi} \tau_0^{2\delta - 1} \\
    &\text{Re} \left\{ e^{i\theta/4} \left[ \frac{S_1}{(e^*)^2} 2 \sin^2 (\theta/2) + i \frac{I_1}{e^*} \sin (\theta) - i \epsilon \right]^{2\delta-1}\right\},\\
    &n_\text{h,1} \big|_\text{TDB} = \frac{ \Gamma (1-2\delta)}{\pi} \tau_0^{2\delta - 1} \\
    &\text{Re} \left\{ e^{-i\theta/4} \left[ \frac{S_1}{(e^*)^2} 2 \sin^2 (\theta/2) + i \frac{I_1}{e^*} \sin (\theta) - i \epsilon \right]^{2\delta-1}\right\},
\end{aligned}
\label{eq:distributions_time-domain}
\end{equation}
with the subscript ``TDB'' the abbreviation of ``time-domain braiding'', the only process allowed in the collision-free limit.
Following Eq.~\eqref{eq:distributions_time-domain}, within this limit,
the particle-hole symmetry remains present in a strongly out-of-equilibrium channel, i.e., $n_\text{p,1}|_\text{TDB}(\epsilon) = n_\text{h,1}|_\text{TDB} (2e^* V_\text{1,eff}^0 - \epsilon)$, with, as the reminder, $ e^* V_\text{1,eff}^0 = I_1 \sin (\theta)/e^*$ 
the new symmetric energy, in terms of
the effective chemical potential $V_\text{1,eff}^0$ [i.e., Eq.~\eqref{eq:veff0}] in the collision-free limit.

Further, when the symmetric energies of both channels coincide, i.e., $e^* V_\text{2,eq} = e^* V_\text{1,eff}^0$, the entire system arrives at the symmetry between effective particle and hole distribution functions, with respect to the same energy.
This symmetry, importantly, is the very reason that leads to the vanishing of $L_\text{IT}^s$ and $L_\text{JV}^s$ [cf. Eqs.~\eqref{eq:it_and_lit} and \eqref{eq:jq_and_lvj} for $s\ =$ neq-eq], and correspondingly the Seebeck and Peltier coefficients [cf. Eq.~\eqref{eq:transport_parameters_definition}].
Indeed, it is known that the generation of thermoelectric behavior requires the breaking of particle-hole symmetry in quantum systems.
Within the framework of chiral Luttinger liquid, due to the linear dispersion of quasiparticles (leading to a constant in-channel single-particle density of states for the linear chiral branch), the symmetry between effective particle-hole distribution functions is equivalent to the particle-hole symmetry in quantum transport, thus leading to the vanishing of thermoelectric coefficients.
Noticeably, the missing of the thermoelectric coefficients further implies the missing of interplay between charge transport and thermal bias, as well as that between thermal transport and electric bias.
This is the very reason that in Sec.~\ref{sec:free_parameters} one can identify the effective equilibrium point by sequentially tuning the voltage ($V_\text{2,eq}$) and temperature ($T_\text{2,eq}$) of the equilibrium channel, 2.

As another crucial consequence of this particle-hole symmetry, when ambient temperatures equal zero, $S_1 = e^* I_1$, the system contains only one relevant energy scale, i.e., $V_\text{2,eq} = V_\text{1,eff}^0 \propto I_1/(e^*)^2$.
The existence of a single energy scale is the very reason that results into the universal features of (charge and heat) conductances [see Eq.~\eqref{eq:transport_parameters_definition}] and also the Lorenz number.
Here the particle-hole symmetry is crucial since otherwise the symmetry-breaking factor, i.e., the direct tunneling and collision of non-equilibrium anyons, will inevitably introduce another relevant energy scale (cf. Sec.~\ref{sec:ph_symm_breaking} for more details), thus sabotaging the universal feature of effective linear response coefficients.

We have thus arrived at the conclusion that for the setup of Fig.~\ref{fig:three_setups}(a), within the collision-free limit where the central QPC allows only time-domain braiding tunneling processes, the system displays the particle-hole symmetry when the effective chemical potential of channel 1 coincides with the chemical potential of channel 2.
This particle-hole symmetry further induces two important features, i.e., (i) the absence of thermoelectric coefficients, and (ii) the existence of a single relevant energy scale within our consideration.
The latter, importantly, further leads to the universal feature of a zoo of quantum transport coefficients, which further provide us extra powerful instruments in the detection of a fractional anyonic braiding phase (cf. Sec.~\ref{sec:phase_three_games}).
Such relation between the particle-hole symmetry, the presence of a single energy scale, and the universal feature agrees perfectly with Ref.~\cite{PavlovKiselevPRL26}, which focuses instead on equilibrium systems.
Our theory is however beyond Ref.~\cite{PavlovKiselevPRL26}, considering the involvement of both the non-equilibrium nature, as well anyonic braiding within our setup.

Before ending this section, we stress again that the conclusions of this section, concerning the relation between particle-hole symmetry, the absence of thermoelectric coefficients, as well as the universal features of quantum transport coefficients (including the effective equilibrium parameters and effective linear response coefficients), apply perfectly to all three setups of Fig.~\ref{fig:three_setups}, although we are taking the correlation functions of the mostly focused setup, i.e., that of Fig.~\ref{fig:three_setups}(a), as the example.
Indeed, once within the collision-free limit, the effective equilibrium of all these setups perfectly satisfies the particle-hole symmetry.

\section{Extracting the fractional braiding phase in the collision-free limit}
\label{sec:phase_three_games}

In Sec.~\ref{sec:games}, we have described how to extract the anyonic braiding phase by detecting the quantum transport parameters that characterize the system at, or slightly away from the effective equilibrium.
There, based on the complexity of potential experiments, we have categorized the situation into ``games'' with different difficulties.
Below we further illustrate the protocol to extra this braiding phase in these games.
Briefly, the number of measurements increases, upon increasing the game difficulty.
Before discussing these games explicitly, we emphasize that, within either game, the fractional braiding phase $\theta$ can be inferred from measurements of only the average charge or heat current. In particular, the statistics-sensitive readout does not require collider noise or drain-current cross-correlations, in contrast to the approaches of Refs.~\cite{BartolomeiScience20,PierrePRX23,LeeNature23}.

\subsection{The ``easy'' game}
\label{sec:easy}

As the reminder, by ``easy'', we refer to the situation where $e^* = \nu e$, $\theta = 2\pi\nu$ and $\delta = \nu/2$, meaning that all three anyonic signatures coincide.
This game is denoted as easy, as all anyonic signatures can be obtained simultaneously, by measuring any universal observable.
For instance, of this case, $\theta$ can be obtained by measuring either the charge or heat conductance [see Eq.~\eqref{eq:universal_coefficients}, with functions $\gamma^s$ and $\chi^s$ provided by Eqs.~\eqref{eq:g_f_neq_eq} to \eqref{eq:g_f_eq_eq}] at the collider.
Notice that one can further check the validity of the obtained $\theta$, or to verify whether or not the system is indeed within the ``easy'' scenario, by checking if $\theta$ obtained from different measured quantities, e.g., $G^s$, $\kappa^s$, or the Lorenz numbers $\mathcal{L}^s$, coincide with each other or not.

\subsection{The intermediate games}
\label{sec:intermediate}

Now we consider two intermediate games.
As outlined in Sec.~\ref{sec:games}, here we consider two cases, where either the fractional charge $e^* = q e$ or the fractional scaling dimension $\delta$ is assumed as known.

Firstly, when the fractional charge $e^*$ is pre-assumed as known, it is optimal to extract the braiding phase by detecting the effective chemical potential.
Indeed, as disclosed by Eq.~\eqref{eq:veff0}, within the collision-free limit,
\begin{equation}
    \sin (\theta) = \frac{(e^*)^2 V_\text{1,eff}^0}{I_1},
    \label{eq:intermediate_1}
\end{equation}
such that measuring the ration between the effective chemical potential $V_\text{1,eff}^0$ and the non-equilibrium current $I_1$ immediate discloses the value of $\theta$.
Importantly, for Laughlin quasiparticles, we can focus on the sector where $0 < \theta < \pi$ when solving Eq.~\eqref{eq:intermediate_1}.

As the second intermediate game, the scaling dimension $\delta$ is assumed as known.
Of this case, the braiding phase can be obtained by measuring, in addition to the collision-free effective parameters [$V_\text{1,eff}^0$ and $T_\text{1,eff}^0$, cf. Eqs.~\eqref{eq:veff0} and \eqref{eq:teff0}, respectively], also the Lorenz number, $\mathcal{L}^\text{neq-neq}$ [cf. Eq.~\eqref{lorenz-number-neq-neq}], of the setup of Fig.~\ref{fig:three_setups}(b).
Indeed, for zero ambient temperatures, we arrive at the relation
\begin{equation}
    \sin^2 \left( \frac{\theta}{2} \right) \tan \left( \frac{\theta}{2} \right) = \frac{1-4\delta}{2} \mathcal{L}^\text{neq-neq} \frac{I_1}{V_\text{1,eff}^0} \left( \frac{T_\text{1,eff}^0}{I_1} \right)^2.
    \label{eq:intermediate_2}
\end{equation}
Following Eq.~\eqref{eq:intermediate_2}, $\theta$ can be identified after obtaining the quantities on its right-hand side, i.e., $V_\text{1,eff}^0/I_1$, $T_\text{1,eff}^0/I_1$ and $\mathcal{L}^\text{neq-neq}$, as $\delta$ is assumed as known for this game.
Again, $0 < \theta < \pi$ for Laughlin quasiparticles.

\subsection{The ``hard'' game}
\label{sec:hard}

Finally, we move to consider the hard game, where it is assumed that none of the fractional quantities, or their relations is known apriori.
Of this case, on the right-hand-side of Eq.~\eqref{eq:intermediate_2}, $\delta$ remains as unknown.
This quantity can be obtained via
an extra measurement, e.g., on the Lorenz number $\mathcal{L}^\text{eq-eq}$ [cf. Eq.~\eqref{lorenz-numbers-eq-eq}], corresponding to the setup of Fig.~\ref{fig:three_setups}(c).
This option is taken, since the value of $\mathcal{L}^\text{eq-eq}$ of Eq.~\eqref{lorenz-numbers-eq-eq} depends on only the fractional scaling dimension $\delta$.
Physically, anyonic braiding is irrelevant in the setup of Fig.~\ref{fig:three_setups}(c), due to the equilibrium nature of both channels 1 and 2.
Consequently, the measurement on the corresponding Lorenz number $\mathcal{L}^\text{eq-eq}$ can immediately inform us of the value of $\delta$,
\begin{equation*}
    \mathcal{L}^\text{eq-eq} \longrightarrow \delta.
\end{equation*}
With this measurement, the hard game reduces to the second immediate game discussed in Sec.~\ref{sec:intermediate}, where the braiding phase, $\theta$, can be obtained following Eq.~\eqref{eq:intermediate_2}, schematically represented by the mapping
\begin{equation*}
    (V_\text{1,eff}^0, T_\text{1,eff}^0, \mathcal{L}^\text{neq-neq};\delta) \longrightarrow \theta.
\end{equation*}
With $\delta$ and $\theta$ obtained, the fractional charge number, $q$, can be obtained via e.g., Eq.~\eqref{eq:intermediate_1}, or schematically,
\begin{equation*}
    (V_\text{1,eff}^0, I_1;\theta) \to q.
\end{equation*}
We have thus provided a potential protocol to experimentally obtain all fractional signatures of a Laughlin system, by measuring effective equilibrium and effective linear response parameters.
This protocol, importantly, remains applicable to a ``hard'' game where the relation among anyonic signatures is not \textit{a priori} known.

%%%%%%%%%%%%%%%%%%%%%%%%%
\begin{figure}[!ht]
  \includegraphics[width= 1 \linewidth]{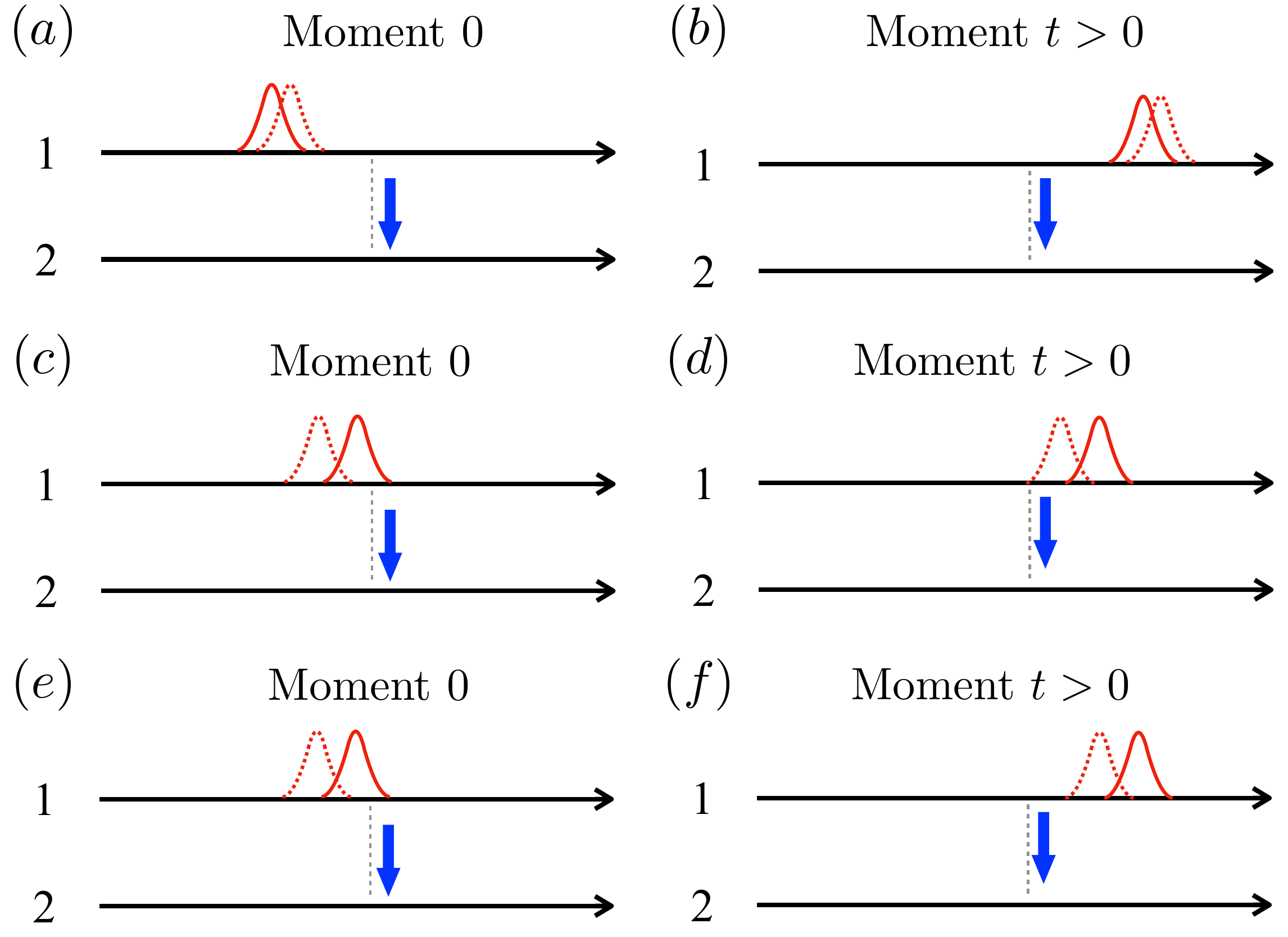}
  \caption{
  \textbf{Three processes that are relevant for evaluating the correlation between two tunneling events (at the central collider), denoted as initial and final states, that occur at moments 0 and $t > 0$, respectively.}
  Here we show only the middle two channels, 1 and 2.
  Channel 1 becomes non-equilibrium, with anyons received from the source channel $S1$ (not shown).
  Non-equilibrium anyons relevant for the initial and final states are
  represented by the solid and dashed red pulses, respectively. The tunneling is shown by the blue arrow at the central collider. (a) and (b) Present the position of non-equilibrium anyons for a time-domain braiding process. (c) and (d) Present the fermion-like processes (cf. Sec.~\ref{sec:direct_processes}), where tunnelings at moments 0 and $t>0$ are both correlated by non-equilibrium anyons. (e) and (f) Present another anyonic unique process (cf. Sec.~\ref{sec:direct_processes}), where only one tunneling events, i.e., the one at moment 0, is correlated with a non-equilibrium anyon.
  }
  \label{fig:time-domain_and_real}
\end{figure}
%%%%%%%%%%%%%%%%%%%%%%%%%

\section{Beyond the collision-free limit}
\label{sec:beyond_tdb}

As shown in the previous section, when the system stays within the collision-free limit, the system displays a symmetry between effective particle and hole distributions, when the system arrives at the effective equilibrium.
This symmetry further results in the universal features of the effective parameters and the effective linear response coefficients, as well as the vanishing of thermoelectric coefficient.
In this section we move beyond the collision-free limit, by introducing the direct tunneling and collision of non-equilibrium anyons, at the central collider.

\subsection{Direct tunneling of non-equilibrium anyons
}
\label{sec:direct_processes}

As the reminder, by ``collision-free'', we refer to the extremely strongly diluted limit $\mathcal{T}_1 \ll 1$ where the tunneling at the central collider allows only time-domain braiding processes.
More specifically, we consider the correlation function between two states, denoted for convenience as initial and final states, where tunneling through the central collider occurs at moments 0 and $t > 0$, respectively.
Crucially, as shown by Figs.~\ref{fig:time-domain_and_real}(a) and \ref{fig:time-domain_and_real}(b), within a time-domain braiding process, tunnelings at the central collider (the blue arrow), at either moments 0 and $t$, is uncorrelated with any non-equilibrium anyon, with the latter represented by the solid and dashed red pulses, for the quasiparticle relevant to the initial and final states, respectively.
Nevertheless, we stress that although the tunneling events appear seemly uncorrelated in Figs.~\ref{fig:time-domain_and_real}(a) and \ref{fig:time-domain_and_real}(b), they are counter-intuitively correlated non-locally in time. This arises from time-domain braiding between the tunnelings at the central collider (the blue arrow), and the non-equilibrium anyons (the red pulses) that bypass the collider.
This is distinct from non-interacting fermionic systems, where these tunneling events at the collider are indeed uncorrelated (in contrast to the anyonic case outlined above), thus leading to a vanishing contribution to observables.

In realistic experimental setups, two extra processes are however inevitable at the central collider. In great contrast to that of time-domain braiding processes, within these two processes the tunneling events at the central collider can correlate, in space-time, with non-equilibrium anyons in channel 1.
To more transparently manifest their difference from the time-domain braiding process, we illustrate the corresponding locations of non-equilibrium anyons of these two processes, in Figs.~\ref{fig:time-domain_and_real}(c) to \ref{fig:time-domain_and_real}(f).

As the first process, following Figs.~\ref{fig:time-domain_and_real}(c) and \ref{fig:time-domain_and_real}(d), for both the initial and final states, the tunneling at the central collider (that occurs at moments 0 and $t$, respectively) is triggered by a tunneling-correlated non-equilibrium anyon.
Here we denote this process as ``fermion-like'', as it is indeed what can only occur in non-interacting fermionic systems.
As its highlighted feature, the non-equilibrium anyon is removed in channel 1 after a ``fermion-like'' direct tunneling at the central collider.
Anyonic braiding will thus not occur afterwards, even after including higher-order tunnelings through the diluter.
This is distinct to time-domain braiding and the other direct tunneling outlined below, where braiding phase can be generated after the resummation over all orders of tunneling at the diluter.
In the correlation function in-time, Eq.~\eqref{eq:neq_correlations}, the fermion-like tunneling contributes to the first line of both correlation functions, $\langle \psi^\dagger_1 (L,t) \psi_1 (L,0) \rangle_\text{neq} $ and $\langle \psi_1 (L,t) \psi^\dagger_1 (L,0) \rangle_\text{neq} $. Indeed, the first line of both correlation functions are basically the product between $\mathcal{T}_1$, the transmission probability through the diluter, and the corresponding equilibrium correlation function, Eq.~\eqref{eq:equilibrium_correlation}, after replacing the characterizing parameters $T_\text{2,eq}$ and $V_\text{2,eq}$ of channel 2 by $T_{S1}$ and $V_{S1}$ of the source channel, $S1$.

Interestingly, in addition to the ``fermion-like'' tunneling process outlined above, a non-equilibrium anyon can directly tunnel through the central collider, through an extra anyonic unique tunneling process.
It can be considered as a combination between the time-domain braiding process, and the fermion-like process.
Indeed, as shown in Figs.~\ref{fig:time-domain_and_real}(e) and \ref{fig:time-domain_and_real}(f), of the initial state, the tunneling through the central collider is correlated with an incoming non-equilibrium anyon (the solid curves), like that in a fermion-like tunneling [see Fig.~\ref{fig:time-domain_and_real}(c)].
As for the final state, the tunneling event is however fully uncorrelated with the non-equilibrium anyon (the dashed curves), mimicking that of a time-domain braiding process [see Fig.~\ref{fig:time-domain_and_real}(b)].
Importantly, since the non-equilibrium anyon of the final state is not in direct connection to a tunneling event (at the central collider), the anyonic braiding phase can be generated, after including higher-order tunneling through the diluter.
Indeed, in the full in-time correlation functions, Eq.~\eqref{eq:neq_correlations}, the term corresponding to this extra tunneling process (i.e., the term proportional to $\mathcal{T}_1 C$) contains the function $g(q,\theta,t)$, where the braiding phase is contained within.
Nevertheless, we stress that the generation of the braiding phase is in-principle different, from the physical perspective, in comparison to the time-domain braiding.
Also notice that in Eq.~\eqref{eq:neq_correlations}, a prefactor $\sin (\theta/2)$ appears in front of $\mathcal{D}_1 C$, indicating that this extra type of direct tunneling is absent in non-interacting fermionic systems, where $\theta = 2\pi$.

With these two extra processes taken into consideration, after the Fourier transformation of Eq.~\eqref{eq:neq_correlations}, the effective particle and hole distribution functions now become
\begin{widetext}
\begin{equation}
\begin{aligned}
    n_{\text{p,1}} & =n_{\text{p,1}}\big|_\text{TDB} + n_{\text{p,1}}^\text{direct} \\
    & \approx \mathcal{T}_1 \left\{ \frac{(2\pi T_{S1} \tau_0)^{2\delta - 1}}{2 \pi \Gamma (2\delta)} \exp\left( \frac{ e^* V_{S1} - \epsilon}{2 T_{S1}} \right) \Bigg| \Gamma\left( \delta + i \frac{ e^* V_{S1} - \epsilon}{2\pi T_{S1}} \right) \Bigg|^2 \right\}  \\
    & +  \frac{\Gamma (1 - 2\delta)}{\pi} \tau_0^{ 2\delta - 1}
    \left( \mathcal{T}_1 C \sin(\theta/2) \, \text{Im}+\text{Re}\right)  \left\{ e^{i\theta/4} \left[ \frac{S_1}{(e^*)^2} \left[1 - \cos (\theta) \right] + i \frac{I_1}{e^*} \sin(\theta)  - i \epsilon \right]^{2\delta-1} \right\}    \\
    n_{\text{h},1} & = n_{\text{h},1}\big|_\text{TDB} + n_{\text{h},1}^\text{direct} \\
      & \approx \mathcal{T}_1 \left\{ \frac{(2\pi T_{S1} \tau_0)^{2\delta - 1}}{2 \pi \Gamma (2\delta)} \exp\left(- \frac{e^* V_{S1} - \epsilon}{2 T_{S1}} \right) \Bigg| \Gamma\left( \delta - i \frac{ e^* V_{S1} - \epsilon}{2\pi T_{S1}} \right) \Bigg|^2 \right\}  \\
    & + \frac{\Gamma (1 - 2\delta)}{\pi} \tau_0^{ 2\delta - 1} 
    \left( \mathcal{T}_1 C \sin(\theta/2)\, \text{Im}+\text{Re}\right)
    \left\{ e^{-i\theta/4} \left\{ \frac{S_1}{(e^*)^2} \left[1 - \cos(\theta) \right] + i \frac{I_1}{e^*} \sin (\theta) - i \epsilon \right\}^{2\delta-1} \right\},
\end{aligned}
\label{eq:full_neq_distributions}
\end{equation}
\end{widetext}
with the superscript ``direct'' referring to the contribution from the direct tunneling and collision of non-equilibrium anyons, manifested by the terms that are explicitly proportional to $\mathcal{T}_1$.
Indeed, Eq.~\eqref{eq:full_neq_distributions} returns to Eq.~\eqref{eq:distributions_time-domain}, the expressions within the collision-free limit, after setting $\mathcal{T}_1$ to zero.
More explicitly, the second line of both $n_\text{p,1}$ and $n_\text{h,1}$ refers to the contribution from fermion-like tunneling process.
The other term explicitly proportional to $\mathcal{T}_1$ instead refers to the last process outlined above.
As anticipated above, only the latter process contains a correction from time-domain braiding, after including higher-order tunnelings through the diluter.
It is absent in the former one, as channel 1 returns to ``equilibrium'', after the non-equilibrium anyon has left channel 1.

%%%%%%%%%%%%%%%%%%%%%%%%%
\begin{figure}[ht!]
  \includegraphics[width= 0.99 \linewidth]{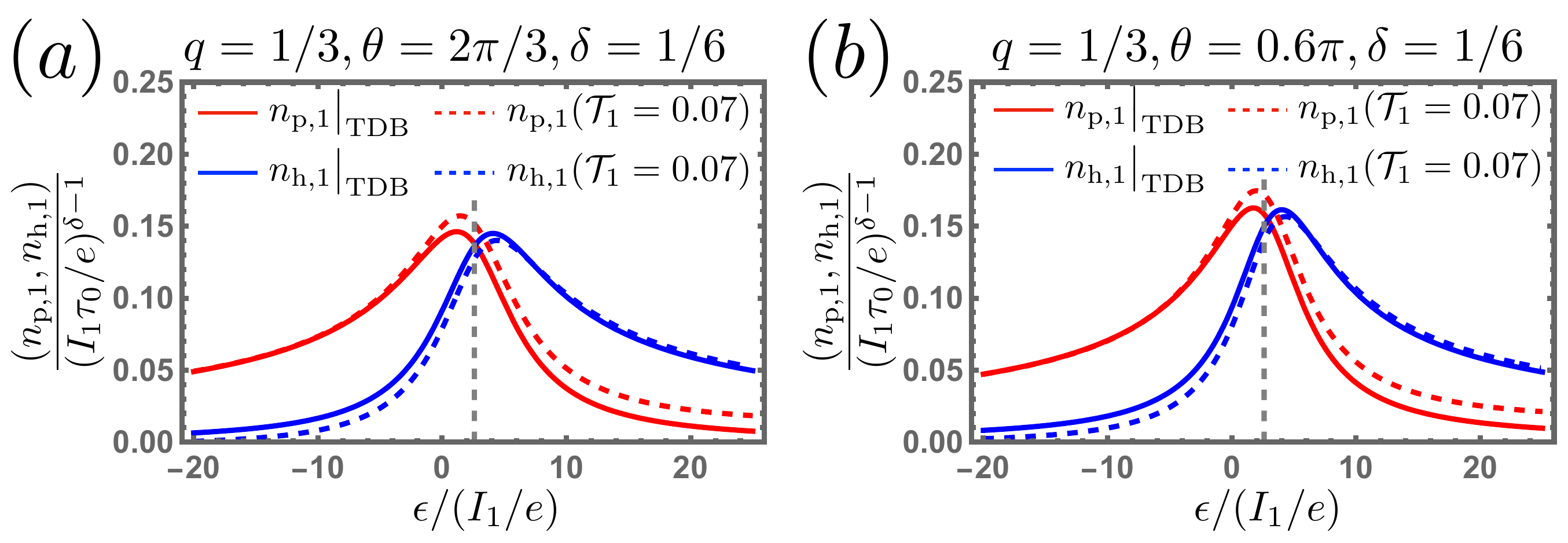}
  \caption{\textbf{Plots of effective particle (the red curves) and hole (the blue curves) distribution functions}. Here the solid and dashed curves correspond to the effective particle and hole distribution functions [cf. Eq.~\eqref{eq:distribution_definition}], for the collision-free case, and the case that is beyond (where $\mathcal{T}_1 = 0.07$), respectively.
  (a) The ideal case where $q = \theta/2\pi = 2\delta = 1/3$.
  (b) The case where $\theta/2\pi = 0.3$ and $q =2 \delta = 1/3$.
  }
  \label{fig:distributions}
\end{figure}
%%%%%%%%%%%%%%%%%%%%%%%%%

\subsection{Particle-hole symmetry breaking by the direct tunneling of non-equilibrium anyons}
\label{sec:ph_symm_breaking}

As the crucial feature of Eq.~\eqref{eq:full_neq_distributions}, when $\mathcal{T}_1$ is finite, these two extra terms mentioned above, corresponding to the direct tunneling of non-equilibrium anyons through the central collider, sabotage the particle-hole symmetry (analyzed in Sec.~\ref{sec:ph_symmetric}) within the collision-free limit.
Indeed, here the second and fifth lines of Eq.~\eqref{eq:full_neq_distributions}, corresponding to the fermion-like contributions described above, are symmetric with respect to $e^* V_{S1}$, the Laughlin surface of the source channel.
Noticeably, this energy $e^* V_{S1}$ deviates significantly from $e^* V_\text{1,eff}^0$, i.e., the symmetric energy within the collision-free limit.
Actually, $V_{S1} > V_\text{1,eff}^0$ due to the dilution at the diluter.
The other extra contribution, corresponding to the imaginary part of the terms within the curly brackets of the third and last lines of Eq.~\eqref{eq:full_neq_distributions}, are instead anti-symmetric with respect to $e^* V_\text{1,eff}^0$.
Either contribution thus suffices to sabotage the particle-hole symmetry that is otherwise present in the collision-free limit.
This result is actually shown in Fig.~\ref{fig:distributions}, where an enhanced $\mathcal{T}_1$ (the dashed curves) clearly drives the effective distribution functions to deviate from the symmetric case (the solid curves).
As will be shown shortly in the next section (Sec.~\ref{sec:collision_seebeck}), 
finite thermoelectric coefficients will ensue be generated, due to the breaking of the particle-hole symmetry.

In addition to the induced particle-hole symmetry breaking, these two terms also bring in multiple energy scales into consideration, even when the ambient temperature of the system equal zero [i.e., when $\mathcal{T}_{S1} = 0$, leading to $S_1 = e^* I_1$ in Eq.~\eqref{eq:full_neq_distributions}].
Indeed, as can be seen from Eq.~\eqref{eq:full_neq_distributions}, now $n_\text{p,1}$ and $n_\text{h,1}$ both become dependent on the bias of the source channel, $V_{S1}$.
This feature, importantly, indicates that the analysis at the end of Sec.~\ref{sec:ph_symmetric}, on the source of universality within the collision-free limit, fails to apply.
As the consequence, now the system quantum transport coefficients, including the charge and heat conductances as well as the Lorenz number, cease to be universal functions of only the anyonic signatures.
Instead, they now depend on also the details of the diluter, manifested by the corresponding transmission probability, $\mathcal{T}_1$.

\subsection{Collision-induced  thermoelectric coefficients}
\label{sec:collision_seebeck}

Now, we extend the effective linear-response theory beyond the collision-free limit, and investigate the role of tunneling and collisions of anyons supplied by diluted non-equilibrium beams.
As discussed already in Sec.~\ref{sec:ph_symm_breaking}, beyond the collision-free limit, the direct tunneling and collision of non-equilibrium anyons bring in two important consequences, i.e., the introduction of an extra relevant energy $e^* V_{S1}$, and the breaking of the particle-hole symmetry.
As anticipated in Sec.~\ref{sec:ph_symm_breaking}, these two consequences ensue lead to two observable characteristics, i.e., the non-universal dependence of transport coefficients, and the generation of finite thermoelectric (Seebeck and Peltier) coefficients, respectively.
Within this section, we present both features anticipated above more explicitly.
Below, we emphasize the inclusion of anyonic collisions by adding a subscript ``coll'' to observables.

\begin{figure}[!t]
  \includegraphics[width= 1 \linewidth]{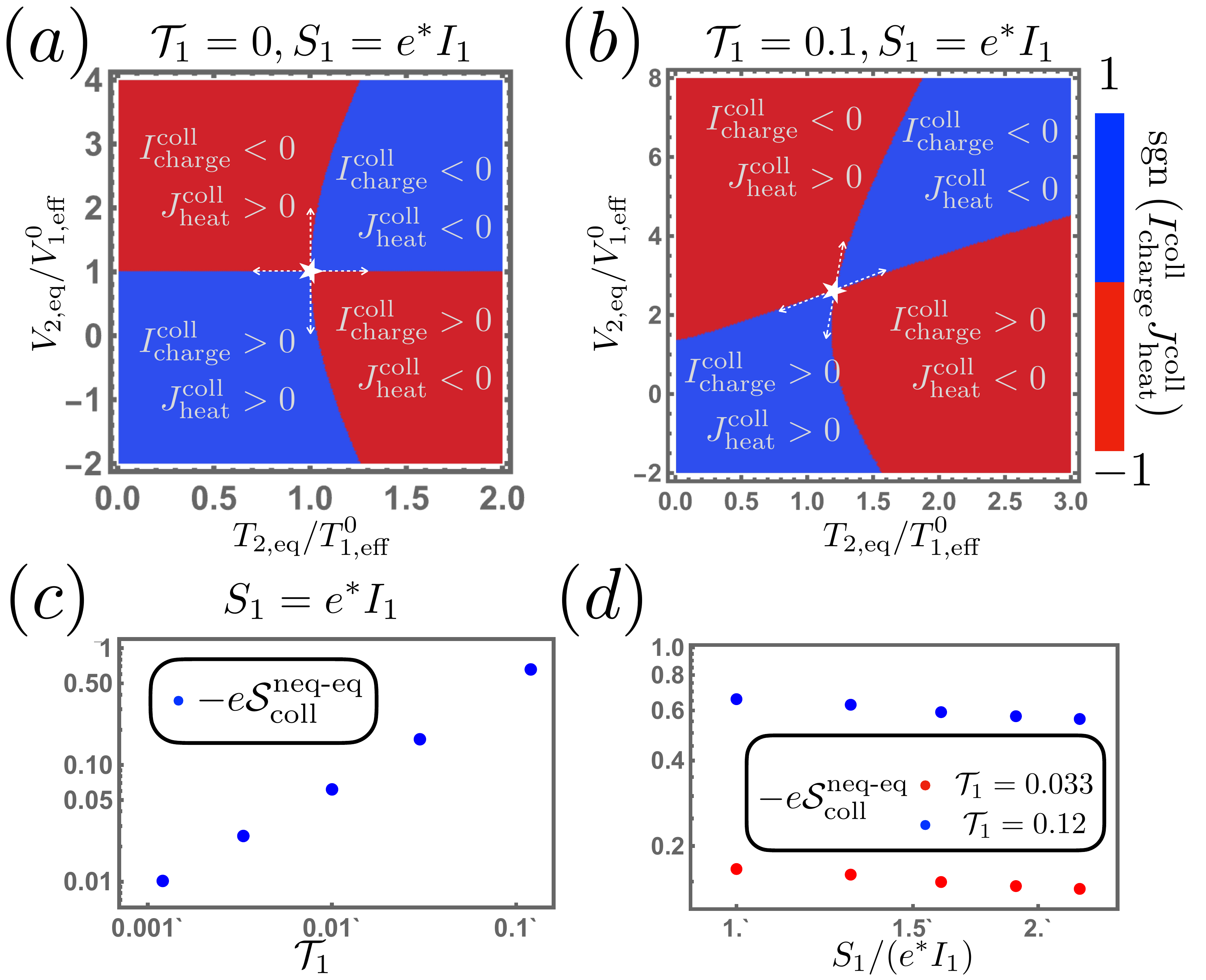}
  \caption{
  \textbf{Effect of the direct tunneling of non-equilibrium anyons, on charge and heat currents and thermoelectric coefficients.}
  Here we present only the results of the ideal case, where $q = \theta/2\pi = 2\delta = \nu$.
  (a) and (b): Diagrams that present the sign (positive in blue areas and negative in red areas) of the product of charge and heat currents, $I_\text{charge}^\text{coll} J_\text{heat}^\text{coll}$, where the subscript ``coll'' highlights the inclusion of anyonic collisions. 
  The crossing point (the white star) between different-color areas denotes the effective equilibrium point. White dashed arrows indicate the directions of boundaries between the blue and red areas at the effective equilibrium point.
  The equilibrium potential and temperature in channel 2 are given in units of $V_{1,\text{eff}}^0 = I_1 \sin (\theta)/(e^*)^2$ 
  [Eq.~\eqref{eq:veff0}] 
  and $T_\text{1,eff}^0 \approx 17.4 I_1/e$ [obtained numerically by solving Eq.~\eqref{eq:eff_temp_definition}], which are the effective chemical potential and effective temperature, respectively, of the non-equilibrium channel 1 in the collision-free limit (hence, superscript ``0'').
  Panel (a) describes the collision-free limit with a vanishingly small $\mathcal{T}_1$, where the white arrows are parallel to either the temperature or bias axis, indicating vanishing Seebeck and Peltier coefficients.
  In panel (b), where $\mathcal{T}_1=0.1$, the boundaries between the areas are tilted, implying non-zero values of Seebeck and Peltier coefficients.
  (c) The Seebeck coefficient as a function of $\mathcal{T}_1$ for $T_{S1} \ll V_{S1}$  (when $S_1 = e^* I_1$) increases linearly as a function of $\mathcal{T}_1$. (d) When $\mathcal{T}_1$ is kept fixed, the dependence of the Seebeck coefficient on the ratio $S_1/(e^* I_1)$ is comparatively much less significant, both for relatively large, $\mathcal{T}_1 = 0.12$ (blue dots), and small, $\mathcal{T}_1 = 0.033$ (red dots), transmission.
  }
  \label{fig:coefficients_collision}
\end{figure}

We begin by presenting the non-universal feature of effective parameters that identify the location of the effective equilibrium.
Different from the collision-free situation, 
after including the direct tunneling and collision of non-equilibrium anyons, the effective parameters can not be obtained analytically, due to the requirement to adjust $V_\text{2,eq}$ and $T_\text{2,eq}$ simultaneously (cf. Sec.~\ref{sec:parameters_beyond} on the corresponding protocol to identify the effective equilibrium).
Indeed, of this case the effective parameters can be readout by checking the sign of the product of currents, $\text{sgn} (I_\text{charge}^\text{coll} J_\text{heat}^\text{coll})$, with the superscript ``coll'' added to highlight the inclusion of the direct tunneling of non-equilibrium anyons (i.e., beyond ``collision-free'').
More explicitly, here we show the sign of the product of currents, $\text{sgn} (I_\text{charge}^\text{coll} J_\text{heat}^\text{coll})$, in Figs.~\ref{fig:coefficients_collision}(a) and \ref{fig:coefficients_collision}(b), for the almost collision-free ($\mathcal{T}_1 \ll 1$) and the collision-involved ($\mathcal{T}_1 = 0.1$) situations, respectively.
Here $I_\text{charge}^\text{coll}$ and $J_\text{heat}^\text{coll}$ are obtained from Eq.~\eqref{eq:currents}, with the expressions of effective distribution functions for in-equilibrium [Eq.~\eqref{eq:distributions_equi} for channel 2] and out-of-equilibrium [Eq.~\eqref{eq:full_neq_distributions} for channel 1] channels, respectively (cf. Appendix~\ref{sec:currents_beyond}).
In both plots, the crossing points (highlighted by white stars) between different-color areas mark out the effective equilibrium points, where
$V_\text{1,eff} = V_\text{2,eq}$ and $T_\text{1,eff} = T_\text{2,eq}$.
As anticipated above, in Fig.~\ref{fig:coefficients_collision}(b), due to the inclusion of the direct tunneling and collision of non-equilibirum anyons (manifested by a finite value of $\mathcal{T}_1$), the location of the effective equilibrium deviates significantly from that [i.e., the white star of Fig.~\ref{fig:coefficients_collision}(a)] within the collision-free limit.
Alternatively, the effective parameters characterizing the effective equilibrium depend non-universally on the detail of the diluter, manifested by the transmission probability through the diluting QPC, $\mathcal{T}_1$.

Now we move to discuss another feature, i.e., the generation of thermoelectric coefficients, based on the plots of Fig.~\ref{fig:coefficients_collision}.
Remarkably, in addition to their distinct locations of effective equilibrium (represented by the white stars), Figs.~\ref{fig:coefficients_collision}(a) and \ref{fig:coefficients_collision}(b) are further contrasted by the different shapes of the boundaries of areas with different colors.
Indeed, for $\mathcal{T}_1 \ll 1$ [Fig.~\ref{fig:coefficients_collision}(a)], time-domain braiding dominates and the system is approximately collision-free. In this case, two white dashed arrows are parallel to either the $T_\text{2,eq}$ or $V_\text{2,eq}$ axis, indicating the vanishing of Peltier and Seebeck coefficients, in agreement with the anticipation and discussion in Sec.~\ref{sec:ph_symmetric}.
For $\mathcal{T}_1 = 0.1$ [Fig.~\ref{fig:coefficients_collision}(b)], collisions of anyons supplied by diluters become important and the white arrows get tilted, indicating finite Seebeck and Peltier coefficients.
This is indeed the case after explicitly performing the evaluation of these coefficients, where, as shown by Fig.~\ref{fig:coefficients_collision}(c), $\mathcal{S}^\text{neq-eq}_\text{coll}$ increases when increasing $\mathcal{T}_1$.
Notice that due to the above-mentioned non-universal feature of effective parameters,
it is more convenient (both analytically and experimentally) to take the derivatives of the currents with respect to the equilibrium voltage or temperature of channel 2:
\begin{align}
L_\text{IV,\text{coll}}^\text{neq-eq} &\equiv - \partial_{V_\text{2,eq}} I_\text{charge}^\text{coll},\quad L^\text{neq-eq}_\text{JT,\text{coll}} \equiv - \partial_{T_\text{2,eq}} J_\text{heat}^\text{coll}, \notag \\ L_\text{JV,\text{coll}}^\text{neq-eq} &\equiv - \partial_{V_\text{2,eq}} J_\text{heat}^\text{coll}, \quad L_\text{IT,\text{coll}}^\text{neq-eq} \equiv - \partial_{T_\text{2,eq}} I_\text{charge}^\text{coll}, 
\label{eq:ls_collision}
\end{align}
although the outcome at effective equilibrium, importantly, exactly agrees with that when defining these parameters within the more standard manner, i.e., Eq.~\eqref{eq:coefficients_definition}, as it should be.

%%%%%%%%%%%%%%%%%%%%%%%%%
%%%%%%%%%%%%%%%%%%%%%%%%%

The resulting Seebeck coefficient, affected by anyonic collisions, is shown in Figs.~\ref{fig:coefficients_collision}(c) and \ref{fig:coefficients_collision}(d).
Following Fig.~\ref{fig:coefficients_collision}(c), where $S_1 = e^* I_1$ (corresponding to zero ambient temperatures in channels $S1$ and 1), the Seebeck coefficient $\mathcal{S}^\text{neq-eq}_\text{coll}$ increases linearly when increasing the transmission probability $\mathcal{T}_1$.
Interestingly, for $\mathcal{T}_1 = 0.1$ (when dilution is still substantial), the Seebeck coefficient $ e \mathcal{S}^\text{neq-eq}_\text{coll}$ is already of order unity. This signifies the efficiency of the Seebeck and Peltier effects as indicators of tunneling and collisions of anyons from diluted beams.
In contrast to this manifest dependence of Seebeck coefficient on $\mathcal{T}_1$, the effect of the deviation of ratio $S_1/( e^* I_1)$ from unity (experimentally realizable by tuning the temperature $T_{S1}$ of source $S1$) is comparatively much less significant, as Fig.~\ref{fig:coefficients_collision}(d) illustrates for $\mathcal{T}_1 = 0.033$ and $\mathcal{T}_1 = 0.12$.
Indeed, after fixing the value of $\mathcal{T}_1$, the noise-to-current ratio $S_1/ e^* I_1$ influences both the conductance ($G^\text{neq-eq}$) and the Seebeck coefficient ($\mathcal{S}^\text{neq-eq}$) simultaneously.
As another feature of the setup of Fig.~\ref{fig:three_setups}(a), both Seebeck and Peltier coefficients are negative, characterizing the difference of landscapes around the Laughlin surfaces for the equilibrium and  non-equilibrium channels.

As a brief summary, when away from the collision-free limit, tunneling of and collisions among non-equilibrium anyons will break the particle-hole symmetry (of the collision-free limit), thus generating finite Seebeck and Peltier coefficients.
At the meantime, effective parameters (voltage and temperature) no-longer universally depend on the non-equilibrium current and noise, respectively.
Instead, they also depend non-universally on the transmission probability through diluter(s).
Both features are perfectly captured by the theoretical anticipation of Sec.~\ref{sec:ph_symm_breaking}.

\subsection{Onsager relation near the effective equilibrium}
\label{sec:onsager_relation}

In Sec.~\ref{sec:collision_seebeck} we have presented the thermoelectric coefficients when the system allows the direct tunneling of non-equilibrium anyons at the central collider.
In Fig.~\ref{fig:coefficients_collision}, we present only the plots of the Seebeck coefficient, as the Peltier coefficient is requested to follow the Onsager relation, $\mathcal{P}^\text{neq-eq} = \bar{T} \mathcal{S}^\text{neq-eq}$ [or, equivalently, $L^\text{neq-eq}_\text{JV} = \bar{T} L^\text{neq-eq}_\text{IT}$, following the definition Eq.~\eqref{eq:transport_parameters_definition}]. 
Actually, as the Onsager relation arises from the celebrated detailed balance, it should be satisfied once the effective equilibrium is well defined.

In this section, we thus prove that the Onsager relation is satisfied when the system is in effective equilibrium, within or beyond the time-domain braiding limit.
Firstly, when the system is within the time-domain braiding limit, both thermoelectric coefficients, including Seebeck and Peltier coefficients, equal zero.
As the consequence, within this limit the Onsager relation is trivially satisfied.

More interestingly, beyond the time-domain braiding limit, both coefficients are non-trivially finite, and the prove of the Onsager relation requires the knowledge of correlation functions.
For the simplicity of the analysis, here we rewrite the correlation function of Eq.~\eqref{eq:neq_correlations} as
\begin{equation}
\begin{aligned}
&\langle \psi_1^\dagger (t) \psi_1 (0)\rangle_\text{neq} \!=\! f_\alpha e^{i e^* V_\text{1,eff}^0 t } \!+\! f_\beta e^{ie^* V_{S1} t } \!+\! if_\gamma e^{i e^* V_\text{1,eff}^0 t },\\
& \langle \psi_1 (t) \psi_1^\dagger (0)\rangle_\text{neq} \!\!=\!\! f_\alpha e^{\!-i e^* V_\text{1,eff}^0 t } \!\!+\!\! f_\beta e^{\!-i e^* V_{S1} t } \!\!-\!\! if_\gamma e^{\!-i e^* V_\text{1,eff}^0 t },\\
&\langle \psi_2^\dagger (t) \psi_2 (0)\rangle_\text{eq} \!\!=\! f_2 e^{i e^* V_\text{2,eq} t } , \langle \psi_2 (t) \psi_2^\dagger (0)\rangle_\text{eq} \!\!=\! f_2 e^{-i e^* V_\text{2,eq} t },
\end{aligned}
\label{eq:general_g_functions}
\end{equation}
in terms of several auxiliary functions
\begin{equation}
\begin{aligned}
    f_\alpha & = \frac{\tau_0^{2\delta - 1}}{2\pi( \tau_0 + it)^{2\delta}}  e^{- \frac{I_1}{e^*} [1 - \cos(\theta) ] |t|},\\
    f_\beta & = \mathcal{D}_1 \frac{1}{2\pi \tau_0} \frac{(\pi T_{S1} \tau_0)^{2\delta}}{\sin^{2\delta} [\pi T_{S1} (\tau_0 + i t)]},\\
    f_\gamma & = \mathcal{D}_1 C \sin \left( \frac{\theta}{2} \right) \text{sgn} (t) \frac{\tau_0^{2\delta - 1}}{2\pi( \tau_0 + it)^{2\delta}}   e^{- \frac{I_1}{e^*} [1 - \cos(\theta) ] |t| },
\end{aligned}
\label{eq:f1s}
\end{equation}
for the non-equilibrium channel 1, and
\begin{equation}
    f_2 = \frac{1}{2\pi \tau_0} \frac{(\pi T_\text{2,eq} \tau_0)^{2\delta}}{\sin^{2\delta} [\pi T_\text{2,eq} (\tau_0 + i t)]},
    \label{eq:f2}
\end{equation}
for the in-equilibrium channel, 2, respectively.
Notice that auxiliary functions of Eqs.~\eqref{eq:f1s} and \eqref{eq:f2} depend only on the temperature, effective or real, but not on the bias.
The effective and real bias-dependence is instead included by the phase factors of Eq.~\eqref{eq:general_g_functions}.

With Eqs.~\eqref{eq:f1s} and \eqref{eq:f2}, the effective linear response coefficients, $L^\text{neq-eq}_\text{IT,coll}$ and $L^\text{neq-eq}_\text{JV,coll}$ [cf. Eq.~\eqref{eq:ls_collision}], in connection to the Seebeck and Peltier coefficients [cf. Eq.~\eqref{eq:transport_parameters_definition}], respectively, can be written explicitly as
\begin{equation}
\begin{aligned}
    & L^\text{neq-eq}_\text{JV,coll} \!=\! -2i e^* \mathcal{T}_C^{(0)} \!\! \!\int \!\! dt \ t \left(\!\frac{\partial f_2}{\partial t} \! \right)\! \! \left\{ f_\alpha \sin [ e^* (V_\text{1,eff}^0 \!- \!V_\text{2,eq})t] \right.\\
    & \left. + f_\beta    \sin [e^* (V_{S1} \!-\! V_\text{2,eq})t]  + f_\gamma  \cos [e^* (V_\text{1,eff}^0 - V_\text{2,eq})t] \right\}\\
   &L^\text{neq-eq}_\text{IT,coll} 
   \!=\! - 2i e^* \mathcal{T}_C^{(0)}\!\!\! \int\!\! dt\! \left(\! \frac{\partial f_2}{\partial T_\text{2,eq}}\!\right) \!\left\{ f_\alpha \sin [e^* (V_\text{1,eff}^0 \!-\! V_\text{2,eq}) t] \right.\\
   & \left.+ f_\beta \sin [e^* (V_{S1} - V_\text{2,eq}) t] + f_\gamma \cos [e^* (V_\text{1,eff}^0 - V_\text{2,eq}) t] \right\}.
\end{aligned}
\label{eq:ljv_lit}
\end{equation}

To proceed, following Eq.~\eqref{eq:f2} of the expression of $f_2$, its differential forms can be written as
\begin{equation}
\begin{aligned}
     &\frac{\partial f_2}{\partial T_\text{2,eq}}  = f_2 \frac{2\delta}{T_\text{2,eq}} \left\{ 1 - \frac{\pi T_\text{2,eq} (\tau_0 + i t)}{\tan \left[ \pi T_\text{2,eq} (\tau_0 + i t) \right]} \right\}, \\
     & \frac{\partial f_2 }{\partial t} = - 2i\pi\delta \ f_2 \frac{ T_\text{2,eq}}{\tan \left[ \pi T_\text{2,eq} (\tau_0 + i t) \right]}.
\end{aligned}
\label{eq:f2_differentiations}
\end{equation}
with Eqs.~\eqref{eq:ljv_lit} and \eqref{eq:f2_differentiations}, we obtain immediately
\begin{equation}
    L^\text{neq-eq}_\text{JV,coll} = \bar{T} L^\text{neq-eq}_\text{IT,coll},
    \label{eq:onsager_main}
\end{equation}
thus leading to the Onsager relation, $\mathcal{P}^\text{neq-eq} = \bar{T} \mathcal{S}^\text{neq-eq}$ [following the definition Eq.~\eqref{eq:transport_parameters_definition}], indicating the satisfactory of the detailed balance of our system at effective equilibrium.

Actually, analysis above is not limited to Eq.~\eqref{eq:f2} for the explicit expression of $f_2$, or the relation of Eq.~\eqref{eq:f2_differentiations}. Indeed, following Appendix~\ref{app:onsager_proof}, derivations above can be directly extended to more general cases where $f_2$ is any function of the temperature, $T_\text{2,eq}$, and the temperature-time product, $T_\text{2,eq} t$.
This requirement actually indicates that in the function of $f_2$, $T_\text{2,eq}$ is the only energy scale. Otherwise the detailed balance may become sabotaged by one (or multiple) extra process that is manifested by an extra relevant energy scale.

\subsection{Experimental relevance of the direct tunneling processes}

Following theoretical analysis and more explicit numerical demonstrations in Secs.~\ref{sec:ph_symm_breaking} and \ref{sec:collision_seebeck}, respectively, we arrive at the conclusion that the permission of non-equilibrium anyons to tunnel and collide at the central QPC brings in two highly non-trivial features that are potentially relevant to practical experimental setups: (i) The particle-hole symmetry breaking that induces finite thermoelectric coefficients that are otherwise zero in the collision-free limit; and (ii) The inclusion of another energy scale, $e^* V_{S1}$, into consideration, which further induces the non-universal feature of thermoelectric coefficients.

Importantly, the latter feature, i.e., this direct-tunneling induced breaking of universality is in-principle present in all experimental efforts that aim to extract the fractional braiding phase, including e.g., Refs.~\cite{BartolomeiScience20,PierrePRX23,LeeNature23} via the developed Fano factor~\cite{RosenowLevkivskyiHalperinPRL16} that aims to extract the braiding phase generated by time-domain braiding.
Indeed, to obtain the universal braiding phase, the insignificance of the non-universal interruptions from the direct tunneling and collision of non-equilibrium anyons must be apriori evaluated and confirmed.

Our work, crucially, provides the thermoelectric coefficients as the option to amplify and quantify the influence of particle-hole symmetry breaking and non-universality, both induced by the direct tunneling and collision of non-equilibrium anyons, in the setup of Fig.~\ref{fig:three_setups}(a).
As the advantage, these thermoelectric coefficients are more sensitive than the simple measurement of the transmission probability at the diluter, $\mathcal{T}_1$.
Indeed, following Fig.~\ref{fig:coefficients_collision}(c), when $\mathcal{T}_1 \approx 0.1$ remains a small quantity, the Seebeck coefficient, $-e\mathcal{S}^\text{neq-eq}$ however has already arrived at one, a rather amplified quantity.
Notice that since $\mathcal{S}^\text{neq-eq}$ is defined as the ratio between $L_\text{IT}^\text{neq-eq}$ and $L_\text{IV}^\text{neq-eq}$, it indicates that when $\mathcal{T}_1 \approx 0.1$, the response of the charge current to a temperature difference has already become comparable to its response to a difference in the electric voltage bias.
From this perspective, the thermoelectric feature, induced by the particle-hole symmetry breaking, already has become significant enough, and thus can not be simply overlooked when $\mathcal{T}_1 \approx 0.1$.

Consequently, our theory indicates that to prepare the system within the collision-free limit, where one can focus on only time-domain braiding, one has to impose a more strict requirement on the transmission probability through the diluter $\mathcal{T}_1$.
In real experiments, following Eq.~\eqref{eq:dilution} for the expression of $\mathcal{T}_1$ (at zero ambient temperature), this requirement is possibly realizable by increasing the value of $V_{S1}$, or by increasing the size of the diluter, via tuning the side gates.
Alternatively, given a finite ambient temperature, one can approach closer to the collision-free limit, by decreasing the ambient temperature.

\section{Discussion}

We consider setups comprising  non-equilibrium
channel(s) which are tunnel-coupled through a ``collider''. This system is tuned to effective equilibrium by requiring that  both the charge and heat currents through
the  collider vanish.
We further define and evaluate the effective linear response coefficients, to characterize the response of charge and heat currents, to a slight deviation of the system from the effective equilibrium point.
Our work reveals that these effective linear response coefficients can both provide extra instruments to identify the anyonic signature of the system, and also disclose key features that are less manifest via alternative measurements.

As an important observation, in the collision-free limit (corresponding to a strong dilution), where time-domain braiding processes dominate,
a non-equilibrium anyonic channel possesses a symmetry between effective particle and hole distributions (solid curves of Fig.~\ref{fig:distributions}).
This symmetry leads to vanishing Seebeck and Peltier coefficients.
In addition, for zero ambient temperatures, a system within the the collision-free limit involves a single energy scale, i.e., $e^* V_\text{1,eff}^0$, thus leading to the universal feature of effective linear response coefficients, including the charge conductance, heat conductance, as well as the Lorenz number [see Eq.~\eqref{eq:universal_coefficients}].
Remarkably, in setups of Figs.~\ref{fig:three_setups}(a) and \ref{fig:three_setups}(b) (influenced by time-domain braiding), the Lorenz number clearly deviates from both $\mathcal{L}_\text{fermion} = \pi^2/3$ for non-interacting fermions and the value of the setup of Fig.~\ref{fig:three_setups}(c) that consists of two equilibrium channels.
Indeed, non-equilibrium anyons are necessary in the generation of anyonic braiding phase.
It is thus absent in the setup of Fig.~\ref{fig:three_setups}(c), due to the missing of non-equilibrium anyons in equilibrium systems.
Experimental measurements of these coefficients, especially the Lorenz numbers, are thus able to reveal the braiding of anyons supplied by the non-equilibrium beams.
Based on the knowledge of anyonic signatures, we define three ``games'' in Sec.~\ref{sec:phase_three_games}, where we outline the protocol to extract the anyonic braiding phase, based on the difficulty of the corresponding game.
Importantly, here protocol of each game involves the measurement on only the charge and heat currents, thus distinguishing them from protocols where correlations are otherwise requested.

However, this symmetry between effective particle and hole distributions is sabotaged by the collisions of non-equilibrium anyons (leading to dashed curves of Fig.~\ref{fig:distributions}).
On one hand, the break of the particle-hole symmetry introduces finite Peltier and Seebeck coefficients [see Figs.~\ref{fig:coefficients_collision}(c) and \ref{fig:coefficients_collision}(d)] that are otherwise zero in the collision-free limit.
On the other hand, this symmetry breaking is introduced due to the involvement of another energy scale, i.e., $e^* V_{S1}$.
Due to the presence of this extra energy, the effective linear response coefficients, which were universal quantities of anyonic braiding signatures in the collision-free limit, now depend non-universally on also the transmission probability through the diluter.
Here both features (the generation of finite thermoelectric coefficients and the non-universal feature of effective linear response coefficients) agree perfectly with that of Ref.~\cite{PavlovKiselevPRL26} in equilibrium systems, thus further supporting the validity of our effective linear response theory, in anyonic-braiding-involved non-equilibrium systems.
As the consequence, to practically obtain the universal braiding phase in real experiments, the insignificance of the non-universal influence, introduced by the direct tunneling and collision of non-equilibrium anyons, should be carefully evaluated and confirmed.
Within this work, we suggest to evaluate this non-universal influence by checking the amplitude of thermoelectric coefficients.

Prospectively, we anticipate extending the study of charge and thermal transport coefficients, defined in our current work for Laughlin quasiparticles, to composite edges, or even non-Abelian systems that are out of equilibrium.
Prospectively, this extension can enable us to extract information on non-Abelian statistics manifested through collisions between non-Abelian anyonic quasiparticles via effective linear-response transport measurements, which appear to be realizable with currently available experimental techniques.

\begin{acknowledgments}
The authors are grateful to  Gabriele Campagnano, Anindya Das, Domenico Giuliano, Moty Heiblum, Andrei Pavlov, Fr\'ed\'eric Pierre, Yuval Ronen, Bernd Rosenow, In\`es Safi, and Oded Zilberberg for fruitful discussions. GZ acknowledges the support from the National Natural Science Foundation of China under Grants No. 12674059, and the startup grant at Nanjing University.
I.G. and Y.G. acknowledge the support by the Deutsche Forschungsgemeinschaft (DFG) through grant No. MI\,$658/10$-$2$.
Y.G. acknowledges support from the
DFG through grant NO. RO\,$2247/11$-$1$, CRC/Transregio 183 and EI 519/71, the US-Israel Binational Science Foundation, and the Minerva Foundation.
\end{acknowledgments}

\begin{appendix}

\section{Microscopic source of the deviation among anyonic signatures}
\label{app:fractional_signatures}

As stated in the main text, we consider three anyonic signatures: (i) the fractional charge $q$, (ii) the fractional statistics $\theta$, and (iii) the fractional scaling dimension $\delta$. Ideally, these quantities are directly related to the filling fraction $\nu$ through $\nu = q/e = \theta/2\pi = 2 \delta$. In practice, however, they can deviate from one another, giving rise to ``games'' of different difficulty, as discussed in Secs.~\ref{sec:games} and \ref{sec:phase_three_games}.
For the convenience of the readers, here we explain the microscopic origins of these deviations and how they enter the transport theory.

We begin with a generalized Gaussian edge, containing $N$ bosonic fields $\boldsymbol{\phi}=(\phi_1,\ldots,\phi_N)^T$ that obey the equal-time commutator
\begin{equation}
[\phi_a(x),\phi_b(x')]=i\pi \mathcal{C}_{ab} \text{sgn}(x-x'),
\label{eq:algebra_general}
\end{equation}
with $\mathcal{C}_{ab}$ the matrix element that characterizes the algebra of the edge.
A quasiparticle of type $\boldsymbol{\ell} = (l_1,\ldots,l_N)$ is represented by the vertex operator
\begin{equation}
\Psi^\dagger_{\boldsymbol{\ell}}=\frac{F_{\boldsymbol{\ell}}}{\sqrt{2\pi \tau_0}}\exp(-i\sum_{a = 1}^N l_a \phi_a ),
\label{eq:quasiparticle_general}
\end{equation}
where $F_{\boldsymbol{\ell}}$ denotes a possible Klein or topological factor.
The system with the algebra of Eq.~\eqref{eq:algebra_general} can be characterized by fractional signatures below.

Firstly, quasiparticle introduced by Eq.~\eqref{eq:quasiparticle_general} can be characterized by the braiding phase,
\begin{equation}
\theta_{\boldsymbol{\ell}} =2 \pi \sum_{a,b = 1}^N l_a \mathcal{C}_{ab} l_b,
\label{eq:theta_quasiparticle}
\end{equation}
which equals that after taking into consideration the contribution of all composing phases.

The second quantity, $q$, characterizing the electromagnetic charge, is determined independently by the charge vector $\boldsymbol{m} = (m_1,\ldots,m_N)$, with
\begin{equation}
    \rho=\frac{e}{2\pi}\sum_{a} m_a \partial_x \phi_a,
\end{equation}
the total charge density, and
\begin{equation}
    q_{\boldsymbol{\ell}} = \sum_{a,b = 1}^N m_a \mathcal{C}_{ab} l_b,
    \label{eq:q_quasiparticle}
\end{equation}
for the quasiparticle that is labeled by $\boldsymbol{\ell}$, respectively.
Like its braiding phase $\theta_{\boldsymbol{\ell}}$ of quasiparticle $\Psi^\dagger_{\boldsymbol{\ell}}$, its charge $q_{\boldsymbol{\ell}}$, given by Eq.~\eqref{eq:q_quasiparticle}, takes into consideration the contributions from all composing bosonic charges.
However, these two quantities are not necessarily proportional to each other, as $\boldsymbol{m}$ and $\boldsymbol{\ell}$ may differ from each other.

Finally, we move to the fractional scaling feature of this quasiparticle, via its correlation function,
\begin{equation}
\begin{aligned}
    \langle\Psi_{\boldsymbol{\ell}}(t)\Psi^\dagger_{\boldsymbol{\ell}}(0)\rangle
& \propto |t|^{- \sum_{a,b=1}^N l_a \mathcal{G}_{ab} l_b
}\\
& = |t|^{-2\delta_{\boldsymbol{\ell}}},
\end{aligned}
\label{eq:scaling_quasiparticle}
\end{equation}
in terms of another matrix $\mathcal{G}$ that can in-principle differ from $\mathcal{C}$ that governs the exchange algebra of Eq.~\eqref{eq:algebra_general}.
With Eq.~\eqref{eq:scaling_quasiparticle}, we can directly obtain the fractional scaling dimension,
\begin{equation}
    \delta_{\boldsymbol{\ell}}= \frac{1}{2} \sum_{a,b=1}^N l_a \mathcal{G}_{ab} l_b.
    \label{eq:delta_quasiparticle}
\end{equation}

Comparison between Eqs.~\eqref{eq:theta_quasiparticle}, \eqref{eq:q_quasiparticle} and \eqref{eq:delta_quasiparticle} leads to the conclusion that the fractional charge ($q_{\boldsymbol{\ell}}$), braiding phase ($\theta_{\boldsymbol{\ell}}$), and scaling dimension ($\delta_{\boldsymbol{\ell}}$) originate from three distinct structures of the low-energy theory: the electromagnetic coupling, the antisymmetric exchange algebra, and the symmetric correlation functions, respectively. In a general edge theory, $\mathcal{C}$ and $\mathcal{G}$ matrices need not coincide, and these three quantities therefore need not obey any fixed relation.

In the Laughlin systems under consideration, each physical edge carries only one chiral channel, where fractional parameters above enter the transport problem through rather different ways.
More specifically, the quasiparticle charge enters the effective equilibrium parameters through $e^*=q e$ [cf. Eqs.~\eqref{eq:veff0} and \eqref{eq:teff0}] and quantifies the charge that is transferred by each tunneling event [cf. Eq.~\eqref{eq:currents_correlations} of $I_\text{charge}$].
The exponent $\delta$ controls the long-time scaling of the tunneling correlators [cf. the denominators of Eqs.~\eqref{eq:equilibrium_correlation} and \eqref{eq:neq_correlations}] and hence the associated renormalization of the transport coefficients [cf. the factor $(\pi \bar{T} \tau_0)^{4\delta - 2}$ of Eq.~\eqref{eq:universal_coefficients}].
Finally, the statistical angle $\theta$ enters through the phase accumulated in time-domain braiding, with a complete winding contributing $\exp (\pm i \theta)$ [cf. the factors $\sin ( \theta)$ and $\cos (\theta)$ of Eq.~\eqref{g-braid}, which enter the correlation Eq.~\eqref{eq:neq_correlations}].
In addition, $\theta$ does not enter the equilibrium correlator, Eq.~\eqref{eq:equilibrium_correlation}, where braiding is absent.
Ideally, these structures collapse to $q=\theta/2\pi=2 \delta=\nu$ for the edge of a Laughlin state with filling fraction $\nu$.
Practically, this relation is however not necessarily valid, due to influences from e.g., interactions or the edge-bulk communication.

\color{black}

\section{The charge and heat tunneling current, beyond the collision-free limit}
\label{sec:currents_beyond}

In the collision-free limit, the tunneling charge and heat current can be evaluated following the first lines of Eqs.~\eqref{eq:it_and_lit} and \eqref{eq:jq_and_lvj}, respectively.
Going beyond the collision-free limit, these two expressions should however be modified, after taking into consideration the tunneling and collision of non-equilibrium anyons.
More specifically, the charge and heat tunneling currents now equal
\begin{equation}
\begin{aligned}
    & I_\text{charge}^\text{coll}
    = 2 i e^* \mathcal{T}_C^{(0)} \int dt \left\{ f_\alpha f_2 \sin [e^* (V_\text{1,eff}^0 - V_\text{2,eq}) t] \right.\\
    & \left. \!+\! f_\beta f_2 \sin [e^* (V_{S1} \!-\! V_\text{2,eq}) t] \!+\! f_\gamma f_2 \cos [e^* (V_\text{1,eff}^0 \!-\! V_\text{2,eq}) t] \right\},\\
    & J_\text{heat}^\text{coll}  = 2i \mathcal{T}_C^{(0)} \!\!\int\! dt  \left[ \frac{\partial f_2 (t)}{\partial t} \right] \! \!\left\{ f_\alpha (t) \cos \left[ e^* (V_\text{1,eff}^0 \!-\! V_\text{2,eq} ) t \right]\right. \\
    & \left. \!+\! f_\beta (t) \cos [e^*  (V_{S1} - V_\text{2,eq}) t]  \!-\! f_\gamma   \sin [e^* (V_\text{1,eff}^0 \!-\! V_\text{2,eq}) t] \right\},
\end{aligned}
\label{eq:currents_connected}
\end{equation}
in terms of auxiliary functions ($f_\alpha$, $f_\beta$, $f_\gamma$ and $f_2$) of Eqs.~\eqref{eq:f1s} and \eqref{eq:f2}.
In Eq.~\eqref{eq:currents_connected}, we have taken $\mu_0 = e^* V_\text{2,eq}$ as the reference of energy when evaluating $J_\text{heat}^\text{coll}$.

\section{Explicit expressions for functions in Eq.~\eqref{eq:universal_coefficients}}
\label{sec:gamma_chi_functions}

Eq.~\eqref{eq:universal_coefficients} contains the setup-specific functions that universally depend on the filling fraction $\nu$.
Here, we provide their explicit expressions. More details can be found in Appendix~\ref{app:chi_gamma}.
We stress that all functions $\gamma^s$ and $\chi^s$  are dimensionless.
Importantly, in addition to the renormalization factor $(\pi \bar{T} \tau_0)^{4\delta - 2}$, the anyonic correlations have a residue effect, as they govern the long-time behavior of the denominators in the integrals determining $\gamma^s$ and $\chi^s$, see, e.g., Eq.~\eqref{eq:g_f_neq_eq} for the $s = $neq-eq setup and Eq.~\eqref{eq:g_f_neq_neq} for the $s =$neq-neq setup.

For our mainly focused neq-eq setup of Fig.~\ref{fig:three_setups}(a), these functions are expressed in terms of the time integrals, as follows:
\begin{equation}
\begin{aligned}
    &\gamma^\text{neq-eq}\left(\theta, \delta \right) \! = \!\frac{i\pi^{2\delta - 2}}{2}\int_{-\infty}^\infty dt' \  \frac{ e^{ - \frac{s_c}{\pi} 2 \sin^2 (\theta/2) | t'| } t' }{(\tau_0' + i t')^{2\delta} \sin^{2\delta}  (\tau_0' + i t')   },\\
    &\chi^\text{neq-eq} \left(\theta,\delta \right) = - \frac{i}{4}\pi^{-2\delta} \partial_r \Omega^\text{neq-eq} (\theta, \delta, r)\Big|_{r = 1},\\
    &\Omega^\text{neq-eq} (\theta,\delta, r) \!\equiv \! 4i\pi^{2\delta} \delta \!\!\!\int_{-\infty}^\infty\!\!\! \! \!dt' \frac{e^{ - \frac{s_c}{\pi} 2 r\! \sin^2 \!(\theta/2) |t'| } \cos  (\tau_0' \!+\! i t') }{(\tau_0' + i t')^{2\delta} \sin^{2\delta +1}  (\tau_0' \!+\! i t') },
\end{aligned}
\label{eq:g_f_neq_eq}
\end{equation}
where $\tau_0' \equiv \tau_0 T_\text{2,eq}$ is a dimensionless infinitesimal, and $t'\equiv \pi T_\text{2,eq} t$ is a dimensionless integral variable.
Here, the response to an infinitesimal temperature difference is introduced by the differentiation with respect to $r = T_\text{1,neq}/T_\text{2,eq}$ (taken to be one after the differentiation).

In the neq-neq setup of Fig.~\ref{fig:three_setups}(b), we have compact analytical expressions of $\gamma^\text{neq-neq}$ and $\chi^\text{neq-neq}$, i.e.,
\begin{equation}
\begin{aligned}
    \gamma^\text{neq-neq}\! (\theta,\delta)& \!\! =\!\! 4^{4\delta\!-\!2}\pi^{-4\delta} \Gamma\! (2 \!-\! 2\delta) \sin (2\pi \delta) \sin^{8\delta\!-\!4} \!\!\left(\!\frac{\theta}{2}\!\right) \!s_c^{4\delta\!-\!2}, \\
    \chi^\text{neq-neq} \! (\theta,\delta)&  \!=\! 4^{4\delta-1} \pi^{-4\delta} \Gamma\! (1\!-\!4\delta) \sin (2\pi\delta) \sin^{8\delta } \!\!\left(\!\frac{\theta}{2}\!\right) \!s_c^{4\delta},
\end{aligned}
\label{eq:g_f_neq_neq}
\end{equation}
with $s_c$ defined in Eq.~\eqref{eq:teff0}, and obtainable from the integral equation, \eqref{eq:eff_temp_definition}.

In the eq-eq setup of Fig.~\ref{fig:three_setups}(c), the expressions are modified into
\begin{equation}
\begin{aligned}
    \gamma^\text{eq-eq} \left( \delta \right) & = \frac{i}{2\pi^2} \int_{-\infty}^\infty d t' \frac{t'}{ \sin^{4\delta} (\tau_0' + i t')  } \\
    & = \frac{1}{4\pi^{3/2}}\cos (2\pi\delta)\,  \Gamma \left( \frac{1}{2} -2\delta \right) \Gamma (2\delta),\\
        \Omega^\text{eq-eq} (\delta, r) & \!\equiv \!\frac{1}{\pi} \int_{-\infty}^\infty \!\!dt' \frac{\cot (\tau_0' + i t') - r \cot[ r(\tau_0' + i t')]  }{\sin^{2\delta}[ r (\tau_0' \!+\! i t')] \sin^{2\delta} (\tau_0'\! +\! i t')}\\
     & = \frac{\sqrt{\pi}\delta^2}{(1 + 4\delta)} \cos (2\pi\delta)\, \Gamma \left( \frac{1}{2 } -2\delta \right) \Gamma (2\delta),\\
    \chi^\text{eq-eq} (\delta ) & = \frac{\pi \delta}{2}  \frac{\partial}{\partial r}  \Omega^\text{eq-eq} (\delta, r)\Big|_{r = 1}\\
    & = \frac{\delta}{4} \int_{-\infty}^\infty dt' \frac{2 (\tau_0' + it') - \sin \left[2  (\tau_0' + i t')\right]}{\sin^{2 + 4\delta}  (\tau_0' + it') }\\
    & = \frac{\sqrt{\pi}\delta^2}{(1 + 4\delta)} \cos (2\pi\delta)\, \Gamma \left( \frac{1}{2 } -2\delta \right) \Gamma (2\delta).
\end{aligned}
\label{eq:g_f_eq_eq}
\end{equation}
\color{black}
Importantly, due to the absence of braiding in equilibrium setups, the cosine factor of $\gamma^\text{eq-eq}$ and $\chi^\text{eq-eq}$ is induced by the $\delta$-dependence of the denominator of the integrals.
For $2 \delta = 1$, both $\gamma^\text{eq-eq}$ and $\chi^\text{eq-eq}$ in Eq.~\eqref{eq:g_f_eq_eq} correctly reduce to their non-interacting fermionic values, $1/2\pi$ and $\pi/6$, respectively.
Following Eq.~\eqref{eq:g_f_eq_eq}, we obtain the Lorenz number for the setup of Fig.~\ref{fig:three_setups}(c), given by Eq.~\eqref{lorenz-numbers-eq-eq}.

\section{Integral forms of effective linear response coefficients}
\label{app:eff_res}

In this section, we follow charge and heat current in Appendix.~\ref{sec:currents_beyond}, to express the effective linear response functions defined in Eq.~\eqref{eq:coefficients_definition}, in terms of auxiliary functions of Eqs.~\eqref{eq:f1s} and \eqref{eq:f2}.

To begin with, with $I_\text{charge}^\text{coll}$ of Eq.~\eqref{eq:currents_connected}, we can obtain two response functions, $L_\text{IV,coll}^\text{neq-eq}$ and $L^\text{neq-eq}_\text{IT,coll} $,
\begin{equation}
\begin{aligned}
    &L_\text{IV,coll}^\text{neq-eq} \equiv -  \frac{\partial}{\partial V_\text{2,eq}} I_\text{charge}^\text{coll} \Bigg|_{V_\text{2,eq} = V_\text{1,eff}, T_\text{2,eq} = T_\text{1,eff}} \!\!\!= G^\text{neq-eq}_\text{coll}\\
    & = 2i (e^*)^2 \mathcal{T}_C^{(0)} \int dt\ t\  f_2 \left\{ f_\alpha  \cos [e^*(V_\text{eff,1}^0 - V_\text{2,eq}) t] \right.\\
    & \left.+ f_\beta  \cos [e^* (V_{S1} - V_\text{2,eq}) t] - f_\gamma  \sin [e^* (V_\text{eff,1}^0 - V_\text{2,eq}) t] \right\} ,\\
    &L^\text{neq-eq}_\text{IT,coll}  \equiv - \frac{\partial}{\partial T_\text{2,eq}} I_\text{charge}^\text{coll} \Bigg|_{V_\text{2,eq} = V_\text{1,eff}, T_\text{2,eq} = T_\text{1,eff}}\\
    & = - 2i e^* \mathcal{T}_C^{(0)} \int dt \left( \frac{\partial f_2}{\partial T_\text{2,eq}}\right) \left\{ f_\alpha \sin [e^* (V_\text{eff,1}^0 - V_\text{2,eq}) t] \right.\\
    &\left. + f_\beta \sin [e^* (V_{S1} - V_\text{2,eq}) t] + f_\gamma \cos [e^* (V_\text{eff,1}^0 - V_\text{2,eq}) t] \right\},
\end{aligned}
\label{eq:current_differentiates}
\end{equation}
as the response of $I_\text{charge}^\text{coll}$ to the difference in voltage and temperature, respectively.
We have several remarks about Eq.~\eqref{eq:current_differentiates}.
Firstly, both charge conductance and Peltier coefficients are defined as differentiation with respect to parameters of the equilibrium channel 2, i.e., $V_\text{2,eq}$ and $T_\text{2,eq}$.
These definitions are taken, since effective parameters $V_\text{1,eff}$ and $T_\text{1,eff}$ depend non-universally on experimental details.
Differentiation with respect to these effective parameters is thus complicated and experimentally less practical.
On the other hand, in the effective linear response regime [where $I_\text{charge}^\text{coll} \approx L_\text{IV,coll}^\text{neq-eq} (V_\text{1,eff} - V_\text{2,eq}) + L^\text{neq-eq}_\text{IT,coll} (T_\text{1,eff} - T_\text{2,eq}) $], differentiation with respect to $-V_\text{2,eq}$ and $-T_\text{2,eq}$
actually leads to the same outcome as that obtained with the more conventional option, i.e., when differentiating with respect to the voltage and temperature differences, $V_\text{1,eff}-V_\text{2,eq}$ and $T_\text{1,eff}-T_\text{2,eq}$.
As another remark, as has been manifestly presented by the Seebeck coefficient $\mathcal{S}$ in Fig.~\ref{fig:coefficients_collision}(c) [and also the Peltier coefficient $\mathcal{P}$, following the Onsager relation of Eq.~\eqref{eq:onsager_main}], the ``off-diagonal'' elements $L^\text{neq-eq}_\text{IT,coll}$ and $L^\text{neq-eq}_\text{JV,coll}$ are vanishing in the collision-free limit [i.e., when $f_\beta = f_\gamma = 0$ and $V_\text{2,eq} = V_\text{1,eff}^0$, cf. Eqs.~\eqref{eq:it_and_lit} and \eqref{eq:jq_and_lvj}], but becomes finite after allowing the direct tunneling of non-equilibrium anyons (when $f_\beta = f_\gamma$ are finite).

Now we move to the heat current, i.e., $J_\text{heat}^\text{coll}$ of Eq.~\eqref{eq:currents_connected}.
Its two response functions can be expressed into the integral form,
\begin{equation}
\begin{aligned}
    &L^\text{neq-eq}_\text{JT,coll} \equiv -\frac{\partial J_\text{heat}^\text{coll}}{\partial T_\text{2,eq}} \\
    & = - 2i \mathcal{T}_C^{(0)} \int\! dt  \left[ \frac{\partial^2 f_2 (t)}{\partial t \partial T_\text{2,eq}} \right] \left\{ f_\alpha (t) \cos \left[ e^* (V_\text{1,eff}^0 \!-\! V_\text{2,eq} ) t \right]\right. \\
    & \left. + f_\beta (t) \cos [e^*  (V_{S1} - V_\text{2,eq}) t]  \!-\! f_\gamma   \sin [e^* (V_\text{1,eff}^0 \!-\! V_\text{2,eq}) t] \right\},\\
   &L^\text{neq-eq}_\text{JV,coll}  \equiv - \frac{\partial J_\text{heat}^\text{coll}}{\partial V_\text{2,eq}}  \\
   &\! =\! -2i \mathcal{T}_C^{(0)} e^* \!\int\! dt \ t\  \left[ \frac{\partial f_2 (t)}{\partial t} \right] \left\{ f_\alpha (t) \sin \left[ e^* (V_\text{1,eff}^0 \!- \! V_\text{2,eq} ) t \right]\right. \\
    & \left. + f_\beta (t) \sin [e^*  (V_{S1} - V_\text{2,eq}) t]  + f_\gamma   \cos [e^* (V_\text{1,eff}^0 \!-\! V_\text{2,eq}) t] \right\},
\end{aligned}
\label{eq:heat_current_differentiates}
\end{equation}
where, akin to the integral expressions of $L_\text{IT,coll}^\text{neq-eq}$ and $L_\text{IV,coll}^\text{neq-eq}$ in Eq.~\eqref{eq:current_differentiates}, $L_\text{JT,coll}^\text{neq-eq}$ and $L_\text{JV,coll}^\text{neq-eq}$ are defined after differentiation with respect to only the real parameters of the equilibrium channel.
Again, since we focus on effective linear response around the effective equilibrium point, differentiation with respect to real parameters leads to the same results, when instead differentiating with respect to the difference between effective and real parameters.
In addition, as $L^\text{neq-eq}_\text{IT,coll}$ in Eq.~\eqref{eq:current_differentiates}, $L^\text{neq-eq}_\text{JV,coll}$ becomes generically finite (even near the effective equilibrium point), after including collisions of non-equilibrium anyons, i.e., when $f_\beta$ and $f_\gamma$ are both finite.

\section{Onsager relation under extended conditions}
\label{app:onsager_proof}

In Sec.~\ref{sec:onsager_relation} we have shown that the Onsager relation is obeyed by the effective equilibrium point of our non-equilibrium system, when channel 2 is described by function $f_2$ of Eq.~\eqref{eq:f2}. In this section, we further show that actually the explicit expression of $f_2$ is not necessary. Instead, effective equilibrium alone suffices to lead to the satisfactory of the Onsager relation.

More specifically, here we consider a more general expression,
\begin{equation}
    f_2^\text{general} = \sum_{j=1}^N f_{2,j} (T_\text{2,eq}, T_\text{2,eq} t),
    \label{eq:general_f2}
\end{equation}
which involves the summation of $N$ functions, $f_{2,i}$, that each depends on $T_\text{2,eq}$, $T_\text{2,eq} t$, or both of them.
Here the superscript ``general'' highlights the fact that we consider a rather general function $f_2$.
The differentiation of $f_2^\text{general}$ of Eq.~\eqref{eq:general_f2} equals,
\begin{equation}
\begin{aligned}
    \frac{\partial f_2^\text{general}}{\partial T_\text{2,eq}} & = \sum_{j=1}^N \left[ \frac{\partial f_{2,j} (r_1,r_2)}{\partial r_1} \Big|_{r_1 = T_\text{2,eq}, r_2 = T_\text{2,eq} t} \right.\\
    &\left. + t \frac{\partial f_{2,j} (r_1,r_2)}{\partial r_2} \Big|_{r_1 = T_\text{2,eq}, r_2 = T_\text{2,eq} t} \right],\\
    \frac{\partial f_2^\text{general}}{\partial t} & = \sum_{j=1}^N T_\text{2,eq} \frac{\partial f_{2,j} (r_1,r_2)}{\partial r_2} \Big|_{r_1 = T_\text{2,eq}, r_2 = T_\text{2,eq} t}. 
\end{aligned}
\label{eq:f2_general_difs}
\end{equation}
With Eq.~\eqref{eq:f2_general_difs}, Eq.~\eqref{eq:heat_current_differentiates} becomes
\begin{widetext}
\begin{equation}
\begin{aligned}
    & L^\text{general}_\text{JV,coll} \!\!=\! -2i e^* \mathcal{T}_C^{(0)} \!\!\!  \int \!\! dt \ t \!\left(\frac{\partial f_2^\text{general}}{\partial t}  \right)\! \left\{ f_\alpha \sin [e^* (V_\text{1,eff}^0 \!-\! V_\text{2,eq})t]  \!+\! f_\beta    \sin [e^* (V_{S1} \!-\! V_\text{2,eq})t] \! +\! f_\gamma  \cos [e^* (V_\text{1,eff}^0 \!-\! V_\text{2,eq})t] \right\}\\
   & = -2i e^* \mathcal{T}_C^{(0)}  T_\text{2,eq} \int dt \sum_{j=1}^N \ t \  \frac{\partial f_{2,j} (r_1,r_2)}{\partial r_2} \Big|_{r_1 = T_\text{2,eq}, r_2 = T_\text{2,eq} t}\\
   \times & \left\{ f_\alpha \sin [e^* (V_\text{1,eff}^0 - V_\text{2,eq})t] + f_\beta    \sin [e^* (V_{S1} - V_\text{2,eq})t] + f_\gamma  \cos [e^* (V_\text{1,eff}^0 - V_\text{2,eq})t] \right\}\\
   & = -2i e^* \mathcal{T}_C^{(0)}  T_\text{2,eq} \int dt \left\{ \frac{\partial f^\text{general}_2}{\partial T_\text{2,eq}} - \sum_{j=1}^N  \frac{\partial f_{2,j} (r_1,r_2)}{\partial r_1} \Big|_{r_1 = T_\text{2,eq}, r_2 = T_\text{2,eq} t} \right\}\\
   \times & \left\{ f_\alpha \sin [e^* (V_\text{1,eff}^0 - V_\text{2,eq})t] + f_\beta    \sin [e^* (V_{S1} - V_\text{2,eq})t] + f_\gamma  \cos [e^* (V_\text{1,eff}^0 - V_\text{2,eq})t] \right\}\\
   & = -2i e^* \mathcal{T}_C^{(0)}  T_\text{2,eq} \int dt \frac{\partial f_2^\text{general}}{\partial T_\text{2,eq}} \left\{ f_\alpha \sin [e^* (V_\text{1,eff}^0 - V_\text{2,eq})t] + f_\beta    \sin [e^* (V_{S1} - V_\text{2,eq})t] + f_\gamma  \cos [e^* (V_\text{1,eff}^0 - V_\text{2,eq})t] \right\}\\
   & = \bar{T} L_\text{LT,coll}^\text{neq-eq}\big|_\text{general},
\end{aligned}
\label{eq:ljv_simplified_general}
\end{equation}
\end{widetext}
where in the last but one line, we have used the fact that
\begin{equation}
\begin{aligned}
    &\int\! dt \frac{\partial f_{2,j} (r_1,r_2)}{\partial r_1} \Bigg|_{r_1\! = \!T_\text{2,eq}, r_2\! = \!T_\text{2,eq} t} \!\!\left\{ f_\alpha \sin [e^* (V_\text{1,eff}^0 \!-\! V_\text{2,eq})t] \right.\\
    & \left. \!+\! f_\beta    \sin [e^* (V_{S1} \!-\! V_\text{2,eq})t] \!+\! f_\gamma  \cos [e^* (V_\text{1,eff}^0 \!-\! V_\text{2,eq})t] \right\}\\
    & = 0.
\end{aligned}
\end{equation}
This integral vanishes since it is proportional to the tunneling charge current [see Eq.~\eqref{eq:currents_connected}], which equals zero at effective equilibrium.

We have thus proven the validity of the Onsager relation in our system, when the function $f_2$ of the equilibrium channel, 2, has the general expression, i.e., $f_2^\text{general}$.
Notice that this conclusion remains valid as long as $f_2^\text{general}$ does not contain a term that depends on a single $t$ (in contrast to the dependence on $T_\text{2,eq}$, and the temperature-time combination, $T_\text{2,eq} t$).
Indeed, when $f_2^\text{general}$ contains a contribution that depends on a single $t$, the system must contain another energy scale, the existence of which can sabotage the Onsager relation.
Importantly, conclusions above are arrived when $\tau_0 \to 0$ is a positive infinitesimal that is simply introduced to guarantee the convergence of the integral. The dimensionless quantity $T_\text{2,eq} \tau_0$ is thus safely negligible.

\section{Channel-resolved conductances in the collision-free limit}
\label{app:chi_gamma}

In the main part of this work, we define only effective linear response transport coefficients, i.e., close to the effective equilibrium point where $T_1 \to T_2$ and $V_1 \to V_2$.
Here, $V_1$ and $T_1$ denote the chemical potential and temperature, whether real or effective, of channel 1, and likewise for $V_2$ and $T_2$.
The subscripts indicating their status, ``eq'' or ``eff'' [e.g., in $V_\text{1,eff}$ for channel 1 in Fig.~\ref{fig:three_setups}(a)], are omitted for notational simplicity and generality.

\begin{figure}[h!]
  \includegraphics[width= 0.8 \linewidth]{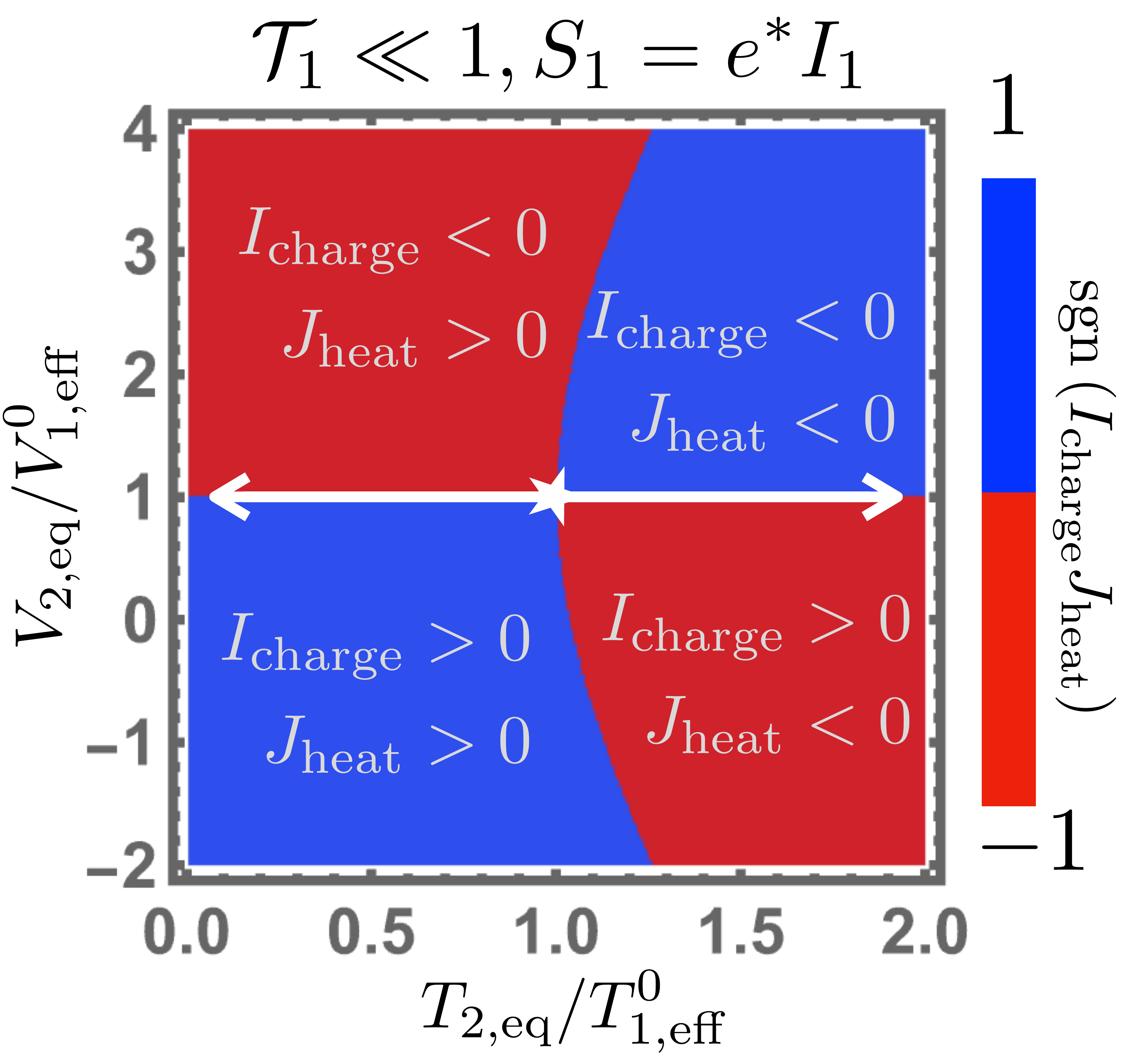}
  \caption{\textbf{Illustration of the regime we consider in Appendix.~\ref{app:chi_gamma}},
  for the setup of Fig.~1(a) (containing one in-equilibrium and one out-of-equilibrium channels) in the collision-free limit.
  Here we require only $V_\text{2,eq} = V_\text{1,eff}^0$, and investigate the contribution to heat conductance from each channel individually, by changing the temperature ratio, $T_\text{2,eq}/T_\text{1,eff}^0$, i.e., along both directions of the white arrow.
  }
  \label{fig:non_equilibrium_regimes}
\end{figure}

This section provides information concerning two topics.
Firstly, we provide more details on the derivation of $\chi_s$ and $\gamma_s$ in the collision-free limit, whose expressions in the integral form have been provided in Appendix~\ref{sec:gamma_chi_functions}.
In addition, we also show that in the collision-free limit, definitions of charge and heat conductances can actually be extended to systems beyond the effective linear response regime.
More specifically, in this section we require only $V_1 \to V_2$.
The equality between temperatures is however relaxed [see Fig.~\ref{fig:non_equilibrium_regimes} for the setup of Fig.~\ref{fig:three_setups}(a) of the main text].
By allowing temperatures of two channels to be different, one can study the contribution of each channel (individually) to transport coefficients.
Importantly, as will be shown shortly, the contribution of an equilibrium channel differs drastically from that of an out-of-equilibrium one.

%%%%%%%%%%%%%%%%%%%%%%%%%
%%%%%%%%%%%%%%%%%%%%%%%%%

Since we focus on the collision-free limit, $L_\text{JV}^s = 0$, such that the heat conductance simply equals the response of heat current to the temperature difference, i.e., $\kappa^s = L_\text{JT}^s = \partial_{T_1 - T_2} J_\text{heat}^s |_\text{eff. equil.}$.
Before visiting three setups separately, we decompose the charge and heat conductance into
\begin{equation}
\begin{aligned}
    G^s & = \frac{1}{2} G_1^s + \frac{1}{2} G_2^s\\
    & \equiv \left(\frac{1}{2}\frac{\partial}{\partial V_1} I_\text{charge}^s\right) + \left(- \frac{1}{2}\frac{\partial}{\partial V_2} I_\text{charge}^s \right),\\
    \kappa^s & =  \frac{1}{2}\kappa_1^s + \frac{1}{2}\kappa_2^s \\
    & \equiv \left(\frac{1}{2}\frac{\partial}{\partial T_1} J_\text{heat}^s\right) + \left(- \frac{1}{2}\frac{\partial}{\partial T_2} J_\text{heat}^s\right),
\end{aligned}
\label{eq:separate_conductances}
\end{equation}
referring to the response of charge ($G_1^s$ and $G_2^s$) and thermal ($\kappa_1^s$ and $\kappa_2^s$) currents to the variation of temperature or bias of each channel separately.
In Eq.~\eqref{eq:separate_conductances}, the factor of one half is added, such that e.g., $\kappa_1^s$ and $\kappa_2^s$ can be interpreted as the response of the heat current $J_\text{heat}^s $ to, \textit{individually}, the temperature of channel 1 and 2, respectively.
A factor one half is likewise introduced to the definition of the charge-conductance contribution from each channel.
Notice that strictly speaking, charge and heat conductances should be defined after differentiating over the parameter differences, i.e., the voltage difference, $V_1 - V_2$, and the temperature difference, $T_1 - T_2$.
We however can arrive at the relatively simpler expression, i.e., Eq.~\eqref{eq:separate_conductances}, after noticing the fact that the temperature and voltage of one channel influence only correlation functions of this channel (not the other one), such that we can treat the differentiation over the temperature difference $T_1 - T_2$, as
\begin{equation}
\begin{aligned}
    & \frac{\partial \left[ n_\text{p,1} (T_1, V_1) n_\text{h,2} (T_2, V_2) \right]}{\partial (T_1 - T_2)}  \\
    = & \frac{1}{2}\! \left (\frac{\partial}{\partial T_1}\! -\! \frac{\partial}{\partial T_2} \right) \left[ n_\text{p,1} (T_1, V_1) n_\text{h,2} (T_2, V_2)  \right],
\end{aligned}
\label{eq:differentiation_transformation}
\end{equation}
and likewise for the differentiation over the voltage difference, $V_1 - V_2$.
Eq.~\eqref{eq:differentiation_transformation} is the reason that we can decompose the contribution to the conductances (either the charge or heat conductance), into that from each channel separately.

\subsection{Tunneling between two non-equilibrium channels}

We begin with setup of Fig.~\ref{fig:three_setups}(b) of the main text, where two non-equilibrium channels communicate through the collider.
This setup is chosen as the starting point, as corresponding derivations are relatively simple.
Of this setup, when $I_1 \to I_2$ (meaning $V_\text{1,eff} \to V_\text{2,eff}$), the charge and heat currents, as well as their corresponding conductances have compact analytical expressions
\begin{equation}
\begin{aligned}
    I_\text{charge}^\text{neq-neq} & = \mathcal{T}_C^{(0)} \frac{\tau_0^{4\delta - 2}}{\pi^2} (1 - 4 \delta) \Gamma (1 - 4\delta)  \sin(2\pi\delta) \sin(4\pi\delta) \\
    & \left[ s_c (T_\text{1,eff} + T_\text{2,eff})  2 \sin^2 \left(\frac{\theta}{2}\right)\right]^{4\delta - 2} (I_1 - I_2),\\
    G^\text{neq-neq} & = \mathcal{T}_C^{(0)} (e^*)^2 \frac{\tau_0^{4\delta - 2}}{\pi^2} \Gamma (2 - 4\delta)  \sin(2\pi\delta) \\
    & \left[s_c (T_\text{1,eff} + T_\text{2,eff})  2 \sin^2 \left(\frac{\theta}{2}\right)\right]^{4\delta - 2},\\
    J_\text{heat}^\text{neq-neq} & = \mathcal{T}_C^{(0)} \frac{\tau_0^{4\delta - 2}}{\pi^2} s_c  \Gamma (1-4\delta) \sin (2\pi\delta) \sin^2 \left(\frac{\theta}{2}\right)\\
    & \left[ s_c (T_\text{eff,1} + T_\text{eff,2}) 2 \sin^2 \left(\frac{\theta}{2}\right) \right]^{ 4\delta - 1 } (T_\text{1,eff} - T_\text{2,eff}),\\
    \kappa^\text{neq-neq} & = \mathcal{T}_C^{(0)} \frac{\tau_0^{4\delta - 2}}{\pi^2} s_c \Gamma (1-4\delta) \sin (2\pi\delta) \sin^2 \left(\frac{\theta}{2}\right) \\
    & \left[ s_c (T_\text{1,eff} + T_\text{2,eff} ) 2 \sin^2 \left(\frac{\theta}{2}\right) \right]^{ 4\delta - 1},
\end{aligned}
\label{eq:current_conductance_neq-neq}
\end{equation}
where $I_1 = I_2$ has been taken for the heat current and the heat conductance.
Eq.~\eqref{eq:current_conductance_neq-neq} correctly reduces to Eqs.~\eqref{eq:universal_coefficients} and \eqref{eq:g_f_neq_neq}, after taking $T_\text{1,eff} = T_\text{2,eff}$, i.e., at effective equilibrium.

We can further figure out contributions to conductances from each channel. Indeed, following their definitions of Eq.~\eqref{eq:separate_conductances}, we arrive at
\begin{equation}
\begin{aligned}
    G_\text{1}^\text{neq-neq} & = G_\text{2 }^\text{neq-neq}= \mathcal{T}_C^{(0)} (e^*)^2 \frac{\tau_0^{4\delta - 2}}{\pi^2} \Gamma (2 - 4\delta)  \sin(2\pi\delta)\\
    &\left[s_c (T_\text{1,eff} + T_\text{2,eff})  2 \sin^2 \left(\frac{\theta}{2}\right)\right]^{4\delta - 2},\\
    \kappa_1^\text{neq-neq} & =  \left[ (4\delta - 1) \frac{T_\text{1,eff} - T_\text{2,eff}}{T_\text{1,eff} + T_\text{2,eff}} + 1\right] \kappa^\text{neq-neq},\\
    \kappa_2^\text{neq-neq} & = \left[ -(4\delta - 1) \frac{T_\text{1,eff} - T_\text{2,eff}}{T_\text{1,eff} + T_\text{2,eff}} + 1\right] \kappa^\text{neq-neq}.
\end{aligned}
\label{eq:conductances_neq-neq}
\end{equation}
In Eq.~\eqref{eq:conductances_neq-neq}, charge conductance from two channels equal, $G_\text{1}^\text{neq-neq} = G_\text{2 }^\text{neq-neq}$, for all values of ratio between two effective temperatures, $T_\text{1,eff}/T_\text{2,eff}$.
This feature, as will be shown, is universally present in all three setups.
Actually, this feature only requires that in the effective linear-response regime, the leading-order electric tunneling current to be proportional to the difference between two chemical potentials.
Such a requirement can be satisfied when the transmission amplitude square at the central collider, $\mathcal{T}_C^{(0)}$, is an energy-independent quantity.

\begin{figure}[h!]
  \includegraphics[width= 0.6 \linewidth]{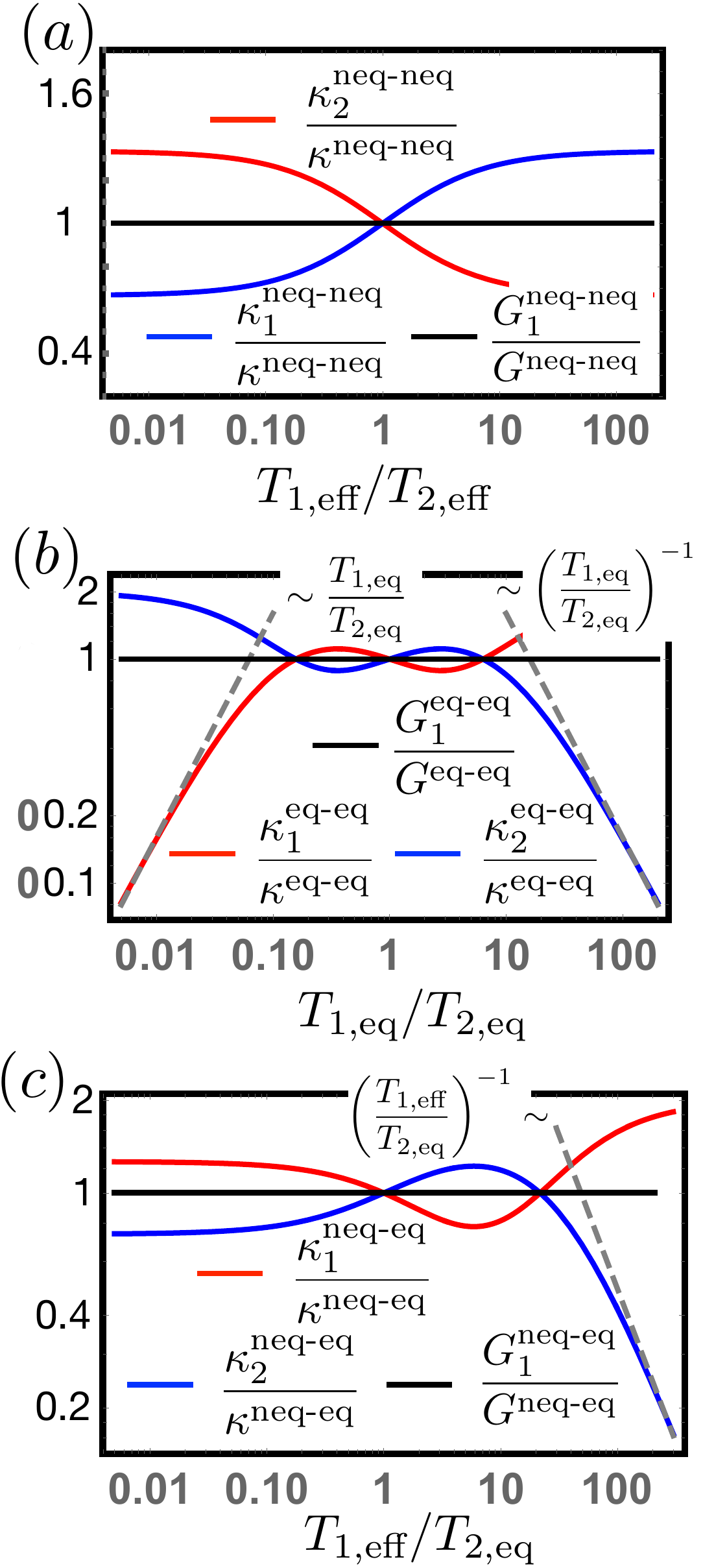}
  \caption{\textbf{Charge and heat conductance from each channel, when $V_1 = V_2$}, for (a) the setup that contains two non-equilibrium channels [Fig.~\ref{fig:three_setups}(b)]; (b) that contains two equilibrium channels [Fig.~\ref{fig:three_setups}(c)], and (c) that contains one equilibrium channel, and one non-equilibrium one [Fig.~\ref{fig:three_setups}(a)]. Without loss of generality, here we take $2\delta = \nu = 1/3$ as the example.
  Major features however also apply to Laughlin systems with other filling fractions.
  Based on these plots, charge conductance contribution from channel 1 always equals that from channel 2 (i.e., $G_\text{1}^s / G^s  = 1$ in all three figures), for all three setups, due to the equal potential ($V_1 = V_2$) requirement.
  In all three setups, their contributions to the heat conductance only equal (where $\kappa_1^s/\kappa^s = \kappa_2^s/\kappa^s = 1$), after further enforcing the equal temperature requirement, i.e., $T_1 = T_2$.
  For an equilibrium channel of (b) or (c), its heat-conductance contribution becomes negligibly small, when its temperature is much smaller than that of the other channel.
  This feature greatly contrasts that of a non-equilibrium channel [in (a) or (b)], which has a finite contribution to the heat conductance, even when its effective temperature is small.
 }
  \label{fig:single_channel_contribution}
\end{figure}

The situation is more complicated for the heat conductance, where both $\kappa_1^\text{neq-neq}/\kappa^\text{neq-neq}$ and $\kappa_2^\text{neq-neq}/\kappa^\text{neq-neq}$ depend on the ratio between effective temperatures $T_\text{1,eff}/T_\text{2,eff}$ of two non-equilibrium channels.
This fact can be more intuitively seen from Fig.~\ref{fig:single_channel_contribution}(a).
However, despite of their temperature-dependence, we emphasize that $\kappa_1^\text{neq-neq}/\kappa_2^\text{neq-neq}$ is between $2\delta/(1-2\delta)$ and $(1-2\delta)/2\delta$.
This fact indicates that conductance contribution from each channel is always of the same order, for all values of the temperature ratio.
This fact, together with the equality between two electric conductance values $G_\text{1}^\text{neq-neq} = G_\text{2 }^\text{neq-neq}$, indicate that for the setup where two non-equilibrium anyonic channels communicate via the collider [i.e., that of Fig.~\ref{fig:three_setups}(b)], conductance contributions from each channel are of the same order.

Before moving to other setups, we notice that in comparison to equilibrium systems, the tuning of effective parameters of a non-equilibrium channel is experimentally more demanding.
Indeed, as shown, for a non-equilibrium channel in the collision-free limit, the effective chemical potential and effective temperature are determined by the non-equilibrium current and tunneling noise, respectively.
Tuning either parameter individually (while keeping the other one fixed) thus intuitively requires fine tuning in practical experiments.
A fine-tuning free protocol is provided by the Supplementary Sec.~S5.

%%%%%%%%%%%%%%%%%%%%%%%%%
%%%%%%%%%%%%%%%%%%%%%%%%%

\subsection{Tunneling between two equilibrium channels}

Now we move to the setup of Fig.~\ref{fig:three_setups}(c) of the main text, where the collider connects two equilibrium channels.
Of this setup, general expressions of tunneling currents, and their corresponding conductances equal
\begin{equation}
\begin{aligned}
&I_\text{charge}^\text{eq-eq} =
2 i e^* \mathcal{T}_C^{(0)} \frac{\tau_0^{4 \delta - 2}}{4\pi^2} (\pi^2 T_\text{1,eq} T_\text{2,eq})^{2\delta}\\
&\int d t \frac{\sin[e^* (V_\text{1,eq} - V_\text{2,eq}) t]}{ \sin^{2\delta} \left[ \pi T_{1,\text{eq}} (\tau_0 + i t) \right] \sin^{2\delta} \left[ \pi T_{2,\text{eq}} (\tau_0 + i t) \right]  },\\
&G^\text{eq-eq} \!=\!\! (e^*)^2 \mathcal{T}_C^{(0)} \! \frac{(\pi^2 T_\text{1,eq} T_\text{2,eq}\tau_0^2 )^{2\delta} }{(\pi T_\text{2,eq} \tau_0 )^2} \gamma^\text{eq-eq} \!\!\left(\!\delta, \frac{T_\text{1,eq}}{T_\text{2,eq}}\!\right),\\
&J_\text{heat}^\text{eq-eq} = 2\delta\  \pi \mathcal{T}_C^{(0)} \frac{\tau_0^{4\delta-2} }{4\pi^2} (\pi^2 T_{1,\text{eq}} T_{2,\text{eq}})^{2\delta} \\
& \int dt \frac{T_\text{2,eq} \cot [\pi T_\text{2,eq} (\tau_0 + i t)] - \cot [\pi T_\text{1,eq} (\tau_0 + i t)]    }{\sin^{2 \delta} [\pi T_\text{1,eq} (\tau_0 + i t)] \sin^{2 \delta} [\pi T_\text{2,eq} (\tau_0 + i t)]}, \\
&\kappa^\text{eq-eq} \!\!=\!\!  \mathcal{T}_C^{(0)}\! (\pi^2 T_\text{1,eq} T_\text{2,eq} \tau_0^2 )^{2\delta}\!\!\frac{T_\text{1,eq} \!\!+\! \!T_\text{2,eq}}{2 (\pi T_\text{2,eq} \tau_0)^2} \chi^\text{eq-eq} \!\!\left(\!\!\delta, \frac{T_\text{1,eq}}{T_\text{2,eq}}\!\right),\\
& = \frac{1}{2} \left( \frac{\partial}{\partial T_\text{1,eq}}  - \frac{\partial}{\partial T_\text{2,eq}} \right) J_\text{heat}^\text{eq-eq}.
\end{aligned}
\end{equation}
More specifically, 
\begin{equation}
\begin{aligned}
    &\gamma^\text{eq-eq} \left( \delta, r \right)  = \frac{i}{2\pi^2} \int d t' \frac{t'}{ \sin^{2\delta} \left[ r (\tau_0' + i t') \right] \sin^{2\delta}  (\tau_0' + i t')   } ,\\
    &\chi^\text{eq-eq} (\delta, r) = \frac{\pi \delta}{2}\left( \frac{2\delta}{r} \frac{1-r}{1 + r} +  \frac{\partial}{\partial r} \right) \Omega^\text{eq-eq} (\delta, r),\\
    &\Omega^\text{eq-eq} (\delta, r) \equiv \frac{1}{\pi} \int dt' \frac{\cot (\tau_0' + i t') - r \cot[r(\tau_0' + i t')]  }{\sin^{ 2\delta}[ r (\tau_0' + i t')] \sin^{2\delta} (\tau_0' + i t')},
\end{aligned}
\end{equation}
where $r \equiv T_\text{1,eq}/T_\text{2,eq}$, referring to the ratio between two real temperatures, and $\Omega^\text{eq-eq} (\delta, r) $ is a dimensionless quantity that depends on only the temperature ratio $r$, and the filling fraction $\nu$.
In comparison to Eq.~\eqref{eq:g_f_eq_eq}, here $\gamma^\text{eq-eq} (\delta, r)$ and $\chi^\text{eq-eq} (\delta, r)$ are functions of both $\delta$, as well as the temperature ratio $r = T_\text{1,eq}/T_\text{2,eq}$.
Indeed, in this section, we require only the particle symmetry, present as long as $V_\text{1,eq} = V_\text{2,eq}$.
At effective equilibrium, $r = 1$, these functions reduce to Eq.~\eqref{eq:g_f_eq_eq}.
To explicitly present contribution of each channel to both charge and heat conductances, we plot $\kappa_1^\text{eq-eq}/\kappa^\text{eq-eq}$,
$\kappa_2^\text{eq-eq}/\kappa^\text{eq-eq}$, and
$G_\text{1}^\text{eq-eq}/G^\text{eq-eq}$ in Fig.~\ref{fig:single_channel_contribution}(b).

Based on Fig.~\ref{fig:single_channel_contribution}(b), the same as the setup [i.e., Fig.~\ref{fig:single_channel_contribution}(a)] that contains two non-equilibrium channels, contribution to the charge conductance is the same [the black line of Fig.~\ref{fig:single_channel_contribution}(b)] for each channel.
The heat conductance, however, displays distinct features between these two setups.
Indeed, $\kappa_1^\text{eq-eq} \gg \kappa_2^\text{eq-eq}$ when $T_\text{1,eq} \gg T_\text{2,eq}$; instead, $\kappa_1^\text{eq-eq} \ll \kappa_2^\text{eq-eq}$ when $T_\text{1,eq} \ll T_\text{2,eq}$.
This fact indicates that for the setup of Fig.~\ref{fig:three_setups}(c), where both involved channels are in-equilibrium, the channel with a much larger temperature dominates the contribution to the heat conductance.
Especially, in the limit where $T_\text{1,eq} \ll T_\text{2,eq}$, $\kappa_1^\text{eq-eq} / \kappa_2^\text{eq-eq} \sim T_\text{1,eq} / T_\text{2,eq}$, indicating the polynomial decay of $\kappa_1^\text{eq-eq}$ when $T_\text{1,eq}$ vanishes.
This feature greatly contrasts that of Fig.~\ref{fig:single_channel_contribution}(a) of the setup that contains two non-equilibrium channels, where instead thermal-conductance contribution of each channel is always of the same order.

\subsection{Tunneling between an equilibrium channel, and an out-of-equilibrium one}

Finally, we consider the setup of Fig.~\ref{fig:three_setups}(a), where the collider bridges an equilibrium channel and an out-of-equilibrium one.
Of this setup, currents and conductances equal
\begin{widetext}
\begin{equation}
\begin{aligned}
    I_\text{charge}^\text{neq-eq} & = 2i e^* \mathcal{T}_C^{(0)} \frac{\tau_0^{4\delta-2} (\pi T_\text{2,eq})^{2\delta}}{4\pi^2} \int dt   \frac{ \exp \left\{ -2 s_c T_\text{1,eff} \sin^2 (\theta/2) | t|   \right\} }{(\tau_0 + i t)^{2\delta} \sin^{2\delta} \left[ \pi T_\text{2,eq} (\tau_0 + i t) \right]  } \sin\left[ e^* \left( V_\text{1,eff}^0 - V_\text{2,eq} \right) t \right],\\
    G^\text{neq-eq}  & = (e^*)^2 \mathcal{T}_C^{(0)} \pi^{2\delta-2} ( T_\text{2,eq} \tau_0)^{4\delta-2}  \gamma^\text{neq-eq}\left(\delta, \frac{T_\text{1,eff}}{T_\text{2,eq}} \right)\\
    & = 2i (e^*)^2 \mathcal{T}_C^{(0)} \frac{\tau_0^{4\delta-2} (\pi T_\text{2,eq})^{2\delta}}{4\pi^2} \int dt \ t\  \frac{ \exp \left\{ -2 s_c T_\text{1,eff} \sin^2 (\theta/2) | t|   \right\} }{(\tau_0 + i t)^{2\delta} \sin^{2\delta} \left[ \pi T_\text{2,eq} (\tau_0 + i t) \right]  } ,\\
    J_\text{heat}^\text{neq-eq} & = -i \mathcal{T}_C^{(0)} \frac{\tau_0^{4\delta - 2}}{4\pi^2} (\pi T_\text{2,eq})^{2\delta} \int dt \frac{ \exp \left\{ -2 s_c T_\text{1,eff} \sin^2 (\theta/2) | t|   \right\} }{(\tau_0 + i t)^{2\delta} \sin^{2\delta} \left[ \pi T_\text{2,eq} (\tau_0 + i t) \right]  } \\
    \times & \left\{ \frac{-2i\delta}{\tau_0 + i t} - \text{sgn}(t) s_c 2 \sin^2(\theta/2) T_\text{1,eff} + \frac{2i \pi \delta \ T_\text{2,eq}}{\tan [\pi T_\text{2,eq} (\tau_0 + i t)]} \right\},\\
    \kappa^\text{neq-eq} & = \mathcal{T}_C^{(0)} (\pi T_\text{2,eq}\tau_0)^{4\delta - 2} \frac{T_\text{1,eff} + T_\text{2,eq}}{2} \chi^\text{neq-eq}\left(\delta, \frac{T_\text{1,eff}}{T_\text{2,eq}} \right) = \frac{1}{2} \left( \frac{\partial}{\partial T_\text{1,eff}}  - \frac{\partial}{\partial T_\text{2,eq}} \right) J_\text{heat}^\text{neq-eq},
\end{aligned}
\label{eq:setup_b_conductances}
\end{equation}
\end{widetext}
where, as the reminder, we have taken equal potential, i.e., $V_\text{2,eq} = V_\text{1,eff}^0 = I_1 \sin(\theta)/(e^*)^2 $ when evaluating the heat current $J_\text{heat}^\text{neq-eq}$ and the corresponding heat conductance $\kappa^\text{neq-eq}$.
Here we did not provide the analytical expression of $\kappa_1^\text{neq-eq}$ and $\kappa_2^\text{neq-eq}$ to avoid complexity.
In Eq.~\eqref{eq:setup_b_conductances}, we have defined two dimensionless quantities
\begin{equation}
\begin{aligned}
    &\gamma^\text{neq-eq}\left(\delta, r \right) = \frac{i \pi^{2\delta-2}}{2}\! \int\! dt' \!\  \frac{ \exp \left\{ -\frac{s_c}{\pi} 2  \sin^2 (\theta/2) r| t'|   \right\} t' }{(\tau_0' + i t')^{2\delta} \sin^{2\delta}  (\tau_0' + i t')  },\\
    &\chi^\text{neq-eq} \left(\delta, r \right) = - \frac{i}{4} \pi^{-2\delta} \left( \frac{\partial}{\partial r} - \frac{4\delta}{1 + r} \right) \Omega^\text{neq-eq} (\delta, r),\\
    &\Omega^\text{neq-eq} (\delta, r) \equiv \pi^{2\delta-1} \int dt' \frac{\exp \left[ -\frac{s_c}{\pi} 2 r \sin^2 (\theta/2) |t'| \right]}{(\tau_0' + i t')^{2\delta} \sin^{2\delta}  (\tau_0' + i t') } \\
    & \left\{ \frac{-2i \pi \delta}{\tau_0' + i t'} -2 r \text{sgn} (t') s_c  \sin^2 (\theta/2) + \frac{2i\pi\delta}{\tan (\tau_0' + it')} \right\},
\end{aligned}
\end{equation}
where again these functions depend also on the temperature ratio, $r = T_\text{1,neq}/T_\text{2,eq}$. And they reduce to Eq.~\eqref{eq:g_f_neq_eq} when $r = 1$.
We further show
charge and heat conductance, contributed by each channel, in  Fig.~\ref{fig:single_channel_contribution}(c).
Firstly, charge conductance from channel 1 equals that from channel 2, for different values of the temperature ratio, $T_\text{1,eff}/T_\text{2,eq}$.
This feature is shared by also the other two setups [cf. Figs.~\ref{fig:single_channel_contribution}(a) and \ref{fig:single_channel_contribution}(b)], as now $V_\text{1,eff} = V_\text{2,eq}$, implying that $I_\text{charge}^\text{neq-eq} \propto (V_\text{1,eff} - V_\text{2,eq})$ in the linear response regime.
Secondly, when $T_\text{1,eff}\ll T_\text{2,eq}$, $\kappa^\text{neq-eq}_1/\kappa^\text{neq-eq}$ and $\kappa^\text{neq-eq}_2/\kappa^\text{neq-eq}$ both saturate to constant values that are of the same order, indicating that in this limit both channels significantly participate in the generation of heat conductance.
In the opposite limit $T_\text{1,eff}\gg T_\text{2,eq}$, on the contrary, $\kappa^\text{neq-eq}_1 \gg \kappa^\text{neq-eq}_2$, meaning that the equilibrium channel, with a comparatively much smaller temperature (than that of the non-equilibrium channel), contributes to a vanishingly small amount of heat conductance.
This observation, together with that of Fig.~\ref{fig:single_channel_contribution}(b), where two equilibrium channels communicate through the collider, indicates that heat conductance contributed by an equilibrium channel is negligible, once its temperature is much smaller than that (either effective or real) of the other channel.
We emphasize that the conclusion is totally different for a non-equilibrium channel [either channel of the setup of Fig.~\ref{fig:three_setups}(b), and channel 1 of the setup of Fig.~\ref{fig:three_setups}(a)], which contributes to a significant amount of heat conductance, even if its temperature is comparatively negligible.

To understand this observation, we can express the heat current in a general (i.e., setup un-specified) form, when $T_2 \gg T_1$,
\begin{equation}
    J^s_\text{heat} \!\propto \!T_2^{4\delta} \! \left[ 1 \!+\! \zeta_1 \frac{T_1}{T_2} \!+\! \zeta_2\left( \frac{T_1}{T_2} \right)^2 \!+\! O \left( \frac{T_1}{T_2} \right)^3 \right],
\end{equation}
where we have only kept contributions to the leading order of $T_1/T_2$, and $\zeta_1$ and $\zeta_2$ are the prefactors of the linear and quadratic contributions, respectively.
Heat conductance from channel 1, which has a comparatively much smaller temperature, then has the feature
\begin{equation}
    \kappa^s_1 = \frac{\partial}{\partial T_1} J^s_\text{heat} \propto T_2^{4\delta-1} \left( 0 + \zeta_1 + 2\zeta_2 \frac{T_1}{T_2}  \right).
\end{equation}

For a non-equilibrium channel, $T_\text{1,eff}$ appears in the exponential factor, i.e., $\exp \left\{ -2 s_c T_\text{1,eff} \sin^2 (\theta/2) | t|   \right\}$. Linear-order expansion then leads to a finite $\zeta_1$: the reason that a non-equilibrium channel contributes to a significant amount of heat conductance, even when its temperature is vanishingly small.

In great contrast, for an equilibrium channel, its temperature $T_\text{1,eq}$ appears in the denominator of correlation functions, in terms of a triangular function, i.e., $\sin^{-2\delta} [\pi T_\text{1,eq} (\tau_0 + it)]$.
When expanding triangular functions, the linear term vanishes, meaning that $\zeta_1 = 0$ for an equilibrium channel, when $T_\text{1,eq}\ll T_2$. As the consequence, for an equilibrium channel, the leading-order correction is proportional to the small temperature ratio, i.e., $\kappa_1^s \propto T_\text{1,eq}/ T_2$ (here $s =$ ``eq-neq'' or ``eq-eq'').
Its contribution to the heat conductance is thus vanishingly small in this limit.

\section{Effective parameters after including the direct tunneling of non-equilibrium anyons}
\label{app:eff_para_direct}

In this section, we show how to obtain the effective parameters ($V_\text{1,eff}$ and $T_\text{1,eff}$) of channel 1, when it is beyond the collision-free limit.
Before turning to the mathematical details, we first explain how direct tunneling of non-equilibrium anyons modifies these effective parameters.

Firstly, in the collision-free limit, $f_\beta = f_\gamma = 0$ in Eq.~\eqref{eq:currents_connected}, such that both the tunneling charge and heat currents vanish, when $V_\text{2,eq} = V_\text{1,eff}^0$ and $T_\text{2,eq} = T_\text{1,eff}^0$.
In this case, the tunneling from channel 1 to channel 2, induced by time-domain braiding, is compensated by the reverse tunneling from channel 2 to channel 1, driven by the voltage bias in channel 2.
However, when $\mathcal{T}_\text{A}$ becomes non-negligible, direct tunneling from channel 1 to channel 2 can no longer be neglected, generating finite $I^\text{dir}_\text{charge}$ and $J^\text{dir}_\text{heat}$. Restoring the modified effective-equilibrium condition, $I_\text{charge}^\text{coll} = J_\text{heat}^\text{coll} =0$, therefore requires a compensating change in the time-domain-braiding contribution, which can be achieved through shifts in the effective parameters $V_\text{1,eff}$ and $T_\text{1,eff}$. Consequently, $V_\text{1,eff} - V_\text{1,eff}^0$ and $T_\text{1,eff} - T_\text{1,eff}^0$ become finite.

%%%%%%%%%%%%%%%%%%%%%%%%%
\begin{figure*}[ht]
  \includegraphics[width= 1 \textwidth]{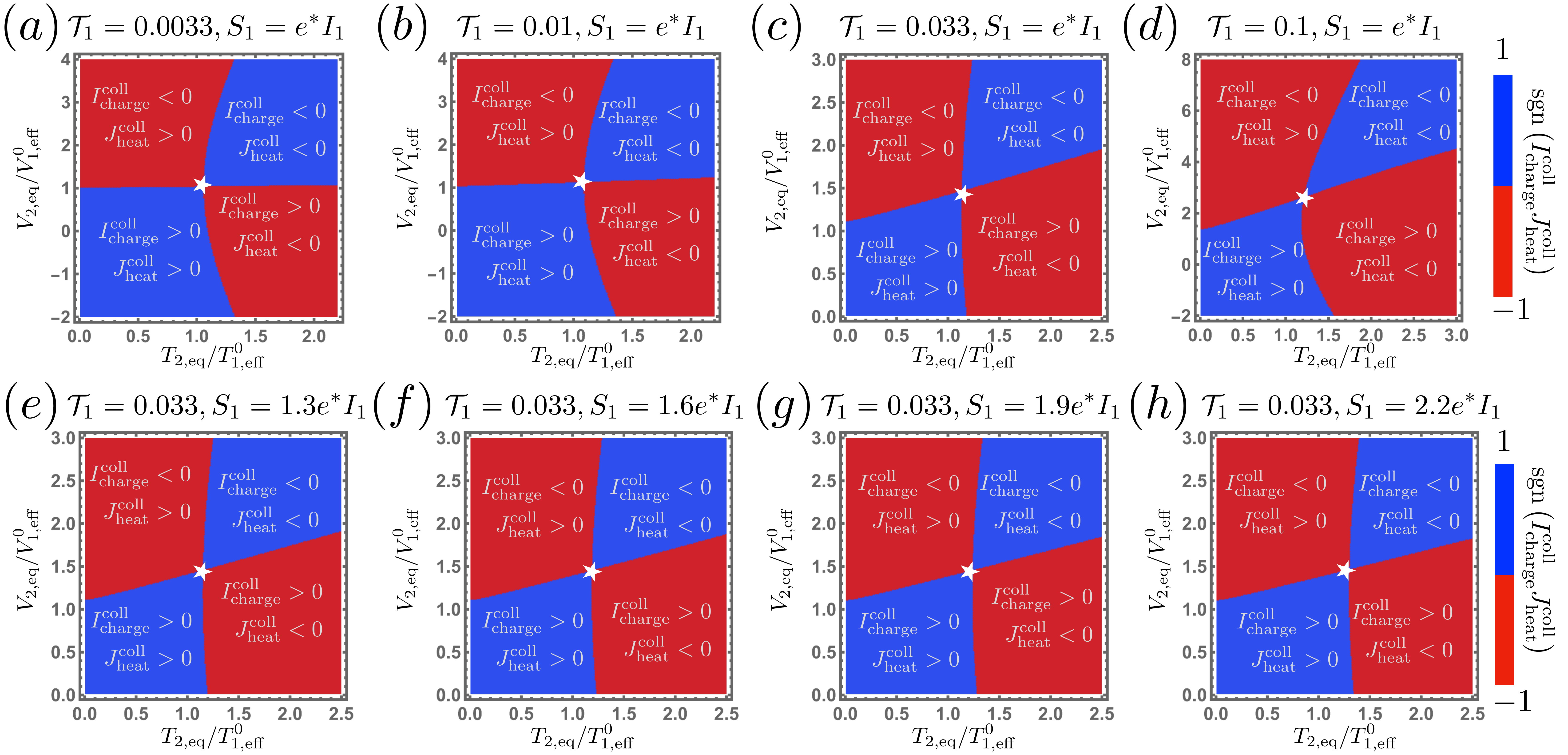}
  \caption{\textbf{Plots of the sign of current-product, i.e., $\text{sgn} (I_\text{charge}^\text{coll} J_\text{heat}^\text{coll})$, of different cases.}
  In these plots, the white star highlights the effective equilibrium point, where $V_\text{2,eq} = V_\text{1,eff}$ and $T_\text{2,eq} = T_\text{1,eff}$.
  (a) to (d): Plots of $\text{sgn} (I_\text{charge}^\text{coll} J_\text{heat}^\text{coll})$, when $S_1 = e^* I_1$, but with different values of transmission probability, $\mathcal{T}_1$, through the diluter.
  (e) to (h): Plots of $\text{sgn} (I_\text{charge}^\text{coll} J_\text{heat}^\text{coll})$, when $\mathcal{T}_1 = 0.033$. Here we instead change the ambient source temperature $T_{S1}$, which correspondingly modifies the noise-current ratio $S_1/(e^* I_1)$ (see the Supplementary Fig.~S2).
  Based on these figures, effective parameters $T_\text{1,eff}$ and $V_\text{1,eff}$ non-universally depend on details of the diluter.
 }
  \label{fig:effective_parameters_collision}
\end{figure*}
%%%%%%%%%%%%%%%%%%%%%%%%%

Mathematically, this modification of effective parameters, accompanied by the direct tunneling of non-equilibrium anyons, can be written into a matrix form,
\begin{equation}
\begin{pmatrix}
V_\text{1,eff} - V_\text{1,eff}^0 \\
T_\text{1,eff} - T_\text{1,eff}^0
\end{pmatrix}
= -\mathcal{R}^{-1} \begin{pmatrix}
I_\text{charge}^\text{dir} \\
J_\text{heat}^\text{dir} 
\end{pmatrix},
\end{equation}
with the response matrix $\mathcal{R}$ defined as
\begin{equation}
\mathcal{R} = 
\begin{pmatrix}
 \partial_{V_\text{2,eq}} I^\text{TDB}_\text{charge}  & \partial_{T_\text{2,eq}} I^\text{TDB}_\text{charge} \\
   \partial_{V_\text{2,eq}} J^\text{TDB}_\text{heat}  & \partial_{T_\text{2,eq}} J^\text{TDB}_\text{heat}
\end{pmatrix} \Bigg|_\text{eff. equil.},
\label{eq:response_matrix}
\end{equation}
whose matrix elements are evaluated in the collision-free limit, near effective equilibrium.
In Eq.~\eqref{eq:response_matrix}, $I^\text{TDB}_\text{charge}$ and $J^\text{TDB}_\text{heat}$ refer to the charge current and heat current, respectively, that are generated by time-domain braiding processes.
From this perspective, near the collision-free limit, the change of the effective parameters are induced by the charge and heat currents ($I_\text{charge}^\text{dir}$ and $J_\text{heat}^\text{dir}$), due to the direct tunneling of non-equilibrium anyons.
In Eq.~\eqref{eq:response_matrix}, the off-diagonal elements vanish due to particle-hole symmetry at the effective-equilibrium point in the collision-free limit.
Notably, although Eq.~\eqref{eq:response_matrix} describes linear response, the matrix element $\partial_{T_\text{2,eq}} J^\text{TDB}_\text{heat}$ scales as $\mathcal{T}_1^{-4\delta}$ at fixed $I_1$. Consequently, the shift in $T_\text{1,eff} - T_\text{1,eff}^0$ is parametrically smaller than that in $V_\text{1,eff} - V_\text{1,eff}^0$, as shown in Fig.~\ref{fig:effective_parameters_collision}.

The inclusion of the direct tunneling of non-equilibrium anyons brings in an extra complexity in determining the effective parameters.
As the reminder, as shown by Eqs.~\eqref{eq:it_and_lit} and \eqref{eq:jq_and_lvj} of the main text, in the collision-free limit, both the Peltier and Seebeck coefficients vanish at $V_\text{2,eq} = V_\text{1,eff}^0$. One can therefore determine $V_\text{1,eff}^0$ and $T_\text{1,eff}^0$ sequentially: first imposing a vanishing charge current to determine $V_\text{1,eff}^0$, and then a vanishing heat current to determine $T_\text{1,eff}^0$.
\color{black}
In great contrast, with the direct tunneling of non-equilibrium anyons, thermoelectric transport coefficients, including Seebeck and Peltier coefficients given by Eqs.~\eqref{eq:current_differentiates} and \eqref{eq:heat_current_differentiates}, become generically finite.
The finiteness of Peltier and Seebeck coefficients is actually in connection to the fact that two effective parameters, i.e., effective chemical potential and effective temperature,
\begin{equation}
    V_\text{1,eff} = V_\text{1,eff} ( S_1 , I_1 , \mathcal{T}_1), \ T_\text{1,eff} = T_\text{1,eff} ( S_1 , I_1 , \mathcal{T}_1).
    \label{eq:nonuniversal_parameters}
\end{equation}
now depend non-universally on the detail of the diluter, which is manifested by its transmission probability.

To understand this fact, similar as Figs.~\ref{fig:coefficients_collision}(a) and \ref{fig:coefficients_collision}(b) of the main text, we plot the sign function of the product of currents, i.e., $\text{sgn} (I_\text{charge}^\text{coll} J_\text{heat}^\text{coll})$, as a function of $V_\text{2,eq}$ and $T_\text{2,eq}$ in Fig.~\ref{fig:effective_parameters_collision}.
Here points locating at boundaries of areas with different colors are featured by a vanishing charge or heat current.
The crossing point of the boundaries, highlighted by the white star, thus refers to the effective equilibrium point where both charge and heat current vanish.

In Fig.~\ref{fig:effective_parameters_collision}(a), the transmission probability through the diluter $\mathcal{T}_1$, and the source temperature $T_{S1}$ are both negligible (more specifically, $\mathcal{T}_1 = 0.0033 \ll 1$ and $S_1 = e^* I_1$).
Of this case, the effective parameters, $V_\text{1,eff}\approx V_\text{1,eff}^0 $ and $T_\text{1,eff} \approx T_\text{1,eff}^0$ [see Eqs.~\eqref{eq:veff0} and \eqref{eq:teff0}], agree nicely with Ref.~\cite{LandscapePRL25}, on effective parameters in the collision-free limit.
In addition, in Fig.~\ref{fig:effective_parameters_collision}(a), parallel directions of the interface (between different-color areas) are parallel to either axis (the $V_\text{2,eq} $ or $T_\text{2,eq}$ axis), indicating the missing of both Peltier and Seebeck coefficients.
When $\mathcal{T}_1$ increases [Figs.~\ref{fig:effective_parameters_collision}(b) to \ref{fig:effective_parameters_collision}(d)], effective parameters clearly begin to deviate from their free-of-collision values [i.e., that given by Fig.~\ref{fig:effective_parameters_collision}(a)].
The deviation becomes increasingly manifest, when $\mathcal{T}_1$ increases.
In addition, in Figs.~\ref{fig:effective_parameters_collision}(b) to \ref{fig:effective_parameters_collision}(d), parallel directions of the interface have finite intersection angles with both axes [a fact that is especially apparent in Fig.~\ref{fig:effective_parameters_collision}(d)], indicating the generation of finite Peltier and Seebeck coefficients by the tunneling and collisions of non-equilibrium anyons.

In Figs.~\ref{fig:effective_parameters_collision}(e) to \ref{fig:effective_parameters_collision}(h), we instead fix the value of $\mathcal{T}_1$ at $\mathcal{T}_1 = 0.033$, and change the temperature $T_{S1}$, leading to different values of the noise-current ratio $S_1/(e^* I_1)$ [see the Supplementary Fig.~S2 for the relation between $T_{S1}$, and the noise-current ratio $S_1/(e^* I_1)$].
Following these figures, modification of $S_1$ can correspondingly influence both effective parameters, $V_\text{1,eff}$ and $T_\text{1,eff}$.
This fact is in great contrast to the situation where one takes into consideration only time-domain braiding effect, where the modification on $S_1$ influences only $T_\text{1,eff}$, but not $V_\text{1,eff}$~\cite{LandscapePRL25}.

\end{appendix}

\newpage

\clearpage

\appendix

\renewcommand{\bibnumfmt}[1]{[S#1]}
\renewcommand{\citenumfont}[1]{S#1}
\global\long\def\theequation{S\arabic{equation}}
\global\long\def\thefigure{S\arabic{figure}}
\renewcommand{\thesection}{\Roman{section}}
\renewcommand{\thesubsection}{\Alph{subsection}}
\setcounter{equation}{0}
\setcounter{figure}{0}

\begin{widetext}
\begin{center}
\textbf{\Large Supplemental Materials for \\``Can out-of-equilibrium linear response reveal anyonic statistics?''}\\
 \vspace{15pt}
\end{center}
\vspace{10pt}

This Supplementary Material file contains more detailed information on five topics: (i) The leading-order (in the tunneling through the diluter) non-equilibrium contribution to the correlation function of the non-equilibrium channel 1; (ii) The non-equilibrium correlation functions, after summing over higher-order tunneling through the diluter; (iii) The non-equilibrium correlation functions, when the source temperature $T_{S1}$ is finite; 
(iv) How to define and measure effective linear response coefficients in terms of the modification of effective parameters of the non-equilibrium channel, and finally, (v) The protocol to experimentally measure effective linear response coefficients, when both channels are out-of-equilibrium [cf. the setup of Fig.~1(b) of the main text].
For simplicity, here we take the Boltzmann constant and the reduced Planck constant as one, $k_B = \hbar = 1$.

\section*{S1.~Correlation functions to the leading order of dilution}

In Secs.~S1 to S3, we derive the non-equilibrium correlation function of Eq.~(16) of the main text.
We begin, in this section, by deriving this equation at zero temperature, when keeping only leading order of dilution.
In Sec.~S2 we include higher-order tunneling through the diluter, and obtain Eq.~(16) of the main text, at zero source temperature.
In Sec.~S3, we further extend this equation to the situation where the source temperature is finite.
To simplify the derivation, we adopt the ideal Laughlin-edge relations $q = \theta/2\pi = 2\delta =\nu$.
The results for practical situations can be obtained straightforwardly by replacing the corresponding factors.

As stated in the main text, in Eq.~(16), the first term of the second and fourth lines correspond to contributions of the so-called time-domain braiding, where tunneling and collisions of non-equilibrium anyons are neglected.
The corresponding collision-free limit is actually the focused topic of most fractional Quantum Hall based anyonic braiding literatures (e.g., Refs.~\cite{SRosenowLevkivskyiHalperinPRL16,SSimNC16,SLeeSimNC22,SschillerPRL23,SLeeNature23}).
In our work, we go beyond the collision-free limit, by including the tunneling and collisions of non-equilibrium anyons, leading to, i.e., terms proportional to $\mathcal{D}_1$ of Eq.~(16) of the main text.
Similar analysis has been carried out by Ref.~\cite{SOneHalfPRB24}, however for only the $\nu = 1/2$ Hong-Ou-Mandel (HOM) interferometer.
Since we are deriving correlation function of a non-equilibrium channel, in Secs.~S1 to S3, we consider the existence of two channels: one source, and one channel that becomes non-equilibrium after receiving a current from the source (via the diluter).
Without loss of generality, we mark the source as $S1$, and the non-equilibrium channel as 1 [cf. the upper two channels of Figs.~1(a) and 1(b) of the main text].

As stated, in Sec.~S1, we begin by considering leading-order expansion at the diluter (in this section), for the zero-temperature situation.
Without loss of generality, we begin with the situation where $t > 0$ is a positive quantity.

\subsection*{S1A.~Different integral regions}

Dealing with non-equilibrium anyons involves integral over their arguments in time.
We begin by dividing these integrals into different regions, for later convenience.
To begin with, to the leading order of dilution ($|\xi_1|^2 = \mathcal{T}_1^{(0)}$), correlation function of channel 1 equals,
\begin{equation}
\begin{aligned}
    \big\langle e^{-i \sqrt{\nu}\phi_1 (t^-,L)} e^{i \sqrt{\nu}\phi_1 (0^+, L)} \big\rangle_\text{neq,2} = -\frac{\mathcal{T}_1^{(0)}}{(2\pi \tau_0)^2} \int_{-\infty}^\infty ds_1 \int_{-\infty}^\infty ds_2 \sum_{\eta_1 \eta_2} \eta_1 \eta_2 e^{-i\nu e V (s_1 - s_2)}\\
    \times \big\langle e^{-i \sqrt{\nu}\phi_1 (t^-,L)} e^{i \sqrt{\nu}\phi_1 (0^+, L)} e^{-i \sqrt{\nu}\phi_1 (s_1^{\eta_1},0)} e^{i \sqrt{\nu}\phi_1 (s_2^{\eta_2},0)} \big\rangle_0 \big\langle e^{i \sqrt{\nu}\phi_{S1} (s_1^{\eta_1},0)} e^{-i \sqrt{\nu}\phi_{S1} (s_2^{\eta_2},0)} \big\rangle_0,
\end{aligned}
\label{eq:leading_order_expansion}
\end{equation}
where $\tau_0$ is the short-time cutoff, $\eta_1$ and $\eta_2$ refer to Keldysh indexes of non-equilibrium anyonic operators, with $s_1$ and $s_2$ their corresponding time arguments.
In Eq.~\eqref{eq:leading_order_expansion}, the subscripts ``neq'' and ``2'' of the left side of the equation highlight the non-equilibrium feature (of channel 1), and the fact that we have taken the second order of the tunneling amplitude $\xi_1$ (of the Hamiltonian at the diluter, $H_\text{dilute} \propto \xi_1 \exp[i\sqrt{\nu} ( \phi_{s1} - \phi_1)] + h.c.$, and $|\xi_1|^2 = \mathcal{T}_1^{(0)}$), respectively.
Another subscript ``0'', appearing in the second line of Eq.~\eqref{eq:leading_order_expansion}, instead indicates to evaluate the correlation functions after turning off both quantum point contacts.
With equilibrium correlation of anyonic vertex operators, Eq.~\eqref{eq:leading_order_expansion} becomes modified into
\begin{equation}
\begin{aligned}
& \big\langle e^{-i \sqrt{\nu}\phi_1 (t^-,L)} e^{i \sqrt{\nu}\phi_1 (0^+, L)} \big\rangle_\text{neq,2}
=  - \mathcal{T}_1^{(0)} \frac{\tau_0^{2\nu}}{(2\pi \tau_0)^2 }\sum_{\eta_1\eta_2} 
\int_{-\infty}^\infty d s_1
\int_{-\infty}^\infty ds_2 \,  \frac{ \eta_1\eta_2   \, e^{-i \nu e V_{S1} (s_1 - s_2)}}{ [l_c + i v  (s_1 - s_2)\chi_{\eta_1,\eta_2} (s_1 - s_2) ]^{2\nu}}
\\
\times & \frac{\tau_0^\nu}{(\tau_0 + i t )^\nu}
\frac{[l_c + i v ( t  - s_1 - L/v ) \chi_{-,\eta_1}(t-s_1)]^{\nu}\, [l_c - i v(  s_2 + L/v) \chi_{+,\eta_2}(- s_2)]^{\nu}}
{[l_c + i v( t - s_2 - L/v ) \chi_{-,\eta_2}(t - s_2 )]^{\nu}\,
[l_c - i v (s_1 + L/v) \chi_{+,\eta_1}(-s_1)]^{\nu} } \\
& =  - |\xi_1|^2 \frac{\tau_0^{2\nu}}{(2\pi \tau_0)^2 } \frac{\tau_0^\nu}{(\tau_0 + i t )^\nu}  \sum_{\eta_1\eta_2} 
\int_{-\infty}^\infty ds
\int_{-\infty}^\infty d\tilde{s}_1 \,  \frac{ \eta_1\eta_2   \, e^{-i \nu eV_{S1} s}}{ [l_c + i v  s\chi_{\eta_1,\eta_2} (s) ]^\nu}
\mathcal{R}_{\eta_1\eta_2} (\tilde{s}_1 -L/v,\tilde{s}_1 - s - L/v, t),
\end{aligned}
\label{eq:second_order_correlation}
\end{equation}
where $l_c \equiv v \tau_0$ refers to the short-distance cutoff, and $L$ refers to the distance between two quantum point contacts, i.e., the diluter and the collider. In the last line, we have 
defined, for later convenience, $s\equiv s_1 - s_2$, $\tilde{s}_1 \equiv s_1 + L /v$, and $\tilde{s}_2 \equiv s_2 + L /v$.
By doing so, $\tilde{s}_1$ and $\tilde{s}_2$ refer to the moments at which a non-equilibrium anyon, injected from the source, arrives at the central collider.
Simplification of the last line of Eq.~\eqref{eq:second_order_correlation} takes into the assumption that the diluter-collider distance, $L$, is much larger than other relevant length scales.
Actually, when $L$ becomes comparable to the typical value of $vt$ (with $v$ the particle velocity), the influence of Klein factors should be further included (see e.g., Ref.~\cite{SGuyonPRB02}) and carefully analyzed.

%%%%%%%%%%%%%%%%%%%%%%%%%
\begin{figure}[h!]
  \includegraphics[width= 0.6 \linewidth]{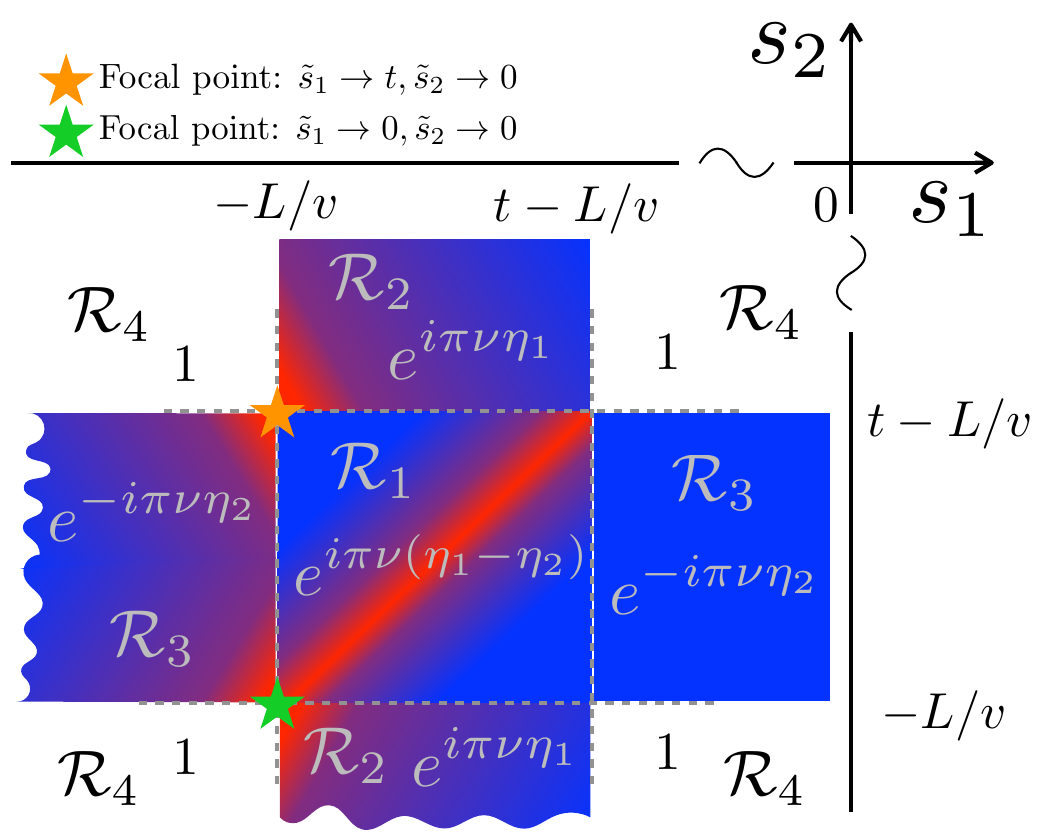}
  \caption{All regions, $\mathcal{R}_1$, $\mathcal{R}_2$,  $\mathcal{R}_3$ and $\mathcal{R}_4$ that are relevant to the integral of Eq.~\eqref{eq:second_order_correlation}.
  The former three regions have finite contributions to the non-equilibrium correlation function, $\big\langle e^{-i \sqrt{\nu}\phi_1 (t^-,L)} e^{i \sqrt{\nu}\phi_1 (0^+, L)} \big\rangle_\text{neq,2}$, to the leading order of dilution $|\xi_1|^2$. The phase factor of $\mathcal{R}_{\eta_1\eta_2} (s_1,s_2)$, cf. Eq.~\eqref{eq:r_eta_value}, has been marked out for each corresponding region.
  Regions $\mathcal{R}_1$, $\mathcal{R}_2$ and $\mathcal{R}_3$ are colored in blue and red, to highlight the fact that $\mathcal{R}_{\eta_1\eta_2} (s_1,s_2)$ has a finite phase in these regions.
  Within these regions, areas with most significant contribution to the integral are highlighted in red.
  The focal points that dominate integrals in Regions $\mathcal{R}_2$ and $\mathcal{R}_3$ are highlighted by orange and green stars, respectively.
 }
  \label{fig:three_regions}
\end{figure}
%%%%%%%%%%%%%%%%%%%%%%%%%

Eq.~\eqref{eq:second_order_correlation} contains the factor
\begin{equation}
\begin{aligned}
    \mathcal{R}_{\eta_1\eta_2} (s_1,s_2,t) & \equiv \frac{ [ l_c + i (vt - vs_1 - L ) \eta_1 ]^\nu \, [ l_c - i ( vs_2 + L ) \eta_2 ]^\nu }{ [ l_c + i (vt - vs_2 - L ) \eta_2  ]^\nu \, [ l_c - i (v s_1 + L ) \eta_1  ]^\nu }\\
    & = \frac{[l_c + i v ( t  -\tilde{s}_1 ) \eta_1]^{\nu}\, [l_c + i v(s - \tilde{s}_1) \eta_2]^{\nu}}
{[l_c + i v( t + s -\tilde{s}_1 ) \eta_2]^{\nu}\,
[l_c - i v \tilde{s}_1 \eta_1]^{\nu} },
\end{aligned}
\label{eq:r_etas}
\end{equation}
which refers to the interplay between non-equilibrium anyons (from the diluter), and those that tunnel at the collider.
Importantly, the phase of $\mathcal{R}_{\eta_1\eta_2} (s_1, s_2,t)$ determines whether the point $(s_1,s_2)$ influences the integral outcome.
More specifically, it can be alternatively written as the product between its norm, and its phase factor, i.e.,
\begin{equation}
    \mathcal{R}_{\eta_1\eta_2} = \big| \mathcal{R}_{\eta_1\eta_2} \big| \times\!
        \left\{ 
       \begin{array}{cc}
      e^{i\pi \nu (\eta_1 - \eta_2) },& \mathcal{R}_1: \quad -\frac{L}{v}<s_1<t-\frac{L}{v}\ \ \text{and}\ \, -\frac{L}{v}<s_2<t-\frac{L}{v},\\[0.1cm]
      e^{i\pi \nu \eta_1},& \mathcal{R}_2: \quad -\frac{L}{v}<s_1<t-\frac{L}{v}\ \ \text{and}\ \, \left(t-\frac{L}{v}<s_2<0 \ \text{or}  \ s_2<-\frac{L}{v}\right),\\[0.1cm]
      e^{-i\pi \nu \eta_2},& \mathcal{R}_3: \quad -\frac{L}{v}<s_2<t-\frac{L}{v}\ \ \text{and}\ \, \left(t-\frac{L}{v}<s_1<0 \ \text{or}  \ s_1<-\frac{L}{v}\right),\\[0.1cm]
      1, &\mathcal{R}_4: \quad \text{otherwise},
      \end{array}
      \right.
      \label{eq:r_eta_value}
\end{equation}
in four different areas (see Fig.~\ref{fig:three_regions}).
Notice that Eq.~\eqref{eq:r_eta_value} is strictly valid, when the distances from branch cuts are much larger than the short-distance cutoff, $l_c$.
Eq.~\eqref{eq:r_eta_value} is nevertheless a reasonable simplification. Actually, as will be shown shortly, relevant integrals over time are contributed by a wide range of time [e.g., the last but one line of Eq.~\eqref{eq:d1} and the second line of Eq.~\eqref{eq:d2_d3_first}, both bounded by $\sim 1/\nu e V_{S1}$]; The contribution of $t\to 0$, where $l_c$ becomes important, is however negligible.

As the starting point, notice that $\mathcal{R}_{\eta_1\eta_2} = |\mathcal{R}_{\eta_1\eta_2}|$ for points belonging to region $\mathcal{R}_4$ [the last line of Eq.~\eqref{eq:r_eta_value}].
These points thus do not contribute to the integral of Eq.~\eqref{eq:second_order_correlation}, as $\sum_{\eta_1\eta_2} \eta_1\eta_2 = 0$, after the summation over $\eta_1$ and $\eta_2$ in Eq.~\eqref{eq:second_order_correlation}.
Below we thus visit only the other three regions, $\mathcal{R}_1$, $\mathcal{R}_2$ and $\mathcal{R}_3$, one after the other.

\subsection*{S1B.~Region $\mathcal{R}_1$}

Contribution of this region actually corresponds to the so-called time-domain braiding discussed by Refs.~\cite{SLeeSimNC22,SschillerPRL23}.
Indeed, as shown by Fig.~\ref{fig:three_regions}, in this region the shifted moments of non-equilibrium anyons, $\tilde{s}_1 $ and $\tilde{s}_2$, corresponding to the moments of their arrival at the collider [as the reminder, see their definitions after Eq.~\eqref{eq:second_order_correlation}], are between the moments of two tunneling events, i.e., at time $0$ and $t$, respectively.
These non-equilibrium anyons thus have no direct correlation with the tunneling events.
Their correlation is instead established via the time-domain braiding that occurs nonlocally in time.

In this region, Eq.~\eqref{eq:r_eta_value} simplifies into
\begin{align}
    &\mathcal{R}_{\eta_1\eta_2}(s_1,s_2,t)
    = \left\{\frac{|vt - vs_1 - L|\,|vs_2 + L| }{|vt - vs_2 - L| \, | vs_1 + L|}\right\}^\nu \,  e^{i\pi \nu (\eta_1-\eta_2)}.
    \label{eqtangling}
\end{align}    
With this expression, correlation function Eq.~\eqref{eq:second_order_correlation} becomes (as a reminder, $s\equiv s_1 - s_2$):
\begin{equation}
\begin{aligned}
&\big\langle e^{-i\sqrt{\nu}\phi_1 (t^-,L)} e^{i \sqrt{\nu}\phi_1 (0^+, L)} \big\rangle_\text{neq,2}^{\mathcal{R}_1}
=  -  \frac{ \mathcal{T}_1^{(0)} \tau_0^{2\nu}}{(2\pi \tau_0 )^2 }  \frac{\tau_0^\nu}{(\tau_0 + i t)^\nu} \\
&  \times \sum_{\eta_1\eta_2}  \eta_1\eta_2 e^{i\pi \nu (\eta_1-\eta_2)}\,
\int_{0}^t d\tilde{s}_1 
\int_{\tilde{s}_1 -t}^{\tilde{s}_1 }\!\! ds \,
\frac{   \, e^{-i \nu eV_{S1} s}}
{ [\tau_0 + i  s\chi_{\eta_1,\eta_2} (s)]^{2\nu}}
\, 
\left\{\frac{(t - \tilde{s}_1)\,(\tilde{s}_1-s) }
    {(t + s - \tilde{s}_1) \, \tilde{s_1}}\right\}^\nu \\
& =   \! -  \frac{ \mathcal{T}_1^{(0)} \tau_0^{2\nu}}{(2\pi \tau_0 )^2 } \frac{\tau_0^\nu}{(\tau_0 \!+\! i t)^\nu}
 \!\sum_{\eta_1\eta_2}\!  \eta_1\eta_2 e^{i\pi \nu (\eta_1-\eta_2)}
\!\!\int_{0}^t \!\! d\tilde{s}_1 \!
\left[\!\int_{0}^{\tilde{s}_1}\!\! ds  
\frac{   \, e^{-i \nu eV_{S1} s}}{ (\tau_0 \!+\! i s\, \eta_2)^{2\nu}}
\!+\!\! \int_{\tilde{s}_1-t}^0 \!\!\!ds 
\frac{   \, e^{-i \nu eV_{S1} s}}{ (\tau_0 \!-\! i   s\, \eta_1 )^{2\nu}}
\!\right] 
\!\left\{\frac{(t\!-\! \tilde{s}_1)\,(\tilde{s}_1\!-\!s) }
    {(t \!+\! s \!-\! \tilde{s}_1) \, \tilde{s_1}}\right\}^\nu
    \\
&  = -  \frac{ \mathcal{T}_1^{(0)} \tau_0^{2\nu}}{(2\pi \tau_0 )^2 } \frac{\tau_0^\nu}{(\tau_0 + i t)^\nu} 2i \sin(\pi\nu)
\int_{0}^t d\tilde{s}_1 
\left[ \int_{0}^{\tilde{s}_1} ds \, e^{-i\nu eV_{S1} s} 
\left[\frac{e^{-i\pi\nu}}{ (\tau_0 + i  s)^{2\nu}} - \frac{e^{i\pi\nu}}{( \tau_0 - i  s )^{2\nu}}\right]
\left\{\frac{(t - \tilde{s}_1)\,(\tilde{s}_1-s) }
    {(t + s - \tilde{s}_1) \, \tilde{s_1}}\right\}^\nu\right. \\
&+ \left.\int_{\tilde{s}_1-t}^0 ds 
\, e^{-i \nu eV_{S1} s} 
\left[\frac{e^{-i\pi\nu}}{ (\tau_0 + i  s)^{2\nu}} - \frac{e^{i\pi\nu}}{( \tau_0 - i s )^{2\nu}}\right]
\,\left\{\frac{(t - \tilde{s}_1)\,(\tilde{s}_1-s) }
    {(t + s - \tilde{s}_1) \, \tilde{s_1}}\right\}^\nu\ \right] \\
& =    -  \frac{ \mathcal{T}_1^{(0)} \tau_0^{2\nu}}{(2\pi \tau_0 )^2 } \frac{\tau_0^\nu}{(\tau_0 + i t)^\nu} 2i \sin(\pi\nu) \int_{0}^t d\tilde{s}_1 
\int_{\tilde{s}_1 -t}^{\tilde{s}_1 }\!\! ds 
\, e^{-i \nu eV_{S1} s} 
\left[\frac{e^{-i\pi\nu}}{ (\tau_0 + i  s)^{2\nu}} - \frac{e^{i\pi\nu}}{( \tau_0 - i s )^{2\nu}}\right]
\,\left\{\frac{(t - \tilde{s}_1)\,(\tilde{s}_1-s) }{(t + s - \tilde{s}_1) \, \tilde{s}_1}\right\}^\nu\\
& \approx - \frac{ \mathcal{T}_1^{(0)} \tau_0^{2\nu}}{(2\pi \tau_0 )^2 } \frac{\tau_0^\nu}{(\tau_0 + i t)^\nu}\frac{2\pi}{\Gamma (2\nu)} \left( 1 - e^{2 i\pi\nu} \right) (\nu e V_{S1})^{ 2\nu - 1 } t = -\frac{\tau_0^\nu}{(\tau_0 + i t)^\nu} \left( 1 - e^{2 i\pi\nu} \right) \frac{I_1}{\nu e} t,
\end{aligned}
\label{eq:d1}
\end{equation}
where the superscript $\mathcal{R}_1$ of the correlation function of the first line highlights the integral region.
In the last line of Eq.~\eqref{eq:d1}, it is assumed that $|\tilde{s}_1 |, |\tilde{s}_1 - t| \gg 1/\nu e V_{S1}$, to fully capture the contribution of the branch cut at $s \to 0$.
Notice that in Region $\mathcal{R}_1$, contribution from the branch cut at $s\to -t$, $\tilde{s}_1\to 0$ can not be captured.
Indeed, after taking $s\to -t$, terms inside the square brackets of the last but one line of Eq.~\eqref{eq:d1} will cancel out each other.
As the consequence, the major contribution of region $\mathcal{R}_1$ comes from the branch cut $\tilde{s}_1 \to \tilde{s}_2$, when $0< \tilde{s}_1 \sim \tilde{s}_2 < t$.
This contribution is alternatively addressed as ``disconnected'' in Ref.~\cite{SOneHalfPRB24}, as for non-interacting fermionic systems, it corresponds to disconnected diagrams that do not contribute to normal-ordered correlations (since $\exp[i\pi\nu (\eta_1 - \eta_2)] = 1$ when $\nu = 1$).
For anyonic systems, strictly speaking, tunnelings at moments 0 and $t$ are correlated due to time-domain braiding, thus leading to finite normal-ordered correlations. In Ref.~\cite{SOneHalfPRB24}, this process is nevertheless addressed as disconnected, to emphasize its similarity to disconnected diagrams in non-interacting fermionic systems, and to address the fact that in these diagrams, non-equilibrium anyons do not participate in tunneling or collisions at the collider (such that tunnelings at the central collider are ``disconnected'' from non-equilibrium anyons through the diluter).

In Eq.~\eqref{eq:d1}, we have also used the approximate expression, valid to the leading order of dilution $\mathcal{T}_1^{(0)}$, of non-equilibrium current in channel 1, i.e.,
\begin{equation}
    I_1 = \mathcal{T}_1^{(0)} \frac{ \nu e \sin(2\pi\nu) \Gamma (1-2\nu) (\nu e V_{S1} \tau_0)^{2\nu-1}}{2\pi^2 \tau_0}.
\end{equation}
As stated above, the major contribution of Eq.~\eqref{eq:d1} comes from the branch cut $|s|\equiv |s_1 - s_2| \sim 1/\nu e V_{S1} \ll |\tilde{s}_1 - t|, |\tilde{s}_1|$, i.e., when two involved non-equilibrium anyons are uncorrelated to two tunneling events at the collider.
The other branch-cut contribution at $\tilde{s}_1 \to 0$ and $\tilde{s}_2 \to t$ however vanishes in region $\mathcal{R}_1$, as terms inside the square brackets of (the last but one line of) Eq.~\eqref{eq:d1} cancels out when $s\to -t$ (with $t \gg \tau_0 > 0$).

\subsection*{S1C.~Region $\mathcal{R}_2$}

Now we move to region $\mathcal{R}_2$ of Fig.~\ref{fig:three_regions}.
Within this region, Eq.~\eqref{eq:second_order_correlation}, with its expression given by the second line of Eq.~\eqref{eq:r_eta_value}, simplifies into
\begin{equation}
\begin{aligned}
&\big\langle e^{-i\sqrt{\nu}\phi_1 (t^-,L)} e^{i \sqrt{\nu}\phi_1 (0^+, L)} \big\rangle_\text{neq,2}^{\mathcal{R}_2}
=  -  \frac{ \mathcal{T}_1^{(0)} \tau_0^{2\nu}}{(2\pi \tau_0 )^2 } \frac{\tau_0^\nu}{(\tau_0 + i t)^\nu}\, \sum_{\eta_1\eta_2} \eta_1\eta_2\,  e^{i\pi \nu\eta_1} \, \int_{-L/v}^{t-L/v} ds_1 \, \left(\frac{vt - vs_1 - L}{vs_1 + L}\right)^\nu
\\
&\qquad \times
\left(\int_{-\infty}^{-L/v} ds_2+\int_{t-L/v}^{0} ds_2\right)\, 
\frac{ e^{-i \nu eV_{S1} (s_1 - s_2)}}{[\tau_0 \!+\! i (s_1 \!-\! s_2) \chi_{\eta_1,\eta_2} (s_1 \!-\! s_2) ]^{2\nu}}\,
\left(\frac{|vs_2 + L| }{|vt - vs_2 - L|}\right)^{\nu}
\\
&=
-  \frac{ \mathcal{T}_1^{(0)} \tau_0^{2\nu}}{(2\pi \tau_0 )^2 } \frac{\tau_0^\nu}{(\tau_0 + i t)^\nu}\,  \int_{-L/v}^{t-L/v} ds_1 \, \left(\frac{vt - vs_1 - L}{vs_1 + L} \right)^\nu e^{-i \nu eV_{S1} s_1}\, \sum_{\eta_1\eta_2} \eta_1\eta_2\,  e^{i\pi \nu \eta_1} \,
\\
&\qquad \times
\left[\!\int_{-\infty}^{-L/v}\!\! ds_2
\frac{ e^{i \nu eV_{S1}  s_2}}{[\tau_0 \!+\! i (s_1 \!-\! s_2) \,\eta_2 ]^{2\nu}}\,
\left(\frac{-vs_2 -L }{vt - vs_2 - L}\right)^\nu
\!+\!\int_{t-L/v}^{0}\!\! ds_2
\frac{ e^{i \nu eV_{S1} s_2}}{[\tau_0 \!-\! i (s_1 \!-\! s_2) \,\eta_1]^{2\nu}}\,
\left(\frac{vs_2 \!+\! L }{vs_2-v t + L}\right)^\nu\!
\right]\,\\
& =
-  \frac{ \mathcal{T}_1^{(0)} \tau_0^{2\nu}}{(2\pi \tau_0 )^2 } \frac{\tau_0^\nu}{(\tau_0 + i t)^\nu}\,  \int_{0}^{t} d\tilde{s}_1 \, \left(\frac{vt - v\tilde{s}_1}{v\tilde{s}_1}\right)^\nu e^{-i\nu eV_{S1} \tilde{s}_1}\, \sum_{\eta_1\eta_2} \eta_1\eta_2\,  e^{i\pi \nu \eta_1} \,
\\
&\qquad \times
\left[\int_{-\infty}^{0}\!\! d\tilde{s}_2
\frac{ e^{i \nu eV_{S1}  \tilde{s}_2}}{[\tau_0 \!+\! i (\tilde{s}_1 \!-\! \tilde{s}_2) \,\eta_2]^{2\nu} }\,
\left(\frac{-v\tilde{s}_2 }{vt - v\tilde{s}_2}\right)^\nu
+\,\underbrace{\int_{t}^{L/v}\!\! d\tilde{s}_2
\frac{ e^{i \nu eV_{S1} \tilde{s}_2}}{[\tau_0 \!-\! i (\tilde{s}_1 \!-\! \tilde{s}_2) \,\eta_1]^{2\nu}}\,
\left(\frac{v\tilde{s}_2}{v\tilde{s}_2-v t}\right)^\nu}_{\text{gives zero after summation over}\ \eta_2}\,
\right].
\end{aligned}
\label{eq:d2}
\end{equation}
where the last term of the last line is independent of the Keldysh index, $\eta_2$. It thus vanishes after summation over $\eta_2$.
Similar to Eq.~\eqref{eq:d1} of Region $\mathcal{R}_1$, here the superscript $\mathcal{R}_2$ of Eq.~\eqref{eq:d2} highlights the region where we perform the integral.

We continue and sum over two Keldysh indexes of the first term, leading to
\begin{equation}
\begin{aligned}
    \big\langle e^{-i\sqrt{\nu}\phi_1 (t^-,L)} e^{i \sqrt{\nu}\phi_1 (0^+, L)} \big\rangle_\text{neq,2}^{\mathcal{R}_2}& =
-  \frac{ \mathcal{T}_1^{(0)} \tau_0^{2\nu}}{(2\pi \tau_0 )^2 } \frac{\tau_0^\nu}{(\tau_0 + i t)^\nu}2 i\sin(\pi\nu)  \int_{0}^{t} d\tilde{s}_1 \, \left(\frac{vt - v\tilde{s}_1}{v\tilde{s}_1}\right)^\nu e^{-i\nu eV_{S1} \tilde{s}_1} \,
\notag 
\\
&\times
\int_{-\infty}^{0}\!\! d\tilde{s}_2
e^{i \nu eV_{S1}  \tilde{s}_2} \left\{\frac{1}{[\tau_0 + i (\tilde{s}_1 - \tilde{s}_2) ]^{2\nu} } - \frac{1}{[\tau_0 - i (\tilde{s}_1 - \tilde{s}_2) ]^{2\nu} } \right\}\,
\left(\frac{-v\tilde{s}_2 }{vt - v\tilde{s}_2}\right)^\nu.
\end{aligned}
\end{equation}
Due to the oscillating factors, $\exp(-i\nu e V_{S1} \tilde{s}_1)$ and $\exp(i\nu e V_{S1} \tilde{s}_2)$, $\tilde{s}_1$ and $\tilde{s}_2$ are both bounded by $(\nu e V_{S1})^{-1}$.
Assuming $(\nu e V_{S1})^{-1} \ll t$, integral of region $\mathcal{R}_2$ simplifies into
\begin{equation}
\begin{aligned}
   \big\langle e^{-i\sqrt{\nu}\phi_1 (t^-,L)} e^{i \sqrt{\nu}\phi_1 (0^+, L)} \big\rangle_\text{neq,2}^{\mathcal{R}_2}& =  -  \frac{ \mathcal{T}_1^{(0)} \tau_0^{2\nu}}{(2\pi \tau_0 )^2 } \frac{\tau_0^\nu}{(\tau_0 + i t)^\nu} 2 i\sin(\pi\nu)  \int_{0}^{t} d\tilde{s}_1 \, \left(\frac{1}{v\tilde{s}_1}\right)^\nu e^{-i\nu eV_{S1} \tilde{s}_1}\\
   &\times
\int_0^\infty\!\! d\tilde{s}_2
e^{-i \nu eV_{S1}  \tilde{s}_2} \left\{\frac{1}{[\tau_0 + i (\tilde{s}_1 + \tilde{s}_2) ]^{2\nu} } - \frac{1}{[\tau_0 - i (\tilde{s}_1 + \tilde{s}_2) ]^{2\nu} } \right\}\,
(v\tilde{s}_2)^\nu\\
& \approx  -  \frac{ \mathcal{T}_1^{(0)} \tau_0^{2\nu}}{(2\pi \tau_0)^2  } \frac{\tau_0^\nu}{(\tau_0 + i t)^\nu} 4 \sin^2(\pi\nu) \int_{0}^{\infty} d\tilde{s}_1 \,  \frac{e^{-i\nu eV_{S1} \tilde{s}_1}}{\tilde{s}_1^\nu}
\int_0^\infty\!\! d\tilde{s}_2
  \frac{e^{-i \nu eV_{S1}  \tilde{s}_2} \tilde{s}_2^\nu}{(\tilde{s}_1 + \tilde{s}_2)^{2\nu}},
\end{aligned}
\label{eq:d2_continue}
\end{equation}
where we have replaced $\tilde{s}_2$ by $-\tilde{s}_2$, and the fact that at the singular point $\tilde{s}_1 + \tilde{s}_2 = 0 $, terms within the curly brackets perfectly cancel out.
We have also extended the integral over $\tilde{s}_1$ to infinity (in the last line), in the $(\nu e V_{S1})^{-1} \ll t$ limit, to fully capture the contribution of the $\tilde{s}_1 \to 0$ branch cut.
This extension of the range of integral is the only approximation in Eq.~\eqref{eq:d2_continue}.
Notice that the integral outcome of Eq.~\eqref{eq:d2_continue} is a complex number, which is different from the $\nu = 1/2$ situation of Ref.~\cite{SOneHalfPRB24}, where the integral outcome is purely imaginary.
We will perform the integral explicitly, after combining Eqs.~\eqref{eq:d2_continue} and \eqref{eq:d3}: the latter integral, which will be analyzed shortly (in the next subsection), comes from the contribution of region $\mathcal{R}_3$.
Actually, following Fig.~\ref{fig:three_regions}, integral outcomes corresponding to both focal points (cf. the green and orange points) involve both regions $\mathcal{R}_2$ and $\mathcal{R}_3$.

\subsection*{S1D.~Region $\mathcal{R}_3$}

Finally we move to region $\mathcal{R}_3$.
Following the third line of Eq.~\eqref{eq:r_eta_value}, correlation function to the leading-order of dilution $|\xi_1|^2$, i.e., Eq.~\eqref{eq:second_order_correlation}, simplifies into
\begin{equation}
\begin{aligned}
&\left\langle e^{i \sqrt{\nu} \phi_1(L,t)}\, e^{-i \sqrt{\nu}\phi_1(L,0)}\right\rangle_\text{neq,2}^{\mathcal{R}_3}=
 -  \frac{ \mathcal{T}_1^{(0)} \tau_0^{2\nu}}{(2\pi \tau_0 )^2 } \frac{\tau_0^\nu}{(\tau_0 + i t)^\nu}\, \sum_{\eta_1\eta_2} \eta_1\eta_2\,  e^{-i\pi \nu \eta_2} \, \int_{-L/v}^{t-L/v} ds_2 \, 
\left(\frac{|vs_2 + L| }{|vt - vs_2 - L|}\right)^\nu
\\
&\qquad \times
\left(\int_{-\infty}^{-L/v} ds_1+\int_{t-L/v}^{0} ds_1\right)\, 
\frac{ e^{-i\nu eV_{S1} (s_1 - s_2)}}{[\tau_0 \!+\! i(s_1 \!-\! s_2) \chi_{\eta_1,\eta_2} (s_1 \!-\! s_2)]^{2\nu} }\,
\left(\frac{|vt - vs_1 - L|}{|vs_1 + L|}\right)^\nu\\
& = -  \frac{ \mathcal{T}_1^{(0)} \tau_0^{2\nu}}{(2\pi \tau_0 )^2 } \frac{\tau_0^\nu}{(\tau_0 + i t)^\nu}\,  
\int_{0}^{t} d\tilde{s}_2 \, 
\left(\frac{v\tilde{s}_2}{vt - v\tilde{s}_2 }\right)^\nu e^{i \nu eV_{S1} \tilde{s}_2}\, \sum_{\eta_1\eta_2} \eta_1\eta_2\,  e^{-i\pi \nu \eta_2} \,
\\
&\qquad \qquad\times
\left[\int_{-\infty}^{0}\!\! d\tilde{s}_1
\frac{ e^{-i\nu eV_{S1}  \tilde{s}_1}}{[\tau_0 \!-\! i (\tilde{s}_1 \!-\! \tilde{s}_2) \,\eta_1]^{2\nu}}\,
\left(\frac{vt - v\tilde{s}_1}{-v\tilde{s}_1}\right)^\nu
+\,\underbrace{\int_{t}^{L/v}\!\! d\tilde{s}_1
\frac{ e^{-i \nu eV_{S1} \tilde{s}_1}}{[\tau_0 \!+\! i (\tilde{s}_1 \!-\! \tilde{s}_2) \,\eta_2]^{2\nu}}\,
\left(\frac{-vt +v\tilde{s}_1}{v\tilde{s}_1}\right)^\nu}_{\text{gives zero after summation over}\ \eta_1}\,
\right]\\
& = -  \frac{ \mathcal{T}_1^{(0)} \tau_0^{2\nu}}{(2\pi \tau_0)^2 } \frac{\tau_0^\nu}{(\tau_0 + i t)^\nu}\,  [-2i\sin (\pi\nu)] 
\int_{0}^{t} d\tilde{s}_2 \, 
\left(\frac{v\tilde{s}_2}{vt - v\tilde{s}_2 }\right)^\nu e^{i \nu eV_{S1} \tilde{s}_2} \,
\\
&\qquad \qquad\times
\int_{-\infty}^{0}\!\! d\tilde{s}_1 \left\{
\frac{ e^{-i\nu eV_{S1}  \tilde{s}_1}}{[l_c \!-\! i v(\tilde{s}_1 \!-\! \tilde{s}_2) ]^{2\nu}} - \frac{ e^{-i\nu eV_{S1}  \tilde{s}_1}}{[\tau_0 \!+\! i (\tilde{s}_1 \!-\! \tilde{s}_2)]^{2\nu}} \right\}\,
\left(\frac{vt - v\tilde{s}_1}{-v\tilde{s}_1}\right)^\nu\\
& = -  \frac{ \mathcal{T}_1^{(0)} \tau_0^{2\nu}}{(2\pi \tau_0)^2 } \frac{\tau_0^\nu}{(\tau_0 + i t)^\nu}\,  [-2i\sin (\pi\nu)] 
\int_{0}^{t} d\tilde{s}_2 \, 
\left(\frac{v\tilde{s}_2}{vt - v\tilde{s}_2 }\right)^\nu e^{i \nu eV_{S1} \tilde{s}_2} \,
\\
&\qquad \qquad\times
\int_0^{\infty}\!\! d\tilde{s}_1 e^{i\nu eV_{S1}  \tilde{s}_1} \left\{
\frac{ 1}{[\tau_0 + i (\tilde{s}_1 + \tilde{s}_2) ]^{2\nu}} - \frac{1}{[\tau_0 - i (\tilde{s}_1 + \tilde{s}_2)]^{2\nu}} \right\}\,
\left(\frac{vt + v\tilde{s}_1}{v\tilde{s}_1}\right)^\nu\\
& = \frac{ \mathcal{T}_1^{(0)} \tau_0^{2\nu}}{(2\pi \tau_0 )^2 } \frac{\tau_0^\nu}{(\tau_0 + i t)^\nu}\,  4\sin^2 (\pi\nu) 
\int_{0}^{t} d\tilde{s}_2 \, 
\left(\frac{\tilde{s}_2}{t - \tilde{s}_2 }\right)^\nu e^{i \nu eV_{S1} \tilde{s}_2} \,
\int_0^{\infty}\!\! d\tilde{s}_1 e^{i\nu eV_{S1}  \tilde{s}_1} \frac{1}{(\tilde{s}_1 + \tilde{s}_2)^{2\nu}}\,
\left(\frac{t + \tilde{s}_1 }{\tilde{s}_1}\right)^\nu,
\end{aligned}
\label{eq:d3}
\end{equation}
where the last term of the fourth line is independent of $\eta_1$, and thus vanishes after summation over $\eta_1$.
To proceed, we notice that when integrate over $\tilde{s}_2$, the last line of Eq.~\eqref{eq:d3} contains two branch-cut contributions, i.e., when $\tilde{s}_1 \to 0$, $\tilde{s}_2 \to t$; and that when $\tilde{s}_1, \tilde{s}_2 \to 0$.

When we take the contribution at the branch cut $\tilde{s}_1, \tilde{s}_2 \to 0$, Eq.~\eqref{eq:d3} simplifies into
\begin{equation}
\begin{aligned}
    &\left\langle e^{i \sqrt{\nu} \phi_1(L,t)}\, e^{-i \sqrt{\nu}\phi_1(L,0)}\right\rangle_\text{neq,2}^{\mathcal{R}_3}\Big|_{\tilde{s}_1, \tilde{s}_2 \to 0}\\
    \approx & \frac{ \mathcal{T}_1^{(0)} \tau_0^{2\nu}}{(2\pi \tau_0 )^2 } \frac{\tau_0^\nu}{(\tau_0 + i t)^\nu}\,  4\sin^2 (\pi\nu) 
\int_{0}^\infty d\tilde{s}_2 \, 
\tilde{s}_2^\nu e^{i \nu eV_{S1} \tilde{s}_2} \,
\int_0^{\infty}\!\! d\tilde{s}_1 e^{i\nu eV_{S1}  \tilde{s}_1} \frac{1}{(\tilde{s}_1 + \tilde{s}_2)^{2\nu}}\,
\frac{1}{\tilde{s}_1^\nu},
\end{aligned}
\label{eq:d3_first}
\end{equation}
where, the same as the treatment of Eq.~\eqref{eq:d2}, here the integral over $\tilde{s}_2$ has been extended to infinity, under the assumption $(\nu e V_{S1})^{-1} \ll t$.
We can combine Eq.~\eqref{eq:d2}, obtained from Region $\mathcal{R}_2$, and Eq.~\eqref{eq:d3_first}, leading to
\begin{equation}
\begin{aligned}
    &\left[\left\langle e^{i \sqrt{\nu} \phi_1(L,t)}\, e^{-i \sqrt{\nu}\phi_1(L,0)}\right\rangle_\text{neq,2}^{\mathcal{R}_2} + \left\langle e^{i \sqrt{\nu} \phi_1(L,t)}\, e^{-i \sqrt{\nu}\phi_1(L,0)}\right\rangle_\text{neq,2}^{\mathcal{R}_3}\right]\Bigg|_{\tilde{s}_1, \tilde{s}_2 \to 0}\\
    \approx & \frac{ \mathcal{T}_1^{(0)} \tau_0^{2\nu}}{(2\pi \tau_0 )^2 } \frac{\tau_0^\nu}{(\tau_0 + i t)^\nu}\,  4\sin^2 (\pi\nu) 2 i\  \text{Im} \left\{
\int_{0}^\infty d\tilde{s}_1 \frac{e^{i \nu eV_{S1} \tilde{s}_1} }{\tilde{s}_1^\nu} \,
\int_0^{\infty}\!\! d\tilde{s}_2 e^{i\nu eV_{S1}  \tilde{s}_2} \frac{\tilde{s}_2^\nu}{(\tilde{s}_1 + \tilde{s}_2)^{2\nu}} \right\}\\
= & i  \mathcal{T}_1^{(0)} \frac{2\nu}{\pi} \Gamma (2-2\nu) \sin^2 (\pi\nu) (\nu e V_{S1} \tau_0)^{2\nu - 2} \frac{\tau_0^\nu}{(\tau_0 + i t)^\nu} =\frac{\tau_0^\nu}{(\tau_0 + i t)^\nu} i \sin(\pi\nu) C_1(\nu) \frac{I_1}{ e^2 V_{S1} } \\
= & \frac{\tau_0^\nu}{(\tau_0 + i t)^\nu} i \sin (\pi\nu) \mathcal{T}_1',
\end{aligned}
\label{eq:d2_d3_first}
\end{equation}
where
\begin{equation}
    C_1(\nu) \equiv \frac{2  \pi (1-2\nu) }{\nu \cos(\pi\nu)}
    \label{eq:c1}
\end{equation}
is a dimensionless universal factor. With this factor, the addressed transmission probability, $\mathcal{T}_1'$, of the last line of Eq.~\eqref{eq:d2_d3_first}, equals
\begin{equation}
    \mathcal{T}_1' = C_1 (\nu) \frac{I_1}{e^2 V_{S1}} \propto \mathcal{T}_1^{(0)}.
    \label{eq:t1_prime}
\end{equation}

To obtain the final result of Eq.~\eqref{eq:d2_d3_first}, we have replaced the integral variables $\tilde{s}_1$ and $\tilde{s}_2$, by their summation (defined as $\tilde{s} \equiv \tilde{s}_1 + \tilde{s}_2 $ in equation below) and difference ($s \equiv \tilde{s}_1 - \tilde{s}_2 $), leading to
\begin{equation}
\begin{aligned}
    & \int_{0}^\infty d\tilde{s}_1 \frac{e^{i \nu eV_{S1} \tilde{s}_1} }{\tilde{s}_1^\nu} \,
    \int_0^{\infty}\!\! d\tilde{s}_2 e^{i\nu eV_{S1}  \tilde{s}_2} \frac{\tilde{s}_2^\nu}{(\tilde{s}_1 + \tilde{s}_2)^{2\nu}}\\
    =& \frac{1}{2} \int_0^\infty d\tilde{s} \frac{e^{i\nu e V_{S1} \tilde{s}}}{\tilde{s}^{2\nu}} \int_{-\tilde{s}}^{\tilde{s}} ds \left(\frac{\tilde{s} - s}{\tilde{s} + s} \right)^\nu = \frac{\pi\nu}{\sin(\pi\nu)} \int_0^\infty d\tilde{s} \frac{e^{i\nu e V_{S1} \tilde{s}}}{\tilde{s}^{2\nu }} s.
\end{aligned}
\end{equation}
Its imaginary part then equals
\begin{equation}
\begin{aligned}
    &2 i \text{Im} \left[ \int_{0}^\infty d\tilde{s}_1 \frac{e^{i \nu eV_{S1} \tilde{s}_1} }{\tilde{s}_1^\nu} \,
    \int_0^{\infty}\!\! d\tilde{s}_2 e^{i\nu eV_{S1}  \tilde{s}_2} \frac{\tilde{s}_2^\nu}{(\tilde{s}_1 + \tilde{s}_2)^{2\nu}} \right]\\
    = & \frac{\pi\nu}{\sin(\pi\nu)} \frac{1}{-2i\sin(\pi\nu)}  \left\{ \int_0^\infty d\tilde{s}\ \tilde{s} \left[ \frac{1}{(\tau_0 + i \tilde{s})^{2\nu}} - \frac{1}{(\tau_0 - i\tilde{s})^{2\nu}} \right] e^{i\nu e V_{S1} \tilde{s}} \right\}\\
    & - \frac{\pi\nu}{\sin(\pi\nu)} \frac{1}{2i\sin(\pi\nu)}  \left\{ \int_0^\infty d\tilde{s}\ \tilde{s} \left[ \frac{1}{(\tau_0 - i \tilde{s})^{2\nu}} - \frac{1}{(\tau_0 + i\tilde{s})^{2\nu}} \right] e^{-i\nu e V_{S1} \tilde{s}} \right\}\\
    = & -\frac{\pi\nu}{2i \sin^2 (\pi \nu)} \int_{-\infty}^\infty d\tilde{s}\  |\tilde{s}|\  \left[ \frac{1}{(\tau_0 + i \tilde{s})^{2\nu}} - \frac{1}{(\tau_0 - i \tilde{s})^{2\nu}} \right] e^{i\nu e V_{S1} \tilde{s}}\\
    = & \frac{\pi\nu}{2i \sin^2 (\pi \nu)} \left\{\frac{\partial}{\partial \lambda}\int_{-\infty}^\infty d\tilde{s}\  \left[ \frac{1}{(\tau_0 + i \tilde{s})^{2\nu}} - \frac{1}{(\tau_0 - i \tilde{s})^{2\nu}} \right] e^{i\nu e V_{S1} \tilde{s}-\lambda |s|}\right\} \Bigg|_{\lambda = 0}\\
    = & \frac{\pi\nu}{2i \sin^2 (\pi \nu)}2 \Gamma(2 - 2\nu) (\nu e V_{S1})^{2 \nu - 2} [-2 \sin^2 (\pi\nu)] = 2i\pi\nu \Gamma(2-2\nu) (\nu e V_{S1})^{2\nu - 2},
\end{aligned}
\end{equation}
where in the last but one line, we differentiate with respect to $\lambda$ first, and send it to zero afterwards.

Now we go back to Eq.~\eqref{eq:d3}, and study the contribution from the other branch cut, i.e., when $\tilde{s}_1 \to 0$, $\tilde{s}_2 \to t$. Of this case, Eq.~\eqref{eq:d3} simplifies into
\begin{equation}
\begin{aligned}
    &\left\langle e^{i \sqrt{\nu} \phi_1(L,t)}\, e^{-i \sqrt{\nu}\phi_1(L,0)}\right\rangle_\text{neq,2}^{\mathcal{R}_3}\Big|_{\tilde{s}_1 \to 0, \tilde{s}_2 \to t}\\
    & \approx  \frac{ \mathcal{T}_1^{(0)} \tau_0^{2\nu}}{(2\pi \tau_0 )^2} \frac{\tau_0^\nu}{(\tau_0 + i t)^\nu}\,  4\sin^2 (\pi\nu) e^{i\nu e V_{S1} t}
\int_{0}^{t} d\tilde{s}_2' \, 
\left(\frac{t - \tilde{s}_2'}{ \tilde{s}_2' }\right)^\nu e^{-i \nu eV_{S1} \tilde{s}_2'} \,
\int_0^{\infty}\!\! d\tilde{s}_1  \frac{e^{i\nu eV_{S1}  \tilde{s}_1}}{(t + \tilde{s}_1 - \tilde{s}_2')^{2\nu}}\,
\left(\frac{t }{\tilde{s}_1}\right)^\nu\\
& = \frac{ \mathcal{T}_1^{(0)} \tau_0^{2\nu}}{(2\pi \tau_0 )^2 } \frac{\tau_0^\nu}{(\tau_0 + i t)^\nu}\,  4\sin^2 (\pi\nu) e^{i\nu e V_{S1} t}
\int_{0}^{t} d\tilde{s}_2' \, 
\frac{e^{-i \nu eV_{S1} \tilde{s}_2'}}{ (\tilde{s}_2' )^\nu}  \,
\int_0^{\infty}\!\! d\tilde{s}_1 
\frac{e^{i\nu eV_{S1}  \tilde{s}_1}}{\tilde{s}_1^\nu}\\
& \approx \frac{ \mathcal{T}_1^{(0)} \tau_0^{2\nu}}{(2\pi \tau_0 )^2 } \frac{\tau_0^\nu}{(\tau_0 + i t)^\nu}\,  4\sin^2 (\pi\nu) e^{i\nu e V_{S1} t}
\int_{0}^\infty d\tilde{s}_2' \, 
\frac{e^{-i \nu eV_{S1} \tilde{s}_2'}}{ (\tilde{s}_2' )^\nu}  \,
\int_0^{\infty}\!\! d\tilde{s}_1 
\frac{e^{i\nu eV_{S1}  \tilde{s}_1}}{\tilde{s}_1^\nu}\\
& =  \frac{ \mathcal{T}_1^{(0)} \tau_0^{2\nu}}{(2\pi \tau_0 )^2 } \frac{\tau_0^\nu}{(\tau_0 + i t)^\nu}\,  4\sin^2 (\pi\nu) \Gamma^2 (1 - \nu) (\nu e V_{S1})^{2\nu - 2} e^{i\nu e V_{S1} t} = \frac{\tau_0^\nu}{(\tau_0 + i t)^\nu} \frac{I_1}{e^2 V_{S1}} C_0(\nu ) e^{i\nu e V_{S1} t} \\
& = \frac{\tau_0^\nu}{(\tau_0 + i t)^\nu} \mathcal{D}_1 e^{i\nu e V_{S1} t},
\end{aligned}
\label{eq:d3_second}
\end{equation}
where $\tilde{s}_2' \equiv t - \tilde{s}_2$ is another quantity that is bounded by $\pm  1/\nu e V_{S1}$, and
\begin{equation}
C_0(\nu) \equiv \tan (\pi\nu) \frac{\Gamma^2(1-\nu)}{\Gamma (1-2\nu) \nu^2},
\label{eq:c}
\end{equation}
is another dimensionless constant that universally depends only on the filling fraction.
With this factor, we introduce another addressed transmission probability, $\mathcal{D}_1$ [of Eq.~\eqref{eq:d3_second} and Eqs.~(16), (17) of the main text], which equals
\begin{equation}
    \mathcal{D}_1 = C_0 (\nu) \frac{I_1}{e^2 V_{S1}} \propto \mathcal{T}_1^{(0)} .
    \label{eq:t1_def}
\end{equation}
Notice that for non-interacting fermions, $\nu = 1$, $\lim_{\nu \to 1} C_0 (\nu) = 2\pi$.
In addition, for $\nu  = 1/2$, two constant parameters, defined by Eqs.~\eqref{eq:c1} and \eqref{eq:c}, equal each other, i.e., $C_1(\nu \to 1/2) = C_0(\nu\to 1/2) = 8$, being consistent with the results of Ref.~\cite{SOneHalfPRB24}.
For general values of $\nu$ they are however different, leading to
\begin{equation}
    \mathcal{T}_1' = \mathcal{D}_1 C (\nu) ,\quad C (\nu) = \frac{2\pi\nu (1-2\nu) \Gamma (1-2\nu)}{\sin(\pi\nu) \Gamma^2 ( 1 - \nu) },
    \label{eq:t1_ratio}
\end{equation}
with the $\nu$-dependent universal factor $C$ the one given by Eq.~(16) of the main text.
Importantly, Eqs.~\eqref{eq:d2_d3_first} and \eqref{eq:d3_second} are both proportional to $I_1/V_{S1}$, a quantity proportional to $\mathcal{T}_1$ [$\propto \mathcal{D}_1$, cf. Eq.~(17) of the main text] and $\mathcal{T}_1^{(0)}$: the same as Eq.~\eqref{eq:d1} that is obtained from Region $\mathcal{R}_1$.
However, as will be shown shortly in Sec.~S2, after resummation over higher-order dilutions, contribution from $\mathcal{R}_1$ recombine with the equilibrium one, into a contribution to the zeroth order of dilution [cf. Eq.~\eqref{eq:d1_resum}].
In the strongly dilute limit, $I_1/\nu e \ll \nu e V_{S1} $, this contribution thus becomes much larger than that obtained in Regions $\mathcal{R}_2$ and $\mathcal{R}_3$, as the latter one is proportional to $\mathcal{T}_1 \sim  I_1/(\nu e^2 V_{S1}) \ll 1$ [cf. Eqs.~\eqref{eq:d3_second} and \eqref{eq:d2_d3_resum}].

Before ending this section, we briefly discuss what happens for a negative $t$.
Basically, following Eq.~\eqref{eq:second_order_correlation}, the influence of the negativity of $t$ can be included by shifting time variables $s_1$ and $s_2$ by $-t$.
After doing so, the only difference between positive and negative $t$ situations grounds in the tangling factor, Eq.~\eqref{eq:r_etas}.
Indeed, when $t$ becomes negative, the phase factor of Eq.~\eqref{eq:r_etas} changes sign.
As the consequence, the phase factor inside of the brackets of Eq.~\eqref{eq:d1}, as well as the prefactor $i\sin (\pi\nu) $ of Eq.~\eqref{eq:d2_d3_first} both change sign.
Eq.~\eqref{eq:d3_second} is however not influenced by a negative $t$, as it is proportional to $[i\sin (\pi\nu)]^2$, which changes sign twice when a positive $t$ becomes negative.

\section*{S2.~Resummation over higher-order dilution contribution}

In Sec.~S1, we have shown zero-temperature correlation functions to the leading order of dilution.
In this section, we go beyond results of Sec.~S1, by including influence of higher-order tunneling through the diluter, which, as will be shown shortly, correspond to braiding among anyons.
Here, extra non-equilibrium anyonic pairs are assumed to belong to region $\mathcal{R}_1$ (thus leading to anyonic braiding), although they influence leading-order pair contributions of all three regions (in Secs.~S2A and S2B, for higher-order influence on contributions of Region $\mathcal{R}_1$, and that of Regions $\mathcal{R}_2$, $\mathcal{R}_3$, respectively).
This option is taken, since the corresponding contribution is of the order of $\mathcal{T}_1 \sim I_1/(\nu^2 e^2 V_{S1})$ (see Ref.~\cite{SOneHalfPRB24} for more discussions), after resummation over dilution to all orders.
Actually, other processes, with extra non-equilibrium anyons in either Regions $\mathcal{R}_2$ and $\mathcal{R}_3$, correspond to co-tunneling at the diluter.
They are thus of at least the second order in $\mathcal{T}_1$, being small in the strongly dilute limit that is visited by our current work.

We begin by considering the next-leading order contribution, i.e., $\propto \left(\mathcal{T}_1^{(0)}\right)^2$, and assume that two extra non-equilibrium anyonic operators have time arguments $s_3$ and $s_4$, and corresponding Keldysh indexes $\eta_3$ and $\eta_4$, respectively.
Here these two extra operators are assumed to belong to Region $\mathcal{R}_1$, such that $-L/v<s_3 \approx s_4< t -L/v$.
This assumption is taken, since the corresponding contribution is larger than other options, where $s_3$ or $s_4$ belongs to the other two Regions.
Indeed, when $s_3$ and $s_4$ are both in Region $\mathcal{R}_1$, the corresponding contribution can be absorbed into leading-order contributions, after performing resummation over dilution to all orders.
In great contrast, when they belong to the other two Regions, the corresponding contributions, which are of higher-order in dilution, can not be re-captured by resummation, thus leading to comparatively much smaller contributions in the strongly diluted limit.
Without the loss of generality, we further begin by assuming that operators with arguments $s_1$ and $s_3$ are creation operators; the other two are instead annihilation ones.
Iteration between different options (to pair-up non-equilibrium anyonic operators) will be discussed and included later.

\subsection*{S2A.~Higher-order dilution correction to integral of region $\mathcal{R}_1$}

To the order of $|\xi_1|^4 = \left[ \mathcal{T}_1^{(0)} \right]^2$, correlation function of operators in channel 1 approximately equal (after ignoring, at the moment, prefactors from Keldysh expansion and number of pairing options)
\begin{equation}
\begin{aligned}
    &\big\langle e^{-i\sqrt{\nu}\phi_1 (t^-,L)} e^{i \sqrt{\nu}\phi_1 (0^+, L)} \big\rangle_\text{neq,4}^{\mathcal{R}_1}
=   \frac{ \left[ \mathcal{T}_1^{(0)} \right]^2 \tau_0^{4\nu}}{(2\pi \tau_0)^4 }  \frac{\tau_0^\nu}{(\tau_0 + i t)^\nu} \\
&  \times \sum_{\eta_1\eta_2}  \eta_1\eta_2 e^{i\pi \nu (\eta_1-\eta_2)}\,
\int_{0}^t d\tilde{s}_1 
\int_{\tilde{s}_1 -t}^{\tilde{s}_1 }\!\! ds \,
\frac{   \, e^{-i \nu eV_{S1} s}}
{ [\tau_0 + i  s\chi_{\eta_1,\eta_2} (s)]^{2\nu}}
\, 
\left\{\frac{(t - \tilde{s}_1)\,(\tilde{s}_1-s) }
    {(t + s - \tilde{s}_1) \, \tilde{s_1}}\right\}^\nu \\
&  \times \sum_{\eta_3\eta_4}  \eta_3\eta_4 e^{i\pi \nu (\eta_3-\eta_4)}\,
\int_{0}^t d\tilde{s}_3 
\int_{\tilde{s}_3 -t}^{\tilde{s}_3 }\!\! ds' \,
\frac{   \, e^{-i \nu eV_{S1} s'}}
{ [\tau_0 + i  s\chi_{\eta_3,\eta_4} (s')]^{2\nu}}
\, 
\left\{\frac{(t - \tilde{s}_3)\,(\tilde{s}_3-s') }
    {(t + s' - \tilde{s}_3) \, \tilde{s_3}}\right\}^\nu \\
& \times \mathcal{I} (s_1^{\eta_1},s_2^{\eta_2};s_3^{\eta_3},s_4^{\eta_4}),
\end{aligned}
\end{equation}
where $\tilde{s}_3 \equiv s_3 + L/v$, $s' \equiv  s_3 - s_4$ are two shifted variables, and the function
\begin{equation}
    \mathcal{I} (s_1^{\eta_1},s_2^{\eta_2};s_3^{\eta_3},s_4^{\eta_4}) \equiv \frac{[\tau_0 + i (s_1-s_3) \chi_{\eta_1,\eta_3} (s_1 - s_3) ]^{2\nu} [\tau_0 + i (s_2-s_4) \chi_{\eta_2,\eta_4} (s_2 - s_4)]^{2\nu}}{[\tau_0 + i (s_1-s_4) \chi_{\eta_1,\eta_4} (s_1 - s_4)]^{2\nu} [\tau_0 + i (s_2-s_3) \chi_{\eta_2,\eta_3} (s_2 - s_3)]^{2\nu}},
    \label{eq:mathcal_i}
\end{equation}
is the factor generated by the ``braiding'' between two non-equilibrium anyonic pairs.
Clearly, with our pairing option, $s_1 \to s_2$ and $s_3 \to s_4$, $\mathcal{I} (s_1^{\eta_1}, s_2^{\eta_2};s_3^{\eta_3},s_4^{\eta_4}) \approx 1$, meaning that non-equilibrium anyonic pairs, at the moments $s_1 \approx s_2$ and $s_3 \approx s_4$, respectively, do not ``braid'' between themselves. As the consequence, the next-leading order correlation of Regime $\mathcal{R}_1$ simplifies into
\begin{equation}
\begin{aligned}
    &\big\langle e^{-i\sqrt{\nu}\phi_1 (t^-,L)} e^{i \sqrt{\nu}\phi_1 (0^+, L)} \big\rangle_\text{neq,4}^{\mathcal{R}_1}
\approx    \frac{\tau_0^\nu}{(\tau_0 + i t)^\nu} \\
&  \times \left( \mathcal{T}_1^{(0)} \frac{\tau_0^{2\nu}}{(2\pi\tau_0 )^2} \sum_{\eta_1\eta_2}  \eta_1\eta_2 e^{i\pi \nu (\eta_1-\eta_2)}\,
\int_{0}^t d\tilde{s}_1 
\int_{\tilde{s}_1 -t}^{\tilde{s}_1 }\!\! ds \,
\frac{   \, e^{-i \nu eV_{S1} s}}
{ [\tau_0 + i s\chi_{\eta_1,\eta_2} (s)]^{2\nu}}
\, 
\left\{\frac{(t - \tilde{s}_1)\,(\tilde{s}_1-s) }
    {(t + s - \tilde{s}_1) \, \tilde{s_1}}\right\}^\nu \right)^2\\
& = \frac{\tau_0^\nu}{(\tau_0 + i t)^\nu} \left[ - \left( 1 - e^{2 i\pi\nu} \right) \frac{I_1}{\nu e} t \right]^2.
\end{aligned}
\label{eq:d1_quadratic}
\end{equation}
Eq.~\eqref{eq:d1_quadratic} discloses the fact that if we focus on region $\mathcal{R}_1$, and assume that all non-equilibrium anyonic pairs are well-separated in space (at any moment), the integral outcome simply equals the product of the prefactor $\tau_0^\nu/(\tau_0 + i t)^\nu$, and $n$ copies of leading-order integral result [cf. Eq.~\eqref{eq:d1}, after ignoring the prefactor $\tau_0^\nu/(\tau_0 + i t)^\nu$], if $n$ is the number of non-equilibrium anyonic pairs.

Now we add back Keldysh expansion prefactors and the number of pairing options, to perform resummation over higher-order processes.
Firstly, for $n$ non-equilibrium anyonic pairs (corresponding to $2n$ anyonic operators), a prefactor $1/(2n!)$ is generated, due to the Keldysh expansion.
To pair up these $2n$ anyonic operators, the number of options equals $C_{2n}^2 C_{2n-2}^2 \cdot\cdot\cdot C_2^2 2^{n}$, where $C_{2n}^2 = 2n (2n -1)/2$ equals the number of options to choose the time arguments of the first pair; and $2^{n}$ comes from the fact that for each anyonic pair, either operator can be the creation one. Finally, the sequence of choosing non-equilibrium anyonic pairs does not matter, leading to an extra prefactor $1/n!$.
After including all these prefactors, resummation over higher-order dilution leads to
\begin{equation}
    \frac{\tau_0^\nu}{(\tau_0 + i t)^\nu} \sum_n \frac{1}{2n!} \frac{C_{2n}^2 C_{2n-2}^2 ... C_4^2 C_2^2 \times  2^n}{n!} \left[ - \left( 1 - e^{2 i\pi\nu} \right) \frac{I_1}{\nu e} t \right]^n =  \frac{\tau_0^\nu}{(\tau_0 + i t)^\nu} \exp\left[ - \left( 1 - e^{2 i\pi\nu} \right) \frac{I_1}{\nu e} t \right].
    \label{eq:d1_resum}
\end{equation}
Importantly, Eq.~\eqref{eq:d1_resum} actually includes the equilibrium contribution, i.e., when $n = 0$.
With this contribution, Eq.~\eqref{eq:d1_resum}, obtained from Region $\mathcal{R}_1$, is of order zero to the dilution.
Nevertheless, contributions from higher-order tunneling through the diluters are important in anyonic time-domain braiding processes.
Indeed, the inclusion of higher-order dilution, importantly, provides the bound of integral over $t$ [due to the exponential factor of Eq.~\eqref{eq:d1_resum}], which is however unclear when considering only the leading-order tunneling through the diluter.

\subsection*{S2B.~Higher-order dilution correction to regions $\mathcal{R}_2$ and $\mathcal{R}_3$}

Now we move to Regions $\mathcal{R}_2$ and $\mathcal{R}_3$, and consider modifications to them induced by contributions of higher order tunneling through the diluter.
These two regimes contain two branch cut contributions: (i) when $\tilde{s}_1, \tilde{s}_2 \to 0$ [i.e., Eq.~\eqref{eq:d2_d3_first}]; and (ii) when $\tilde{s}_1 \to 0$ and $\tilde{s}_2 \to t$ [i.e., Eq.~\eqref{eq:d3_second}].

We first consider contribution (i) above.
Of this contribution, the adding of another non-equilibrium pair of anyons leads to the correlation function
\begin{equation}
\begin{aligned}
    &\left[\left\langle e^{i \sqrt{\nu} \phi_1(L,t)}\, e^{-i \sqrt{\nu}\phi_1(L,0)}\right\rangle_\text{neq,4}^{\mathcal{R}_2} + \left\langle e^{i \sqrt{\nu} \phi_1(L,t)}\, e^{-i \sqrt{\nu}\phi_1(L,0)}\right\rangle_\text{neq,4}^{\mathcal{R}_3} \right] \Big|_{\tilde{s}_1, \tilde{s}_2 \to 0}\\
    \approx & \frac{ \mathcal{T}_1^{(0)} \tau_0^{2\nu}}{(2\pi \tau_0 )^2 } \frac{\tau_0^\nu}{(\tau_0 + i t)^\nu}\,  4\sin^2 (\pi\nu) 2 i\  \text{Im} \left\{
\int_{0}^\infty d\tilde{s}_1 \frac{e^{i \nu eV_{S1} \tilde{s}_1} }{\tilde{s}_1^\nu} \,
\int_0^{\infty}\!\! d\tilde{s}_2 e^{i\nu eV_{S1}  \tilde{s}_2} \frac{\tilde{s}_2^\nu}{(\tilde{s}_1 + \tilde{s}_2)^{2\nu}} \right\}\\
\times & \left[ - \mathcal{T}_1^{(0)} \frac{\tau_0^{2\nu}}{(2\pi\tau_0)^2}\right] \sum_{\eta_3\eta_4}  \eta_3\eta_4 e^{i\pi \nu (\eta_3-\eta_4)}\,
\int_{0}^t d\tilde{s}_3 
\int_{\tilde{s}_3 -t}^{\tilde{s}_3 }\!\! ds' \,
\frac{   \, e^{-i \nu eV_{S1} s'}}
{ [\tau_0 + i s' \chi_{\eta_3,\eta_4} (s')]^{2\nu}}
\, 
\left\{\frac{(t - \tilde{s}_3)\,(\tilde{s}_3-s') }
    {(t + s' - \tilde{s}_3) \, \tilde{s}_3}\right\}^\nu\\
\times  & \mathcal{I} (s_1^{\eta_1},s_2^{\eta_2};s_3^{\eta_3},s_4^{\eta_4}),
\end{aligned}
\label{eq:d2_d3_first_4th}
\end{equation}
where again two extra non-equilibrium anyonic operators have time arguments $s_3$ and $s_4$, with Keldysh indexes $\eta_3$ and $\eta_4$, respectively.
Since $\tilde{s}_1, \tilde{s}_2 \to 0$, $\mathcal{I} (s_1^{\eta_1},s_2^{\eta_2};s_3^{\eta_3},s_4^{\eta_4}) \approx 1$ [see its definition, Eq.~\eqref{eq:mathcal_i}], the same as Eq.~\eqref{eq:d1_quadratic} when evaluating the contribution of Region $\mathcal{R}_1$.
With this simplification, integral Eq.~\eqref{eq:d2_d3_first_4th} simplifies into
\begin{equation}
\begin{aligned}
    &\left[\left\langle e^{i \sqrt{\nu} \phi_1(L,t)}\, e^{-i \sqrt{\nu}\phi_1(L,0)}\right\rangle_\text{neq,4}^{\mathcal{R}_2} + \left\langle e^{i \sqrt{\nu} \phi_1(L,t)}\, e^{-i \sqrt{\nu}\phi_1(L,0)}\right\rangle_\text{neq,4}^{\mathcal{R}_3} \right] \Big|_{\tilde{s}_1, \tilde{s}_2 \to 0}\\
    \approx & \frac{\tau_0^\nu}{(\tau_0 + it)^\nu} i\sin(\pi\nu) \frac{I_1}{e^2 V_{S1} } C_1 (\nu) \left[ - \left( 1 - e^{2 i\pi\nu} \right) \frac{I_1}{\nu e} t \right],
\end{aligned}
\label{eq:d2_d3_first_quadratic}
\end{equation}
where the factor between two square brackets comes from the extra non-equilibrium anyonic pair, which actually coincides with the leading-order result when integrating over $\mathcal{R}_1$.
Consequently, when considering contribution (i), i.e., the branch cut $\tilde{s}_1, \tilde{s}_2 \to 0$, each extra non-equilibrium anyonic pair simply leads to an extra product factor $-\left[ 1 - \exp (2 i\pi\nu) \right] I_1 t/\nu e  $.

Now we evaluate prefactors, when considering contributions from higher orders in dilution.
Without the loss of generality, we expand $H_\text{dilute}$ to the $2n$th order, leading to a prefactor, $1/(2n!)$ for Keldysh expansion.
Among these $2n$ non-equilibrium anyons, two of them (one annihilation and one creation operators) are correlated with the tunneling event at the collider, at the moment zero.
There are $2n (2n - 1) $ options to choose these two operators.

The rest $2n - 2$ operators instead combine into $n - 1$ anyonic pairs, where operators belonging to the same pair have the same argument in time.
The number of options to choose pairing operators equals
\begin{equation}
   \frac{2^{n-1}}{(n-1)!} C_{2n-2}^2 C_{2n-4}^2 \cdot \cdot\cdot C_2^2 = \frac{(2n-2)!}{(n-1)!} ,
\end{equation}
where the factor $2^{n-1}$ equals the number of options to choose $n - 1$ creation operators, from $n-1$ anyonic pairs.
Another factor, $1/(n-1)!$, addresses the sequence to pick out $n-1$ non-equilibrium anyonic pairs.

With all these prefactors included, we obtain the result, after the resummation over higher-order dilutions
\begin{equation}
\begin{aligned}
    &\frac{\tau_0^\nu}{(\tau_0 + i t)^\nu} i\sin(\pi\nu) \frac{I_1}{e^2 V_{S1} } C_1(\nu) \sum_n \frac{1}{2n!} 2n (2n - 1) \frac{(2n - 2)!}{(n-1)!}\left[ - \left( 1 - e^{2 i\pi\nu} \right) \frac{I_1}{\nu e} t \right]^{n-1} \\
    =&  \frac{\tau_0^\nu}{(\tau_0 + i t)^\nu} i\sin(\pi\nu) \frac{I_1}{e^2 V_{S1} } C_1(\nu) \exp\left[ - \left( 1 - e^{2 i\pi\nu} \right) \frac{I_1}{\nu e} t \right].
\end{aligned}
\label{eq:d2_d3_resum}
\end{equation}

Now we move to the contribution (ii) mentioned above, i.e., when $\tilde{s}_1 \to 0$ and $\tilde{s}_2 \to t$.
When considering contributions to the order of $|\xi_1|^4$, the leading-order contribution, Eq.~\eqref{eq:d3_second}, now becomes
\begin{equation}
\begin{aligned}
    &\left\langle e^{i \sqrt{\nu} \phi_1(L,t)}\, e^{-i \sqrt{\nu}\phi_1(L,0)}\right\rangle_\text{neq,4}^{\mathcal{R}_3}\Big|_{\tilde{s}_1 \to 0, \tilde{s}_2 \to t}\\
    & \approx \frac{ \mathcal{T}_1^{(0)} \tau_0^{2\nu}}{(2\pi \tau_0 )^2 } \frac{\tau_0^\nu}{(\tau_0 + i t)^\nu}\,  4\sin^2 (\pi\nu) e^{i\nu e V_{S1} t}
\int_{0}^\infty d\tilde{s}_2' \, 
\frac{e^{-i \nu eV_{S1} \tilde{s}_2'}}{ (\tilde{s}_2' )^\nu}  \,
\int_0^{\infty}\!\! d\tilde{s}_1 
\frac{e^{i\nu eV_{S1}  \tilde{s}_1}}{\tilde{s}_1^\nu}\\
\times & \left[ - \mathcal{T}_1^{(0)} \frac{\tau_0^{2\nu}}{(2\pi\tau_0 )^2}\right] \sum_{\eta_3\eta_4}  \eta_3\eta_4 e^{i\pi \nu (\eta_3-\eta_4)}\,
\int_{0}^t d\tilde{s}_3 
\int_{\tilde{s}_3 -t}^{\tilde{s}_3 }\!\! ds' \,
\frac{   \, e^{-i \nu eV_{S1} s'}}
{ [\tau_0 + i s' \chi_{\eta_3,\eta_4} (s')]^{2\nu}}
\, 
\left\{\frac{(t - \tilde{s}_3)\,(\tilde{s}_3-s') }
    {(t + s' - \tilde{s}_3) \, \tilde{s_3}}\right\}^\nu\\
\times  & \mathcal{I} (s_1^{\eta_1},s_2^{\eta_2};s_3^{\eta_3},s_4^{\eta_4}).
\end{aligned}
\label{eq:d3_second_4th}
\end{equation}
Of this case, $\tilde{s}_1 \to 0 < \tilde{s}_3, \tilde{s}_4 < \tilde{s}_2 \to t$, so that
\begin{equation}
\begin{aligned}
    \mathcal{I} (s_1^{\eta_1},s_2^{\eta_2};s_3^{\eta_3},s_4^{\eta_4}) & \approx \frac{[\tau_0 + i (-s_3) \chi_{\eta_1,\eta_3} ( - \tilde{s}_3) ]^{2\nu} [\tau_0 + i (t-\tilde{s}_4) \chi_{\eta_2,\eta_4} (t - \tilde{s}_4)]^{2\nu}}{[\tau_0 + i (-\tilde{s}_4) \chi_{\eta_1,\eta_4} ( - \tilde{s}_4)]^{2\nu} [\tau_0 + i (t-\tilde{s}_3) \chi_{\eta_2,\eta_3} (t - \tilde{s}_3)]^{2\nu}}\\
    & = \Bigg| \frac{[\tau_0 + i (-s_3) \chi_{\eta_1,\eta_3} ( - \tilde{s}_3) ]^{2\nu} [\tau_0 + i (t-\tilde{s}_4) \chi_{\eta_2,\eta_4} (t - \tilde{s}_4)]^{2\nu}}{[\tau_0 + i (-\tilde{s}_4) \chi_{\eta_1,\eta_4} ( - \tilde{s}_4)]^{2\nu} [\tau_0 + i (t-\tilde{s}_3) \chi_{\eta_2,\eta_3} (t - \tilde{s}_3)]^{2\nu}} \Bigg| e^{i\pi\nu (\eta_4 - \eta_3)}.
\end{aligned}
\label{eq:ifactor_d3_second}
\end{equation}
The phase factor of Eq.~\eqref{eq:ifactor_d3_second} cancels out that of the third line of Eq.~\eqref{eq:d3_second_4th}, after which the term $\propto \left[ \mathcal{T}_1^{(0)} \right]^2$ becomes independent of $\eta_3$ or $\eta_4$.
As the consequence, at this order, correlation function vanishes after summing over Keldysh indexes, $\eta_3$ or $\eta_4$.
Similarly, for even higher orders, the result vanishes, after summing over corresponding Keldysh indexes.
Consequently, when taking the $\tilde{s}_1 \to 0$, $\tilde{s}_2 \to t$ branch cut, contribution from higher-order non-equilibrium anyonic pairs vanish
\begin{equation}
    \left\langle e^{i \sqrt{\nu} \phi_1(L,t)}\, e^{-i \sqrt{\nu}\phi_1(L,0)}\right\rangle_{\text{neq},2n \ge 4}^{\mathcal{R}_3}\Big|_{\tilde{s}_1 \to 0, \tilde{s}_2 \to t} = 0,
    \label{eq:d3_nth}
\end{equation}
so that the only nonzero result comes from the leading-order expansion at the diluter, i.e., Eq.~\eqref{eq:d3_second}.
We emphasize, however, that Eq.~\eqref{eq:d3_nth} only applies to the situation where extra (in addition to the chosen values $\tilde{s}_1$ and $\tilde{s}_2$ above) non-equilibrium anyonic pairs are within the area $\mathcal{R}_1$.
Actually, if moments of the extra pair (say, $s_3$ and $s_4$) satisfy $s_3 \to s_1 \to -L/v$ and $s_4 \to s_2 \to t - L/v$, the situation corresponds to the process where two non-equilibrium anyons simultanously tunnel at the collider.
The final integral result is then finite.
However, in this work we neglect these processes, as the corresponding integral outcome of e.g., the next-leading-order result is proportional to $\mathcal{T}_1^2$, an outcome negligible in the strongly dilute limit (beyond the precision of our current work, $\sim \mathcal{T}_1$).

Combining Eqs.~\eqref{eq:d3_second}, \eqref{eq:d1_resum} and \eqref{eq:d2_d3_resum}, we arrive at zero-temperature correlation functions [with $\mathcal{T}_1$ and $\mathcal{T}_1'$ given by Eqs.~\eqref{eq:t1_def} and \eqref{eq:t1_prime}, respectively, as the reminder]
\begin{equation}
\begin{aligned}
\big\langle T_K \psi^\dagger_1 (L,t^-) \psi_1 (L,0^+) \big\rangle_\text{neq} \Big|_{T_{S1} = 0} =&  \frac{\tau_0^{\nu - 1}}{2\pi( \tau_0 + it)^\nu} \left\{ e^{- \frac{I_1}{\nu e} [1 - \cos(2\pi\nu) ] |t| + i \frac{I_1}{\nu e} \sin(2\pi\nu) t}\right.\\
+&\left. \mathcal{D}_1 e^{i\nu e V_{S1} t} + i \mathcal{T}_1' \text{sgn} (t) \sin(\pi\nu)  e^{- \frac{I_1}{\nu e} [1 - \cos(2\pi\nu) ] |t| +  i \frac{I_1}{\nu e} \sin(2\pi\nu) t} \right\},\\
\big\langle T_K \psi_1 (L,t^-) \psi^\dagger_1 (L,0^+) \big\rangle_\text{neq} \Big|_{T_{S1} = 0} =& \frac{\tau_0^{\nu - 1}}{2\pi( \tau_0 + it)^\nu} \left\{ e^{- \frac{I_1}{\nu e} [1 - \cos(2\pi\nu) ] |t| - i \frac{I_1}{\nu e} \sin(2\pi\nu) t} \right.\\
+&\left. \mathcal{D}_1  e^{-i\nu e V_{S1} t} - i \mathcal{T}_1' \text{sgn} (t) \sin(\pi\nu)  e^{- \frac{I_1}{\nu e} [1 - \cos(2\pi\nu) ] |t| - i \frac{I_1}{\nu e} \sin(2\pi\nu) t} \right\},
\end{aligned}
\label{eq:full_correlations_zero_t}
\end{equation}
which recovers Eq.~(16) of the main text, after assuming finite ambient temperature of channel 1, and replacing $\nu$ with the three anyonic signatures $q$, $\theta/2\pi$ and $2\delta$, as appropriate.

\section*{S3. Correlation functions when the source temperature $T_{S1}$ is finite}

Eq.~\eqref{eq:full_correlations_zero_t} provides the correlation function when two temperatures equal zero: the temperature of channel $S1$, $T_{S1}$, and that of channel 1 before the diluter.
In this section, we further extend Eq.~\eqref{eq:full_correlations_zero_t} to the situation where $T_{S1}$ is finite.
The ambient temperature of channel 1 however remains zero.
In reality, the temperature of channel 1 before the diluter should be much lower than the energy scale set by the inverse of the typical time interval between neighboring anyons, $\nu e/I_1$.
Indeed, otherwise the non-equilibrium feature of channel 1 is smeared out by the thermal fluctuation, as then typically the expected number of non-equilibrium anyons observed during the temperature inverse is much smaller than one (, meaning the absence of non-equilibrium feature).
Notice that the situation is rather different for a finite temperature $T_{S1}$ in the source, as the latter governs the width of non-equilibrium anyons, rather than the typical distance between neighboring anyons.
In addition, $T_{S1}$ is taken to be finite, such that the shot noise $S_1$ and the tunneling current $I_1$ can be considered as two independent variables [cf. discussions after Eq.~\eqref{eq:i1_s1}].
This independence is crucial, as $S_1$ and $I_1$ govern two independent parameters, i.e., the effective temperature and the effective chemical potential, respectively, of an anyonic channel in the collision-free limit.

When $T_{S1}$ is finite, the zero-temperature leading-order correlation, Eq.~\eqref{eq:second_order_correlation}, becomes modified into
\begin{equation}
\begin{aligned}
& \big\langle e^{-i \sqrt{\nu}\phi_1 (t^-,L)} e^{i \sqrt{\nu}\phi_1 (0^+, L)} \big\rangle_\text{neq,2}
\!=\!  -  \frac{\mathcal{T}_1^{(0)}}{(2\pi \tau_0 )^2 }\sum_{\eta_1\eta_2} 
\int_{-\infty}^\infty \! d s_1
\int_{-\infty}^\infty \! ds_2 \,  \frac{ \eta_1\eta_2   \, e^{-i e \nu V_{S1} (s_1 - s_2)} (\pi T_{S1}\tau_0)^{\nu}}{\sin^{\nu}\left\{\pi T_{S1} [\tau_0 + i (s_1 - s_2)\chi_{\eta_1,\eta_2} (s_1 - s_2) ]\right\}}
\\
\times & \frac{\tau_0^\nu}{[\tau_0 \!+\! i (s_1 \!-\! s_2)\chi_{\eta_1,\eta_2} (s_1 \!-\! s_2)]^\nu}\frac{\tau_0^\nu}{(\tau_0 + i t)^\nu}
\frac{[\tau_0 \!+\! i ( t  \!-\! s_1 \!-\! L/v ) \chi_{-,\eta_1}(t\!-\!s_1)]^{\nu}\, [\tau_0 \!-\! i(  s_2 \!+\! L/v) \chi_{+,\eta_2}(- s_2)]^{\nu}}
{[\tau_0 \!+\! i ( t \!-\! s_2 \!-\! L/v ) \chi_{-,\eta_2}(t \!-\! s_2 )]^{\nu}\,
[\tau_0 \!-\! i (s_1 \!+\! L/v) \chi_{+,\eta_1}(-s_1)]^{\nu} } \\
& =  - \mathcal{T}_1^{(0)} \frac{(\pi T_{S1}\tau_0)^{\nu}}{(2\pi \tau_0 )^2 } \frac{\tau_0^\nu}{(\tau_0 + i t)^\nu}  \sum_{\eta_1\eta_2} \eta_1\eta_2
\int_{-\infty}^\infty ds
\int_{-\infty}^\infty d\tilde{s}_1 \, \frac{\tau_0^\nu}{[\tau_0 + i  s \chi_{\eta_1\eta_2} (s)]^\nu} \frac{e^{-i \nu eV_{S1} s}}{ \sin^\nu\left\{ \pi T_{S1}[\tau_0 + i s\chi_{\eta_1,\eta_2} (s) ]\right\}}
\\
& \quad\times 
\frac{[\tau_0 + i ( t  -\tilde{s}_1 ) \chi_{-,\eta_1}(t+L/v-\tilde{s}_1)]^{\nu}\, [\tau_0 + i (s - \tilde{s}_1) \chi_{+,\eta_2}(L/v+s-\tilde{s}_1)]^{\nu}}
{[\tau_0 + i ( t + s -\tilde{s}_1 ) \chi_{-,\eta_2}(t+L/v+s-\tilde{s}_1)]^{\nu}\,
[\tau_0 - i \tilde{s}_1 \chi_{+,\eta_1}(L/v-\tilde{s}_1)]^{\nu} } \\
& =  - \mathcal{T}_1^{(0)} \frac{1}{(2\pi \tau_0 )^2 } \frac{\tau_0^{2\nu}}{(\tau_0 + i t)^\nu}  \sum_{\eta_1\eta_2} \eta_1\eta_2
\int_{-\infty}^\infty ds
\int_{-\infty}^\infty d\tilde{s}_1 \, \frac{e^{-i \nu eV_{S1} s}}{[\tau_0 + i s \chi_{\eta_1\eta_2} (s)]^\nu} \frac{   (\pi T_{S1}\tau_0)^{\nu}}{\sin^\nu\left\{\pi T_{S1} [\tau_0 + i s\chi_{\eta_1,\eta_2} (s) ]\right\}}\\
&\quad \times \frac{[\tau_0 + i ( t  -\tilde{s}_1 ) \eta_1]^{\nu}\, [\tau_0 + i (s - \tilde{s}_1) \eta_2]^{\nu}}
{[\tau_0 + i ( t + s -\tilde{s}_1 ) \eta_2]^{\nu}\,
[\tau_0 - i \tilde{s}_1 \eta_1]^{\nu} },
\end{aligned}
\label{eq:second_order_correlation_finite_ts1}
\end{equation}
with fractional-power factors on the denominator of the zero-temperature case [i.e., Eq.~\eqref{eq:second_order_correlation}] replaced by temperature-dependent sinusoidal functions.
As the reminder, here $s\equiv s_1 - s_2$, $\tilde{s}_1 \equiv s_1 + L /v$, and $\tilde{s}_2 \equiv s_2 + L /v$, cf. Eq.~\eqref{eq:second_order_correlation}.
Noteworthily, as the temperature of channel 1 (before the diluter) is zero, the last line of the leading-order expression, Eq.~\eqref{eq:second_order_correlation_finite_ts1} remains the same as its zero-temperature version, i.e., $\mathcal{R}_{\eta_1\eta_2}$ of Eq.~\eqref{eq:r_etas}.
The conclusion however becomes more complicated when considering higher-order tunnelings through the diluter, where the interplay between non-equilibrium anyonic holes in the source channel $S1$ becomes temperature dependent [cf. Eqs.~\eqref{eq:i_finite} and \eqref{eq:i_finite_d3}].
Below we follow discussions of Sec.~S1, and separately analyze Eq.~\eqref{eq:second_order_correlation_finite_ts1} in three regions $\mathcal{R}_1$, $\mathcal{R}_2$ and $\mathcal{R}_3$ (cf. Fig.~\ref{fig:three_regions}).

\subsection*{S3A. Finite-temperature contribution of region $\mathcal{R}_1$}

When considering the integral Region $\mathcal{R}_1$, the major contribution comes from the part after taking $s\equiv s_1 - s_2\to 0$.
By doing so, the last line of Eq.~\eqref{eq:second_order_correlation_finite_ts1} simplifies into $\exp [ i\pi\nu (\eta_1 - \eta_2) ]$.
Actually, this term remains unchanged, when compared to the zero-temperature expression, Eq.~\eqref{eq:r_etas}.
Indeed, this term is produced when evaluating correlation among operators in channel 1, which has zero temperature before the diluter.
Corresponding integral, to the leading order of dilution, then becomes
\begin{equation}
\begin{aligned}
&\big\langle e^{-i \sqrt{\nu}\phi_1 (t^-,L)} e^{i \sqrt{\nu}\phi_1 (0^+, L)} \big\rangle_\text{neq,2}^{\mathcal{R}_1}  \\
= & - \mathcal{T}_1^{(0)} \frac{1}{(2\pi \tau_0 )^2 } \frac{\tau_0^\nu}{(\tau_0 + i t)^\nu} \frac{\tau_0^\nu}{[\tau_0 + i s \chi_{\eta_1\eta_2} (s)]^\nu}\sum_{\eta_1\eta_2} \eta_1\eta_2 e^{i\pi\nu (\eta_1 - \eta_2)}
\int_{-\infty}^\infty ds
\int_{0}^t d\tilde{s}_1 \,  \frac{  e^{-i \nu eV_{S1} s} (\pi T_{S1}\tau_0)^{\nu}}{\sin^\nu\left\{\pi T_{S1} [\tau_0 + i  s\chi_{\eta_1,\eta_2} (s) ]\right\}}\\
= & - \mathcal{T}_1^{(0)} \frac{t}{(2\pi \tau_0)^2 } \frac{\tau_0^\nu}{(\tau_0 + i t)^\nu} \left\{\! \int_0^\infty \!\!ds \frac{e^{-i\nu e V_{S1} s}}{(\tau_0 + i s )^\nu} \frac{(\pi T_{S1})^\nu}{\sin^\nu \left[\pi T_{S1} ( \tau_0 + i s ) \right]^\nu} 
    +\!  \int_{-\infty}^0 \!\!ds \frac{e^{-i\nu e V_{S1} s}}{(\tau_0 - i s )^\nu} \frac{(\pi T_{S1})^\nu}{\sin^\nu \left[\pi T_{S1} ( \tau_0 - i s ) \right]^\nu}\right.
    \\
+ &  \int_0^\infty ds \frac{e^{-i\nu e V_{S1} s}}{(\tau_0 - i s )^\nu} \frac{(\pi T_{S1})^\nu}{\sin^\nu \left[\pi T_{S1} ( \tau_0 - i s ) \right]^\nu} 
+  \int_{-\infty}^0 ds \frac{e^{-i\nu e V_{S1} s}}{(\tau_0 + i s )^\nu} \frac{(\pi T_{S1})^\nu}{\sin^\nu \left[\pi T_{S1} ( \tau_0 + i s ) \right]^\nu}
\\
- & \left.  \int_{-\infty}^\infty ds \frac{e^{-i\nu e V_{S1} s}}{(\tau_0 + i s )^\nu} \frac{(\pi T_{S1})^\nu}{\sin^\nu \left[\pi T_{S1} ( \tau_0 + i s ) \right]^\nu} e^{-2i\pi\nu}  
-  \int_{-\infty}^\infty ds \frac{e^{-i\nu e V_{S1} s}}{(\tau_0 - i s )^\nu} \frac{(\pi T_{S1})^\nu}{\sin^\nu \left[\pi T_{S1} ( \tau_0 - i s ) \right]^\nu} e^{2i\pi\nu} \right\}\\
= & - \mathcal{T}_1^{(0)} \frac{t}{(2\pi \tau_0 )^2 } \frac{\tau_0^\nu}{(\tau_0 + i t)^\nu} \left\{ \left( 1 - e^{-2i\pi\nu} \right) \int_{-\infty}^\infty \!\!ds \frac{e^{-i\nu e V_{S1} s}}{(\tau_0 + i s )^\nu} \frac{(\pi T_{S1})^\nu}{\sin^\nu \left[\pi T_{S1} ( \tau_0 + i s ) \right]^\nu} \right.\\
&\left. + \left( 1 - e^{2i\pi\nu} \right) \!  \int_{-\infty}^\infty \!\!ds \frac{e^{-i\nu e V_{S1} s}}{(\tau_0 - i s )^\nu} \frac{(\pi T_{S1})^\nu}{\sin^\nu \left[\pi T_{S1} ( \tau_0 - i s ) \right]^\nu} \right\},
\end{aligned}
\label{eq:r1_leading_integral_form}
\end{equation}
where we have used the fact that $0 < \tilde{s}_1 < t$ in Region $\mathcal{R}_1$, and that when $s \to 0$, the integral becomes independent of the value of $\tilde{s}_1$.
As another important feature, the integral over $t$ of Eq.~\eqref{eq:r1_leading_integral_form} can be bounded by two candidate time scales, i.e., the temperature inverse $1/\pi T_{S1}$, and the applied bias inverse $1/\nu e V_{S1}$.
The typical half width of non-equilibrium anyons is, crucially, encoded by the latter quantity.
When $\pi T_{S1} < \nu e V_{S1}$, the integral is bounded by $1/\nu e V_{S1}$.
Of this case, the information of non-equilibrium anyons is fully captured by the integral.
On the contrary, when temperature has become large enough, $\pi T_{S1} > \nu e V_{S1}$, the integral in time becomes instead bounded by the temperature inverse, a quantity that is smaller in comparison to the typical width of non-equilibrium anyons.
As the consequence, for a large enough temperature, the integral over $t$ can not capture the full information of non-equilibrium anyons that tunnel through the diluter.
Nevertheless, mathematically, the integral of Eq.~\eqref{eq:r1_leading_integral_form} can be performed in both regions.

To proceed, we notice that integrals of the last line, and that of the last-but-one line is induced by tunneling from channel 1 to $S1$, and that from $S1$ to 1, respectively.
Based on this observation, we arrive at the finite-temperature, of integral in region $\mathcal{R}_1$
\begin{equation}
\begin{aligned}
   \big\langle e^{-i \sqrt{\nu}\phi_1 (t^-,L)} e^{i \sqrt{\nu}\phi_1 (0^+, L)} \big\rangle_\text{neq,2}^{\mathcal{R}_1} &= \frac{1}{2\pi \tau_0} \frac{\tau_0^\nu}{(\tau_0 + it)^\nu} \left[ -\frac{I_{S1 \to 1}}{\nu e} \left( 1-e^{2i\pi\nu} \right)t -\frac{I_{1 \to S1}}{\nu e} \left( 1-e^{-2i\pi\nu} \right)t  \right]\\
   & = \frac{1}{2\pi \tau_0} \frac{\tau_0^\nu}{(\tau_0 + it)^\nu} \left\{ -\frac{S_1}{\nu^2 e^2} [1 - \cos (2\pi\nu) ]\, t + i \frac{I_1}{\nu e} \sin (2\pi\nu)\, t \right\},
\end{aligned}
\label{eq:d1_finite_t}
\end{equation}
where $I_{S1 \to 1}$ and $I_{1 \to S1}$ refer to the current that tunnels from the source channel $S1$ to the channel 1, and that of the opposite direction, respectively.
More specifically, the current that tunnels through the diluter, $I_1$, and the correspondingly generated noise, $S_1$, can be expressed in terms of the integrals
\begin{equation}
\begin{aligned}
    I_1 & = I_{S1 \to 1} - I_{1 \to S1} = \nu e \mathcal{T}_1^{(0)}  \int_{-\infty}^\infty \!\!ds \left\{ \frac{e^{-i\nu e V_{S1} s}}{(\tau_0 + i s )^\nu} \frac{(\pi T_{S1})^\nu}{\sin^\nu \left[\pi T_{S1} ( \tau_0 + i s ) \right]^\nu} - \frac{e^{-i\nu e V_{S1} s}}{(\tau_0 - i s )^\nu} \frac{(\pi T_{S1})^\nu}{\sin^\nu \left[\pi T_{S1} ( \tau_0 - i s ) \right]^\nu}\right\}\\
    & = -2i \nu e \mathcal{T}_1^{(0)}  \int_{-\infty}^\infty \!\!ds \frac{\sin (\nu e V_{S1} s)}{(\tau_0 + i s )^\nu} \frac{(\pi T_{S1})^\nu}{\sin^\nu \left[\pi T_{S1} ( \tau_0 + i s ) \right]^\nu},\\
    S_1 & = \nu e (I_{S1 \to 1} \!+\! I_{1 \to S1})\! =\! \mathcal{T}_1^{(0)} \nu^2 e^2 \!\!\int_{-\infty}^\infty \!\!ds \left\{ \frac{e^{-i\nu e V_{S1} s}}{(\tau_0 + i s )^\nu} \frac{(\pi T_{S1})^\nu}{\sin^\nu \left[\pi T_{S1} ( \tau_0 + i s ) \right]^\nu} + \frac{e^{-i\nu e V_{S1} s}}{(\tau_0 - i s )^\nu} \frac{(\pi T_{S1})^\nu}{\sin^\nu \left[\pi T_{S1} ( \tau_0 - i s ) \right]^\nu}\right\}\\
    & = 2 \nu^2 e^2 \mathcal{T}_1^{(0)} \int_{-\infty}^\infty \!\!ds \frac{\cos (\nu e V_{S1} s)}{(\tau_0 + i s )^\nu} \frac{(\pi T_{S1})^\nu}{\sin^\nu \left[\pi T_{S1} ( \tau_0 + i s ) \right]^\nu}.
\end{aligned}
\label{eq:i1_s1}
\end{equation}
Notice that when $T_{S1} = 0$, the second term of the first and third lines vanishes after the integral over time, and thus $S_1 = \nu e I_1$.
In general, however, this term is non-zero when $T_{S1}$ is instead finite, where $S_1 \neq \nu e I_1$, due to the influence of thermal fluctuations.
Nevertheless, despite of the received thermal influences, $S_1$ is actually the tunneling current noise (i.e., shot noise).
Within this work, as the ambient temperature in channel 1 (i.e., that before the diluter) equals zero, its noise after the diluter contains only $S_1$, but without contribution from the Nyquist-Johnson noise.
More generally, when the channel 1 contains a finite ambient temperature, the post-diluter noise in channel 1 contains two contributions, i.e., the shot noise $S_1$, and the Nyquist-Johnson noise induced by its ambient temperature.

In Eq.~\eqref{eq:d1_finite_t}, the real and imaginary parts of the factor within the curly brackets are proportional to the noise $S_1$ and non-equilibrium current $I_1$, respectively.
Here both the non-equilibrium current ($I_1$) and the corresponding noise ($S_1$) are explicitly introduced, as their ratio $S_1/I_1$ is now a function of the source temperature $T_{S1}$: such that one can treat $I_1$ and $S_1$ as two independent variables. This is in great contrast to Eq.~\eqref{eq:d1} of the zero temperature situation, where $S_1 / I_1 = \nu e$ is a constant number.

When discussing high-order dilution in a finite-temperature situation, interplay between different non-equilibrium anyonic pairs, Eq.~\eqref{eq:mathcal_i}, becomes modified into
\begin{equation}
\begin{aligned}
    & \mathcal{I}_{T_{S1}} (s_1^{\eta_1},s_2^{\eta_2};s_3^{\eta_3},s_4^{\eta_4})  \equiv \frac{[\tau_0 + i (s_1-s_3) \chi_{\eta_1,\eta_3} (s_1 - s_3) ]^\nu [\tau_0 + i (s_2-s_4) \chi_{\eta_2,\eta_4} (s_2 - s_4)]^\nu}{[\tau_0 + i (s_1-s_4) \chi_{\eta_1,\eta_4} (s_1 - s_4)]^\nu [\tau_0 + i (s_2-s_3) \chi_{\eta_2,\eta_3} (s_2 - s_3)]^\nu}\\
    & \times \frac{\sin^\nu\left\{ \pi T_{S1} [\tau_0 + i (s_1-s_3) \chi_{\eta_1,\eta_3} (s_1 - s_3) ] \right\} \sin^\nu\left\{ \pi T_{S1}[\tau_0 + i (s_2-s_4) \chi_{\eta_2,\eta_4} (s_2 - s_4)]\right\}}{\sin^\nu\left\{ \pi T_{S1}[\tau_0 + i (s_1-s_4) \chi_{\eta_1,\eta_4} (s_1 - s_4)]\right\} \sin^\nu\left\{ \pi T_{S1}[\tau_0 + i (s_2-s_3) \chi_{\eta_2,\eta_3} (s_2 - s_3)]\right\}},
\end{aligned}
\label{eq:i_finite}
\end{equation}
with the subscript ``$T_{S1}$'' highlighting the finite source temperature, $T_{S1}$.
In comparison to its zero-temperature version, Eq.~\eqref{eq:mathcal_i}, the last line of Eq.~\eqref{eq:i_finite} now has an explicit dependence on the source temperature $T_{S1}$, due to correlation between operators from the source.
Nevertheless, Eq.~\eqref{eq:i_finite} becomes approximately equivalent to the zero temperature version.
Indeed, even in the large temperature limit, $\pi T_{S1} > \nu e V_{S1}$, as long as two non-equilibrium anyonic pairs (at moments $s_1 \approx s_2$, and $s_3 \approx s_4$, respectively) are well separated, the last line of Eq.~\eqref{eq:i_finite} becomes approximately $\exp \left\{ \pi T_{S1} \left[ |s_1 - s_3| + |s_2 - s_4| - |s_1 - s_4| - |s_2 - s_3| \right] \right\} \approx 1$, the same as the zero-temperature result.
With this knowledge, for higher-order dilution, if non-equilibrium anyonic pairs are independent from each other (corresponding to the case where these pairs are well-separated in space, in the strongly dilute limit), the integral simply equals the product of contribution from each anyonic pair.
After resummation over higher-order dilutions, we have thus arrived at
\begin{equation}
    \big\langle e^{-i \sqrt{\nu}\phi_1 (t^-,L)} e^{i \sqrt{\nu}\phi_1 (0^+, L)} \big\rangle_\text{neq}^{\mathcal{R}_1} =  \frac{\tau_0^\nu}{(\tau_0 + it)^\nu} \exp \left\{ -\frac{S_1}{\nu^2 e^2} [1 - \cos (2\pi\nu) ]\, t + i \frac{I_1}{\nu e} \sin (2\pi\nu)\, t \right\},
    \label{eq:region_r1_finite_temperature}
\end{equation}
which equals Eq.~(22) of the main text, for region $\mathcal{R}_1$.
Again, akin to discussions after Eq.~\eqref{eq:d1_finite_t}, the finite-temperature result, Eq.~\eqref{eq:region_r1_finite_temperature}, contains both the current $I_1$ and noise $S_1$ on the exponential factor, as $S_1 \neq I_1 \nu e$ when $T_{S1}$ is finite.

\subsection*{S3B. Finite-temperature contribution, of branch cuts at $\tilde{s}_1 \to 0$, $\tilde{s}_2 \to t$}

Now we move to the other two regions.
We first consider the branch cut contribution near $\tilde{s}_1 \to 0$ and $\tilde{s}_2 \to t$.
Near this branch cut, integral of Eq.~\eqref{eq:second_order_correlation_finite_ts1} becomes
\begin{equation}
\begin{aligned}
&\big\langle e^{-i \sqrt{\nu}\phi_1 (t^-,L)} e^{i \sqrt{\nu}\phi_1 (0^+, L)} \big\rangle_\text{neq,2}^{\mathcal{R}_3} \Big|_{\tilde{s}_1 \to 0, \tilde{s}_2 \to t}  \\
= & -  \frac{ \mathcal{T}_1^{(0)} \tau_0^{\nu}}{(2\pi \tau_0)^2 } \frac{(\pi T_{S1} \tau_0^2)^\nu}{(\tau_0 + i t)^\nu}\,  [-2i\sin (\pi\nu)] 
\int_{0}^{t} d\tilde{s}_2 \, 
\left(\frac{\tilde{s}_2}{t - \tilde{s}_2 }\right)^\nu e^{i \nu eV_{S1} \tilde{s}_2} \,
\int_0^{\infty}\!\! d\tilde{s}_1 e^{i\nu eV_{S1}  \tilde{s}_1} \left(\frac{t + \tilde{s}_1}{\tilde{s}_1}\right)^\nu\\
&\times
 \left\{
\frac{ 1}{[\tau_0 + i (\tilde{s}_1 + \tilde{s}_2) ]^\nu \sin^\nu\left\{ \pi T_{S1} [\tau_0 + i (\tilde{s}_1 + \tilde{s}_2) ] \right\}} - \frac{1}{[\tau_0 - i (\tilde{s}_1 + \tilde{s}_2)]^\nu \sin^\nu\left\{ \pi T_{S1} [\tau_0 - i (\tilde{s}_1 + \tilde{s}_2) ] \right\}} \right\}\,
\\
& \xrightarrow{\tilde{s}_1 \to 0, \tilde{s}_2 \to t} -  \frac{ \mathcal{T}_1^{(0)} \tau_0^{\nu}}{(2\pi \tau_0)^2 } \frac{(\pi T_{S1} \tau_0^2)^\nu}{(\tau_0 + i t)^\nu}\,  [-2i\sin (\pi\nu)] \int_{0}^{t} d\tilde{s}_2' \, 
\left(\frac{t}{\tilde{s}_2' }\right)^\nu e^{-i \nu eV_{S1} \tilde{s}_2'} \,
\int_0^{\infty}\!\! d\tilde{s}_1 e^{i\nu eV_{S1}  \tilde{s}_1} \left(\frac{t}{\tilde{s}_1}\right)^\nu\\
& [-2i\sin (\pi\nu) ]\left[\frac{1}{ t} \frac{1}{\sin \left( \pi T_{S1} t\right)}\right]^\nu\\
= & \frac{(\pi T_{S1}\tau_0)^\nu}{\sin^\nu [\pi T_{S1} (\tau_0 + i t) ]^\nu} \frac{I_1}{e^2 V_{S1}} C_0(\nu ) e^{i\nu e V_{S1} t},
\end{aligned}
\label{eq:d2_d3_finite_ts1}
\end{equation}
where $I_1$ is the non-equilibrium current in channel 1, when $T_{S1} = 0$. As the reminder, $\tilde{s}_2' \equiv t- \tilde{s}_2$, cf. Eq.~\eqref{eq:d3_second}.
Following Eq.~\eqref{eq:d2_d3_finite_ts1},  temperature $T_{S1}$ does not influence integrals above, but only modifies the prefactor.
Indeed, $T_{S1}$ influences only the branch cut at $\tilde{s}_1 + \tilde{s}_2 = 0$, but not the branch cut $\tilde{s}_1 \to 0$, $\tilde{s}_2 \to t$ under current consideration.

When including higher-order diultions, the function that describes interplay between non-equilibrium anyon pairs now becomes (taking $\tilde{s}_1 < \tilde{s}_3\sim \tilde{s}_4 < \tilde{s}_2$)
\begin{equation}
\begin{aligned}
    & \mathcal{I}_{T_{S1}} (s_1^{\eta_1},s_2^{\eta_2};s_3^{\eta_3},s_4^{\eta_4})  \equiv \Bigg|\frac{[\tau_0 + i (s_1-s_3) \chi_{\eta_1,\eta_3} (s_1 - s_3) ]^\nu [\tau_0 + i (s_2-s_4) \chi_{\eta_2,\eta_4} (s_2 - s_4)]^\nu}{[\tau_0 + i (s_1-s_4) \chi_{\eta_1,\eta_4} (s_1 - s_4)]^\nu [\tau_0 + i (s_2-s_3) \chi_{\eta_2,\eta_3} (s_2 - s_3)]^\nu}\Bigg|\\
    & \times \Bigg|\frac{\sin^\nu\left\{ \pi T_{S1} [\tau_0 + i (s_1-s_3) \chi_{\eta_1,\eta_3} (s_1 - s_3) ] \right\} \sin^\nu\left\{ \pi T_{S1}[\tau_0 + i (s_2-s_4) \chi_{\eta_2,\eta_4} (s_2 - s_4)]\right\}}{\sin^\nu\left\{ \pi T_{S1}[\tau_0 + i (s_1-s_4) \chi_{\eta_1,\eta_4} (s_1 - s_4)]\right\} \sin^\nu\left\{ \pi T_{S1}[\tau_0 + i (s_2-s_3) \chi_{\eta_2,\eta_3} (s_2 - s_3)]\right\}}\Bigg|\\
    &\times \exp [ i\pi\nu (\eta_4 - \eta_3) ],
\end{aligned}
\label{eq:i_finite}
\end{equation}
which produces the same phase as the zero-temperature formula, Eq.~\eqref{eq:ifactor_d3_second}.
Consequently, following discussions akin to these after Eq.~\eqref{eq:i_finite}, the same as the zero temperature situation, high-order contributions perfectly vanish after summation
over Keldysh indexes $\eta_3$ and $\eta_4$.

As the consequence, the same as the zero-temperature situation, when considering the branch cuts at $\tilde{s}_1 \to 0$, $\tilde{s}_2 \to t$, we only need to visit leading-order contribution.
We emphasize again that this conclusion is only valid when extra non-equilibrium pairs, e.g., these with time arguments $s_3$ and $s_4$, are in region $\mathcal{R}_1$: otherwise higher-order co-tunneling at diluters induces finite contributions: they are however negligible in the strongly-diluted limit and thus not visited by our current work.

\subsection*{S3C. Finite-temperature contribution, of branch cuts at $\tilde{s}_1 \to 0$, $\tilde{s}_2 \to 0$}

Finally, we turn to the last branch cut, at $\tilde{s}_1 \to 0$, $\tilde{s}_2 \to 0$.
When taking this branch cut, zero-temperature integral, Eq.~\eqref{eq:d2_d3_first}, now becomes
\begin{equation}
\begin{aligned}
    & \left[ \left\langle e^{i \sqrt{\nu} \phi_1(L,t)}\, e^{-i \sqrt{\nu}\phi_1(L,0)}\right\rangle_\text{neq,2}^{\mathcal{R}_2} + \left\langle e^{i \sqrt{\nu} \phi_1(L,t)}\, e^{-i \sqrt{\nu}\phi_1(L,0)}\right\rangle_\text{neq,2}^{\mathcal{R}_3} \right] \Bigg|_{\tilde{s}_1, \tilde{s}_2 \to 0}\\
    \approx & \frac{ \mathcal{T}_1^{(0)} \tau_0^{\nu} (\pi T_{S1} \tau_0)^\nu}{(2\pi \tau_0)^2 } \frac{\tau_0^\nu}{(\tau_0 + i t)^\nu}\,  8i\sin^2 (\pi\nu) \  \text{Im} \left\{
\int_{0}^\infty d\tilde{s}_1 \frac{e^{i \nu eV_{S1} \tilde{s}_1} }{\tilde{s}_1^\nu} \int_0^{\infty}\!\! d\tilde{s}_2 \frac{\tilde{s}_2^\nu}{(\tilde{s}_1 + \tilde{s}_2)^\nu} \frac{e^{i\nu eV_{S1}  \tilde{s}_2}}{\sinh^\nu [\pi T_{S1} (\tilde{s}_1 + \tilde{s}_2)]} \right\}\\
= & 4i \frac{ \mathcal{T}_1^{(0)} \tau_0^{\nu} (\pi T_{S1} \tau_0)^\nu}{(2\pi \tau_0)^2 } \frac{\tau_0^\nu}{(\tau_0 + i t)^\nu}\,  \sin^2 (\pi\nu) \ \text{Im} \left\{ \int_0^\infty d\tilde{s} \frac{e^{i\nu e V_{S1} \tilde{s}}}{\tilde{s}^\nu \sinh^\nu (\pi T_{S1} \tilde{s})} \int_{-\tilde{s}}^{\tilde{s}} ds \frac{(\tilde{s} - s)^\nu}{(\tilde{s} + s)^\nu} \right\}\\
= & 8 i\frac{ \mathcal{T}_1^{(0)} \tau_0^{\nu} (\pi T_{S1} \tau_0)^\nu}{(2\pi \tau_0)^2 } \frac{\tau_0^\nu}{(\tau_0 + i t)^\nu}\, \pi\nu \sin (\pi\nu)  \\
& \times \frac{-i}{-2 i\sin(\pi\nu)}\text{Re} \left\{ \int_0^\infty d\tilde{s} \left[ \frac{e^{i\nu e V_{S1} \tilde{s}}}{(\tau_0 + i \tilde{s})^\nu \sin^\nu [\pi T_{S1} (\tau_0 + i\tilde{s})]} -  \frac{e^{i\nu e V_{S1} \tilde{s}}}{(\tau_0 - i \tilde{s})^\nu \sin^\nu [\pi T_{S1} (\tau_0 - i\tilde{s})]} \right]\tilde{s}\right\}\\
= & 2i \frac{ \mathcal{T}_1^{(0)} \tau_0^{\nu} (\pi T_{S1} \tau_0)^\nu}{(2\pi \tau_0)^2 } \frac{\tau_0^\nu}{(\tau_0 + i t)^\nu}\, \pi\nu  \\
& \times \left\{ \int_0^\infty d\tilde{s} \left[ \frac{e^{i\nu e V_{S1} \tilde{s}}}{(\tau_0 + i \tilde{s})^\nu \sin^\nu [\pi T_{S1} (\tau_0 + i\tilde{s})]} + \frac{e^{-i\nu e V_{S1} \tilde{s}}}{(\tau_0 - i \tilde{s})^\nu \sin^\nu [\pi T_{S1} (\tau_0 - i\tilde{s})]} \right. \right.\\
& \left.\left. -  \frac{e^{i\nu e V_{S1} \tilde{s}}}{(\tau_0 - i \tilde{s})^\nu \sin^\nu [\pi T_{S1} (\tau_0 - i\tilde{s})]} - \frac{e^{-i\nu e V_{S1} \tilde{s}}}{(\tau_0 + i \tilde{s})^\nu \sin^\nu [\pi T_{S1} (\tau_0 + i\tilde{s})]} \right]\tilde{s}\right\}\\
= & \! 2i \pi\nu \frac{ \mathcal{T}_1^{(0)} \tau_0^{\nu} (\pi T_{S1} \tau_0)^\nu}{(2\pi \tau_0)^2 } \frac{\tau_0^\nu}{(\tau_0 + i t)^\nu} \!\! \int_{-\infty}^\infty\!\!\! d\tilde{s} \ |\tilde{s}|\  \left\{ \frac{e^{i\nu e V_{S1} \tilde{s}}}{(\tau_0 \!+\! i \tilde{s})^\nu \sin^\nu [\pi T_{S1} (\tau_0\! +\! i\tilde{s})]} \!- \!\frac{e^{-i\nu e V_{S1} \tilde{s}}}{(\tau_0 \!+\! i \tilde{s})^\nu \sin^\nu [\pi T_{S1} (\tau_0 \!+\! i\tilde{s})]} \right\}.
\end{aligned}
\label{eq:d2_d3_first_ts1}
\end{equation}
We can check asymptotic behavior of Eq.~\eqref{eq:d2_d3_first_ts1}, of different limiting cases.
For the low-temperature limit $\pi T_{S1} \ll \nu e V$, $|\tilde{s}| \sim 1/\nu e V$, Eq.~\eqref{eq:d2_d3_first_ts1} behaves $\sim I_1/\nu e V$, of the same form as that of the zero-temperature result, Eq.~\eqref{eq:d2_d3_first}.
In contrast, in the opposite limit, $\pi T_{S1} \gg \nu e V$, the oscillating term that appears at the numerator of the integrand of Eq.~\eqref{eq:d2_d3_first_ts1} becomes unimportant.
Of this case, one can expand these exponential factors, leading to results that are proportional to $V_{S1}$. Of this case, the final result is instead $\sim I_1/\pi T_{S1}$.
In general,
\begin{equation}
\begin{aligned}
    &\left[ \left\langle e^{i \sqrt{\nu} \phi_1(L,t)}\, e^{-i \sqrt{\nu}\phi_1(L,0)}\right\rangle_\text{neq,2}^{\mathcal{R}_2} + \left\langle e^{i \sqrt{\nu} \phi_1(L,t)}\, e^{-i \sqrt{\nu}\phi_1(L,0)}\right\rangle_\text{neq,2}^{\mathcal{R}_3}\right] \Bigg|_{\tilde{s}_1, \tilde{s}_2 \to 0}
    \approx i\sin (\pi\nu) C_2(\nu,V_{S1}, T_{S1}) \frac{I_1}{e^2 V_{S1}},
\end{aligned}
\end{equation}
where we have written the integral in a similar form of the zero-temperature case, Eq.~\eqref{eq:d2_d3_first}, after introducing a temperature-dependent function
\begin{equation}
    C_2(\nu, V_{S1}, T_{S1}) \equiv e V_{S1} \frac{2\pi}{\sin (\pi\nu) }  \frac{\int_{-\infty}^\infty d\tilde{s} \ |\tilde{s}|\  \left\{ \frac{e^{i\nu e V_{S1} \tilde{s}}}{(\tau_0 + i \tilde{s})^\nu \sin^\nu [\pi T_{S1} (\tau_0 + i\tilde{s})]} - \frac{e^{-i\nu e V_{S1} \tilde{s}}}{(\tau_0 + i \tilde{s})^\nu \sin^\nu [\pi T_{S1} (\tau_0 + i\tilde{s})]} \right\}}{\int_{-\infty}^\infty d\tilde{s} \  \left\{ \frac{e^{i\nu e V_{S1} \tilde{s}}}{(\tau_0 + i \tilde{s})^\nu \sin^\nu [\pi T_{S1} (\tau_0 + i\tilde{s})]} - \frac{e^{-i\nu e V_{S1} \tilde{s}}}{(\tau_0 + i \tilde{s})^\nu \sin^\nu [\pi T_{S1} (\tau_0 + i\tilde{s})]} \right\}}.
    \label{eq:c2}
\end{equation}
To obtain an intuitive understanding, we present the plot of Eq.~\eqref{eq:c2} in Fig.~\ref{fig:c2_finite_temperature}, for different temperature/bias ratio, $e V_{S1}/\pi T_{S1}$.

%%%%%%%%%%%%%%%%%%%%%%%%%
\begin{figure}[h!]
  \includegraphics[width= 0.5 \linewidth]{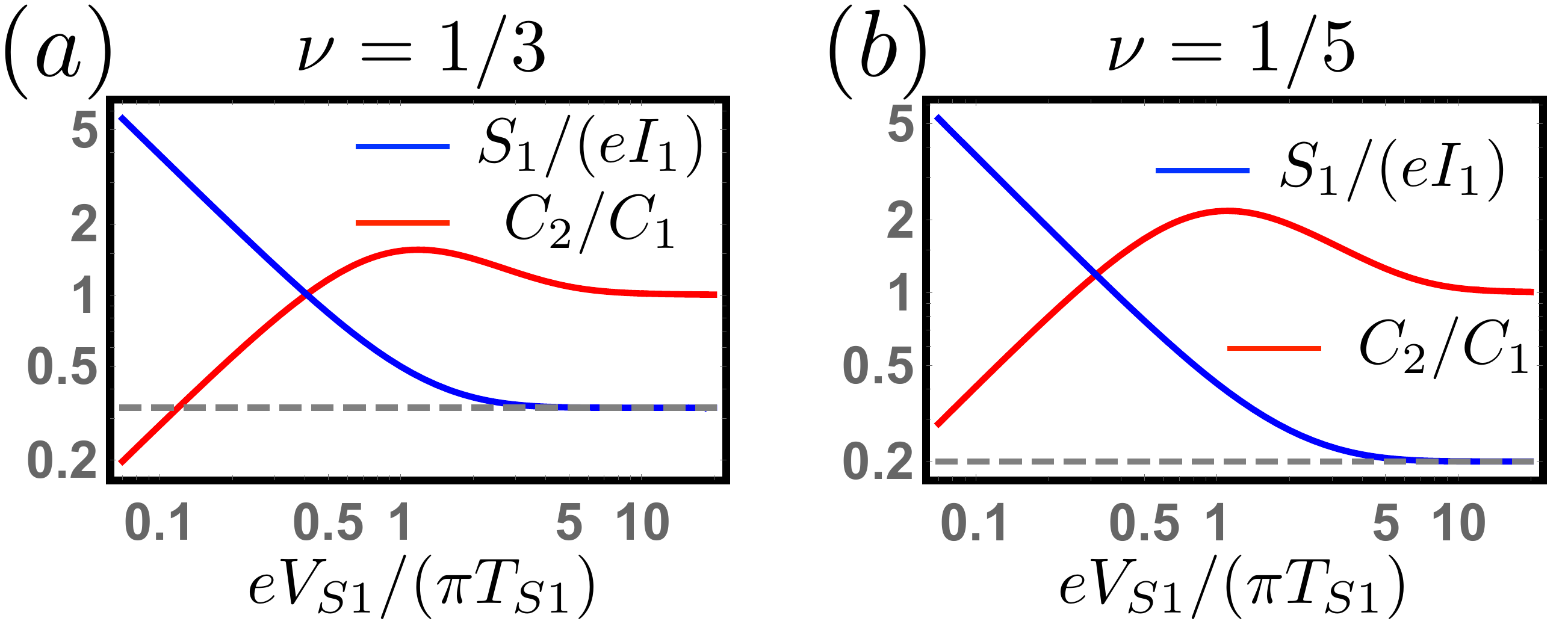}
  \caption{The value of $C_2(\nu, V_{S1}, T_{S1})/C_1 (\nu)$, and the noise-current ratio $S_1/e I_1$, for (a) $\nu = 1/3$ and (b) $\nu = 1/5$. Gray dashed lines refer to the value of $\nu$ in each figure. Here $C_1 (\nu)$ is defined in Eq.~\eqref{eq:c1}.
 }
  \label{fig:c2_finite_temperature}
\end{figure}
%%%%%%%%%%%%%%%%%%%%%%%%%

Now we move to consider corrections from higher-order dilutions.
Again, we start with the next-leading order dilution, where another pair of non-equilibrium pairs transport through the diluter at moments $s_3$ and $s_4$ (with $0< s_3 \sim s_4 < t$, for time-domain braiding), with Keldysh indexes $\eta_3$ and $\eta_4$, respectively.
Since $\tilde{s}_1, \tilde{s}_2 \to 0$, interplay between two non-equilibrium anyonic pairs now become
\begin{equation}
\begin{aligned}
    & \mathcal{I}_{T_{S1}} (s_1^{\eta_1},s_2^{\eta_2};s_3^{\eta_3},s_4^{\eta_4})  \equiv \Bigg|\frac{[\tau_0 + i (s_1-s_3) \chi_{\eta_1,\eta_3} (s_1 - s_3) ]^\nu [\tau_0 + i (s_2-s_4) \chi_{\eta_2,\eta_4} (s_2 - s_4)]^\nu}{[\tau_0 + i (s_1-s_4) \chi_{\eta_1,\eta_4} (s_1 - s_4)]^\nu [\tau_0 + i (s_2-s_3) \chi_{\eta_2,\eta_3} (s_2 - s_3)]^\nu}\Bigg|\\
    & \times \Bigg|\frac{\sin^\nu\left\{ \pi T_{S1} [\tau_0 + i (s_1-s_3) \chi_{\eta_1,\eta_3} (s_1 - s_3) ] \right\} \sin^\nu\left\{ \pi T_{S1}[\tau_0 + i (s_2-s_4) \chi_{\eta_2,\eta_4} (s_2 - s_4)]\right\}}{\sin^\nu\left\{ \pi T_{S1}[\tau_0 + i (s_1-s_4) \chi_{\eta_1,\eta_4} (s_1 - s_4)]\right\} \sin^\nu\left\{ \pi T_{S1}[\tau_0 + i (s_2-s_3) \chi_{\eta_2,\eta_3} (s_2 - s_3)]\right\}}\Bigg|,
\end{aligned}
\label{eq:i_finite_d3}
\end{equation}
being independent of Keldysh indexes $\eta_3$, $\eta_4$.
Consequently, next-leading order correlation, for the branch cut at $\tilde{s}_1 \sim \tilde{s}_2 \to 0$, becomes
\begin{equation}
\begin{aligned}
    & \left[\left\langle e^{i \sqrt{\nu} \phi_1(L,t)}\, e^{-i \sqrt{\nu}\phi_1(L,0)}\right\rangle_\text{neq,4}^{\mathcal{R}_2} + \left\langle e^{i \sqrt{\nu} \phi_1(L,t)}\, e^{-i \sqrt{\nu}\phi_1(L,0)}\right\rangle_\text{neq,4}^{\mathcal{R}_3} \right]\Bigg|_{\tilde{s}_1, \tilde{s}_2 \to 0}\\
    = & 2 \pi\nu \frac{ \mathcal{T}_1^{(0)} \tau_0^{\nu} (\pi T_{S1} \tau_0)^\nu}{(2\pi \tau_0)^2 } \frac{\tau_0^\nu}{(\tau_0 + i t)^\nu} \!\! \int_{-\infty}^\infty\!\!\! d\tilde{s} \ |\tilde{s}|\  \left\{ \frac{e^{i\nu e V_{S1} \tilde{s}}}{(\tau_0 \!+\! i \tilde{s})^\nu \sin^\nu [\pi T_{S1} (\tau_0\! +\! i\tilde{s})]} \!- \!\frac{e^{-i\nu e V_{S1} \tilde{s}}}{(\tau_0 \!+\! i \tilde{s})^\nu \sin^\nu [\pi T_{S1} (\tau_0 \!+\! i\tilde{s})]} \right\}\\
    \times & \!\frac{ \mathcal{T}_1^{(0)} t}{ (2\pi \tau_0)^2 }\! \left\{\! \left( 1 \!-\! e^{-2i\pi\nu} \right) \!\!\int_{-\infty}^\infty \!\!\!ds' \frac{e^{-i\nu e V_{S1} s'}}{(\tau_0 \!+\! i s' )^\nu} \frac{(\pi T_{S1})^\nu}{\sin^\nu\! \left[\pi T_{S1} ( \tau_0 \!+\! i s' ) \right]^\nu}  \!+\! \left( 1 \!-\! e^{2i\pi\nu} \right) \!\!  \int_{-\infty}^\infty \!\!\!ds' \frac{e^{-i\nu e V_{S1} s'}}{(\tau_0 \!-\! i s' )^\nu} \frac{(\pi T_{S1})^\nu}{\sin^\nu \!\left[\pi T_{S1} ( \tau_0 \!-\! i s' ) \right]^\nu}\!\! \right\} \\
    = & -2 \pi\nu \frac{ \mathcal{T}_1^{(0)} \tau_0^{\nu} (\pi T_{S1} \tau_0)^\nu}{(2\pi \tau_0)^2 } \frac{\tau_0^\nu}{(\tau_0 + i t)^\nu} \!\! \int_{-\infty}^\infty\!\!\! d\tilde{s} \ |\tilde{s}|\  \left\{ \frac{e^{i\nu e V_{S1} \tilde{s}}}{(\tau_0 \!+\! i \tilde{s})^\nu \sin^\nu [\pi T_{S1} (\tau_0\! +\! i\tilde{s})]} \!- \!\frac{e^{-i\nu e V_{S1} \tilde{s}}}{(\tau_0 \!+\! i \tilde{s})^\nu \sin^\nu [\pi T_{S1} (\tau_0 \!+\! i\tilde{s})]} \right\}\\
    \times &  \left\{ -\frac{S_1}{\nu^2 e^2} [1 - \cos (2\pi\nu) ]\, t + i \frac{I_1}{\nu e} \sin (2\pi\nu)\, t \right\},
\end{aligned}
\label{eq:d3_ts1_4th}
\end{equation}
where $s'\equiv s_3 - s_4$ is the difference between the moment of the extra non-equilibrium anyonic pair.
Following Eq.~\eqref{eq:d3_ts1_4th}, when considering the branch cut, $\tilde{s}_1 \sim \tilde{s}_2 \to 0$, correction from higher-order dilution, manifested via extra self-contracted non-equilibrium pairs, simply introduces another multiplying factor that equals the leading-order result of regime $\mathcal{R}_1$, as that shown in the last line of Eq.~\eqref{eq:d3_ts1_4th}.
Consequently, following previous argument on the resummation over higher-order terms, we arrive at
\begin{equation}
\begin{aligned}
    & \left[\left\langle e^{i \sqrt{\nu} \phi_1(L,t)}\, e^{-i \sqrt{\nu}\phi_1(L,0)}\right\rangle_\text{neq}^{\mathcal{R}_2} + \left\langle e^{i \sqrt{\nu} \phi_1(L,t)}\, e^{-i \sqrt{\nu}\phi_1(L,0)}\right\rangle_\text{neq}^{\mathcal{R}_3} \right]\Bigg|_{\tilde{s}_1, \tilde{s}_2 \to 0}\\
    = & i\sin (\pi\nu) C_2(\nu, V_{S1}, T_{S1}) \frac{I_1}{e^2 V_{S1}} \exp\left\{ -\frac{S_1}{\nu^2 e^2} [1 - \cos (2\pi\nu) ]\, t + i \frac{I_1}{\nu e} \sin (2\pi\nu)\, t \right\},
\end{aligned}
\label{eq:r2_r3_finite_temperature}
\end{equation}
as the finite-temperature version of correlation functions, when considering the branch cut at $\tilde{s}_1 \sim \tilde{s}_2 \to 0$.
Again, the exponential factor of Eq.~\eqref{eq:r2_r3_finite_temperature} contains both the non-equilibrium current $I_1$ and the corresponding noise $S_1$, as these two quantities are not proportional to each other, but depend on also the source temperature, $T_{S1}$.

\subsection*{S4. Effective linear-response transport coefficients, defined with respect to effective parameters}

Within the main text [cf. Eq.~(35)], effective linear-response transport coefficients are defined by taking derivatives of the electric and heat currents with respect to the real parameters $T_\text{2,eq}$ and $V_\text{2,eq}$ of equilibrium channel 2.
We take these definitions, as effective parameters ($V_\text{1,eff}$ and $T_\text{1,eff}$) of channel 1 non-universally depend on all details of the non-equilibrium channel.
The tuning and control of these effective parameters are thus more demanding in practical experiments.
Nevertheless, in this section we explicitly show how to measure these effective linear-response transport coefficients, if they are defined by differentiating over currents, with respect to effective parameters.
Transport coefficients defined in this way, as will be shown, is of apparently greater complexity (and thus experimentally impractical), in comparison to Eq.~(35) of the main text.

Briefly, following Appendix~E of the main text, effective parameters $V_\text{1,eff}$ and $T_\text{1,eff}$ are both functions of non-universal system details, after including collisions of non-equilibrium anyons.
As the consequence, in the effective linear-response regime (i.e., $V_\text{1,eff} \approx V_\text{2,eq}$ and $T_\text{1,eff} \approx T_\text{2,eq}$), tunneling charge and heat currents can be written as
\begin{equation}
\begin{aligned}
    I_\text{charge}^\text{coll} &\approx L^\text{neq-eq}_\text{IV,coll} [V_\text{1,eff} ( S_1 , I_1 , \mathcal{T}_1) - V_\text{2,eq}] + L^\text{neq-eq}_\text{IT,coll} [T_\text{1,eff} ( S_1 , I_1 , \mathcal{T}_1) - T_\text{2,eq}],\\
    J_\text{heat}^\text{coll} &\approx L^\text{neq-eq}_\text{JV,coll} [V_\text{1,eff} ( S_1 , I_1 , \mathcal{T}_1) - V_\text{2,eq}] + L^\text{neq-eq}_\text{JT,coll} [T_\text{1,eff} ( S_1 , I_1 , \mathcal{T}_1) - T_\text{2,eq}],\\
    L^\text{neq-eq}_\text{IV,coll} & \equiv -\frac{\partial}{\partial V_\text{2,eq}} I_\text{charge}^\text{coll} \Big|_{V_\text{2,eq} = V_\text{1,eff}, T_\text{2,eq} = T_\text{1,eff}} = \frac{\partial}{\partial (V_\text{1,eff} - V_\text{2,eq})} I_\text{charge}^\text{coll} \Big|_{V_\text{2,eq} = V_\text{1,eff}, T_\text{2,eq} = T_\text{1,eff}},\\
    L^\text{neq-eq}_\text{IT,coll} & \equiv -\frac{\partial}{\partial T_\text{2,eq}} I_\text{charge}^\text{coll} \Big|_{V_\text{2,eq} = V_\text{1,eff}, T_\text{2,eq} = T_\text{1,eff}} = \frac{\partial}{\partial (T_\text{1,eff} - T_\text{2,eq})} I_\text{charge}^\text{coll} \Big|_{V_\text{2,eq} = V_\text{1,eff}, T_\text{2,eq} = T_\text{1,eff}},\\
    L^\text{neq-eq}_\text{JV,coll} & \equiv -\frac{\partial}{\partial V_\text{2,eq}} J_\text{heat}^\text{coll} \Big|_{V_\text{2,eq} = V_\text{1,eff}, T_\text{2,eq} = T_\text{1,eff}} = \frac{\partial}{\partial (V_\text{1,eff} - V_\text{2,eq})} J_\text{heat}^\text{coll} \Big|_{V_\text{2,eq} = V_\text{1,eff}, T_\text{2,eq} = T_\text{1,eff}},\\
    L^\text{neq-eq}_\text{JT,coll} & \equiv -\frac{\partial}{\partial T_\text{2,eq}} J_\text{heat}^\text{coll} \Big|_{V_\text{2,eq} = V_\text{1,eff}, T_\text{2,eq} = T_\text{1,eff}} = \frac{\partial}{\partial (T_\text{1,eff} - T_\text{2,eq})} J_\text{heat}^\text{coll} \Big|_{V_\text{2,eq} = V_\text{1,eff}, T_\text{2,eq} = T_\text{1,eff}}.
\end{aligned}
\label{eq:connected_linear_response}
\end{equation}
Briefly, the bi-linear structure of the first lines of Eq.~\eqref{eq:connected_linear_response} is the direct consequence of the definition of effective parameters, i.e., $V_\text{1,eff}$ and $T_\text{1,eff}$.
Indeed, these parameters are defined to equal the values of real parameters, $V_\text{2,eq}$ and $T_\text{2,eq}$, given no charge or heat current that transports between channel 1 (being non-equilibrium) and channel 2 (being equilibrium).
Following this definition, around the effective equilibrium point, any modification of effective parameters ($V_\text{2,eq}$ and $T_\text{2,eq}$), triggered by the detuning (off effective equilibrium) of $I_1$, $S_1$ and/or $\mathcal{T}_1$, can be compensated (i.e., under which the system returns to effective equilibrium) by tuning real parameters $V_\text{2,eq}$ and $T_\text{2,eq}$ of channel 2 by the same amount.
This bi-linear structure of the first two lines of Eq.~\eqref{eq:connected_linear_response}, remarkably, indicates that around effective equilibrium, one can experimentally obtain thermoelectric coefficients (Peltier and Seebeck coefficients), as well as charge and heat conductances, by measuring the response of currents to an infinitesimal modification of real parameters, $V_\text{2,eq}$ and $T_\text{2,eq}$.
Below we discuss how to experimentally obtain the response to effective parameters, $V_\text{1,eff}$ and $T_\text{1,eff}$.

With Eq.~\eqref{eq:connected_linear_response}, we can obtain response of currents to the change of $I_1$ and $S_1$, leading to
\begin{equation}
\begin{aligned}
    \frac{\partial}{\partial I_1} I_\text{charge}^\text{coll} \Big|_{V_\text{2,eq} = V_\text{1,eff}, T_\text{2,eq} = T_\text{1,eff}} & = L^\text{neq-eq}_\text{IV,coll} \frac{\partial}{\partial I_1} V_\text{1,eff} \Big|_{V_\text{2,eq} = V_\text{1,eff}, T_\text{2,eq} = T_\text{1,eff}} + L^\text{neq-eq}_\text{IT,coll} \frac{\partial}{\partial I_1} T_\text{1,eff} \Big|_{V_\text{2,eq} = V_\text{1,eff}, T_\text{2,eq} = T_\text{1,eff}},\\
    \frac{\partial}{\partial S_1} I_\text{charge}^\text{coll} \Big|_{V_\text{2,eq} = V_\text{1,eff}, T_\text{2,eq} = T_\text{1,eff}} & = L^\text{neq-eq}_\text{IV,coll} \frac{\partial}{\partial S_1} V_\text{1,eff} \Big|_{V_\text{2,eq} = V_\text{1,eff}, T_\text{2,eq} = T_\text{1,eff}} + L^\text{neq-eq}_\text{IT,coll} \frac{\partial}{\partial S_1} T_\text{1,eff} \Big|_{V_\text{2,eq} = V_\text{1,eff}, T_\text{2,eq} = T_\text{1,eff}},\\
    \frac{\partial}{\partial I_1} J_\text{heat}^\text{coll} \Big|_{V_\text{2,eq} = V_\text{1,eff}, T_\text{2,eq} = T_\text{1,eff}} & = L^\text{neq-eq}_\text{JV,coll} \frac{\partial}{\partial I_1} V_\text{1,eff} \Big|_{V_\text{2,eq} = V_\text{1,eff}, T_\text{2,eq} = T_\text{1,eff}} + L^\text{neq-eq}_\text{JT,coll} \frac{\partial}{\partial I_1} T_\text{1,eff} \Big|_{V_\text{2,eq} = V_\text{1,eff}, T_\text{2,eq} = T_\text{1,eff}},\\
    \frac{\partial}{\partial S_1} J_\text{heat}^\text{coll} \Big|_{V_\text{2,eq} = V_\text{1,eff}, T_\text{2,eq} = T_\text{1,eff}} & = L^\text{neq-eq}_\text{JV,coll} \frac{\partial}{\partial S_1} V_\text{1,eff} \Big|_{V_\text{2,eq} = V_\text{1,eff}, T_\text{2,eq} = T_\text{1,eff}} + L^\text{neq-eq}_\text{JT,coll} \frac{\partial}{\partial S_1} T_\text{1,eff} \Big|_{V_\text{2,eq} = V_\text{1,eff}, T_\text{2,eq} = T_\text{1,eff}}.
\end{aligned}
\label{eq:coefficients_effective_parameters}
\end{equation}
Following Eq.~\eqref{eq:coefficients_effective_parameters}, we arrive at
\begin{equation}
\begin{aligned}
    L^\text{neq-eq}_\text{IV,coll} & = \frac{\partial_{S_1} T_\text{1,eff} \partial_{I_1}-  \partial_{I_1} T_\text{1,eff} \partial_{S_1}}{\partial_{I_1} V_\text{1,eff} \partial_{S_1} T_\text{1,eff} - \partial_{S_1} V_\text{1,eff} \partial_{I_1} T_\text{1,eff} } I_\text{charge}^\text{coll},\\
    L^\text{neq-eq}_\text{JT,coll} & = -\frac{\partial_{S_1} V_\text{1,eff} \partial_{I_1}-  \partial_{I_1} V_\text{1,eff} \partial_{S_1}}{\partial_{I_1} V_\text{1,eff} \partial_{S_1} T_\text{1,eff} - \partial_{S_1} V_\text{1,eff} \partial_{I_1} T_\text{1,eff} } J_\text{heat}^\text{coll},\\
    L^\text{neq-eq}_\text{IT,coll} & =- \frac{\partial_{S_1} V_\text{1,eff} \partial_{I_1}-  \partial_{I_1} V_\text{1,eff} \partial_{S_1}}{\partial_{I_1} V_\text{1,eff} \partial_{S_1} T_\text{1,eff} - \partial_{S_1} V_\text{1,eff} \partial_{I_1} T_\text{1,eff} } I_\text{charge}^\text{coll},\\
    L^\text{neq-eq}_\text{JV,coll} & = \frac{\partial_{S_1} T_\text{1,eff} \partial_{I_1}-  \partial_{I_1} T_\text{1,eff} \partial_{S_1}}{\partial_{I_1} V_\text{1,eff} \partial_{S_1} T_\text{1,eff} - \partial_{S_1} V_\text{1,eff} \partial_{I_1} T_\text{1,eff} } J_\text{heat}^\text{coll},
\end{aligned}
\label{eq:hard_definitions}
\end{equation}
which discloses the way to experimentally measure effective linear-response transport coefficients, as response to effective parameters of non-equilibrium channel 1.
We emphasize that Eq.~\eqref{eq:hard_definitions} is theoretically equivalent to definitions of transport coefficients of Eq.~(35) of the main text, where they are defined as the response of charge and heat currents to parameters of the equilibrium channel, i.e., $V_\text{2,eq}$ and $T_\text{2,eq}$.
Their equivalence can be seen by taking reverse steps to go back to Eq.~\eqref{eq:connected_linear_response}.
Nevertheless, Eq.~\eqref{eq:hard_definitions} provides an alternative way, other than Eq.~(35) of the main text, to experimentally measure transport coefficients by tuning parameters of the non-equilibrium channel.
In great contrast to Eq.~(35) of the main text where effective parameters are defined as the response to real parameters in the equilibrium channel 2,
Eq.~\eqref{eq:hard_definitions} further requires to know the response of effective parameters $V_\text{1,eff}$ and $T_\text{1,eff}$ to non-equilibrium parameters $I_1$ and $S_1$.
Eq.~\eqref{eq:hard_definitions} is thus experimentally much less practical in obtaining the charge and heat transport coefficients.

\subsection*{S5. The protocol, to experimentally measure effective linear response coefficients, when both channels are out-of-equilibrium}

In this section, we briefly discuss how to realize the experimental measurement of linear response coefficients of the neq-neq setup, i.e., that of Fig.~1(b) of the main text.
This question naturally arises, as in a non-equilibrium channel, the tuning of both the effective temperature and effective chemical potential requires (instead of more direct tuning knobs, e.g., the source potential and temperature) an indirect tuning of the non-equilibrium current and shot noise, e.g., $I_1$ and $S_1$ of channel 1, respectively.
Fine tuning (on the bias and/or temperature) is thus request, if one targets to tune either $I_1$ or $S_1$, while keeping the other quantity invariant.
Here we thus provide an experimentally more accessible protocol, which is free from fine-tuning, to measure the effective linear response coefficients, when both channels are out-of-equilibrium, and in the collision-free limit.

Briefly, the experimental measurement of effective linear response coefficients takes three major steps.
As the starting step, one tunes the system to effective equilibrium, such that no charge or heat current tunnels between two non-equilibrium channels.
As the second step, one tunes the source bias $V_{S1}$, while keeping the temperature in the source, $T_{S1}$, unchanged, thus generating finite tunneling charge and heat currents
\begin{equation}
\begin{aligned}
    I_\text{charge} \big|_{T_{S1}} & = L_\text{IT}^\text{neq-neq} \delta T_\text{1,eff}\big|_{T_{S1}} + L_\text{IV}^\text{neq-neq} \delta V_\text{1,eff} \big|_{T_{S1}} = \frac{1}{ \nu^2 e^2 s_c} L_\text{IT}^\text{neq-neq} \delta S_1\big|_{T_{S1}} + \frac{\sin(2\pi\nu)}{\nu^2 e^2} L_\text{IV}^\text{neq-neq} \delta I_1\big|_{T_{S1}},\\
    J_\text{heat} \big|_{T_{S1}} & = L_\text{JT}^\text{neq-neq} \delta T_\text{1,eff}\big|_{T_{S1}} + L_\text{JV}^\text{neq-neq} \delta V_\text{1,eff} \big|_{T_{S1}} = \frac{1}{ \nu^2 e^2 s_c} L_\text{JT}^\text{neq-neq} \delta S_1\big|_{T_{S1}} + \frac{\sin(2\pi\nu)}{\nu^2 e^2} L_\text{JV}^\text{neq-neq} \delta I_1\big|_{T_{S1}},
\end{aligned}
\label{eq:step_vs1}
\end{equation}
where the subscript $T_{S1}$ highlights the unchanged parameter during this measurement (thus, the parameter $V_{S1}$ is the parameter that is tuned).
More specifically, $I_\text{charge} \big|_{T_{S1}}$ and $J_\text{heat} \big|_{T_{S1}}$ are the tunneling charge and heat currents, induced by tuning $V_{S1}$.
The other quantities,
$\delta T_\text{1,eff}\big|_{T_{S1}}$ and $\delta V_\text{1,eff}\big|_{T_{S1}}$, instead refer to the corresponding modification of the effective temperature and chemical potential.
Quantities $I_\text{charge} \big|_{T_{S1}}$ and $J_\text{heat} \big|_{T_{S1}}$, on the left side of Eq.~\eqref{eq:step_vs1}, can be obtained by measuring the change of charge and heat currents in channel 2, due to the charge and heat flow through the central collider.
Quantities $\delta I_1 |_{T_{S1}}$ and $\delta S_1 |_{T_{S1}}$, on the right side of the equation, can instead be experimentally measured after pinching off the central collider.

Finally, we return to the effective equilibrium in the starting step.
Instead of tuning $V_{S1}$ [cf. Eq.~\eqref{eq:step_vs1}], here we tune the temperature $T_{S1}$ of channel $S1$, leading to another group of modifications on the tunneling charge and heat currents, i.e.,
\begin{equation}
\begin{aligned}
    I_\text{charge} \big|_{V_{S1}} & = L_\text{IT}^\text{neq-neq} \delta T_\text{1,eff}\big|_{V_{S1}} + L_\text{IV}^\text{neq-neq} \delta V_\text{1,eff} \big|_{V_{S1}} = \frac{1}{ \nu^2 e^2 s_c} L_\text{IT}^\text{neq-neq} \delta S_1\big|_{V_{S1}} + \frac{\sin(2\pi\nu)}{\nu^2 e^2} L_\text{IV}^\text{neq-neq} \delta I_1\big|_{V_{S1}},\\
    J_\text{heat} \big|_{V_{S1}} & = L_\text{JT}^\text{neq-neq} \delta T_\text{1,eff}\big|_{V_{S1}} + L_\text{JV}^\text{neq-neq} \delta V_\text{1,eff} \big|_{V_{S1}} = \frac{1}{ \nu^2 e^2 s_c} L_\text{JT}^\text{neq-neq} \delta S_1\big|_{V_{S1}} + \frac{\sin(2\pi\nu)}{\nu^2 e^2} L_\text{JV}^\text{neq-neq} \delta I_1\big|_{V_{S1}},
\end{aligned}
\label{eq:step_ts1}
\end{equation}
where $|_{V_{S1}}$ instead highlights the fact that we tune only $T_{S1}$ (while keeping $V_{S1}$ fixed) within this step.

Under the requirement,
\begin{equation}
    \delta I_1 \big|_{T_{S1}} \delta S_1 \big|_{V_{S1}} = \delta I_1 \big|_{V_{S1}} \delta S_1 \big|_{T_{S1}},
\end{equation}
which can be generically satisfied by controlling the change of $V_{S1}$ or $T_{S1}$, we can obtain all four effective linear response coefficients by solving these four linear equations.
More specifically, $L_\text{IT}^\text{neq-neq}$, $L_\text{IV}^\text{neq-neq}$ can be obtained by solving the first lines of both Eq.~\eqref{eq:step_vs1} and \eqref{eq:step_ts1}, leading to
\begin{equation}
\begin{aligned}
    L_\text{IV}^\text{neq-neq} & = \frac{\nu^2 e^2}{\sin(2\pi\nu)} \frac{I_\text{charge} \big|_{T_{S1}} \delta S_1 \big|_{V_{S1}} - I_\text{charge} \big|_{V_{S1}} \delta S_1 \big|_{T_{S1}}}{\delta I_1 \big|_{T_{S1}} \delta S_1 \big|_{V_{S1}} - \delta I_1 \big|_{V_{S1}} \delta S_1 \big|_{T_{S1}}},\\
    L_\text{IT}^\text{neq-neq} & = \nu^2 e^2 s_c \frac{I_\text{charge} \big|_{T_{S1}} \delta I_1 \big|_{V_{S1}} - I_\text{charge} \big|_{V_{S1}} \delta I_1 \big|_{T_{S1}}}{\delta S_1 \big|_{T_{S1}} \delta I_1 \big|_{V_{S1}} - \delta S_1 \big|_{V_{S1}} \delta I_1 \big|_{T_{S1}}}.
\end{aligned}
\end{equation}
Instead, $L_\text{JT}^\text{neq-neq}$, $L_\text{JV}^\text{neq-neq}$ can be obtained by solving the second lines of these two equations, leading to
\begin{equation}
\begin{aligned}
    L_\text{JV}^\text{neq-neq} & = \frac{\nu^2 e^2}{\sin(2\pi\nu)} \frac{J_\text{heat} \big|_{T_{S1}} \delta S_1 \big|_{V_{S1}} - J_\text{heat} \big|_{V_{S1}} \delta S_1 \big|_{T_{S1}}}{\delta I_1 \big|_{T_{S1}} \delta S_1 \big|_{V_{S1}} - \delta I_1 \big|_{V_{S1}} \delta S_1 \big|_{T_{S1}}},\\
    L_\text{JT}^\text{neq-neq} & = \nu^2 e^2 s_c \frac{J_\text{heat} \big|_{T_{S1}} \delta I_1 \big|_{V_{S1}} - J_\text{heat} \big|_{V_{S1}} \delta I_1 \big|_{T_{S1}}}{\delta S_1 \big|_{T_{S1}} \delta I_1 \big|_{V_{S1}} - \delta S_1 \big|_{V_{S1}} \delta I_1 \big|_{T_{S1}}}.
\end{aligned}
\end{equation}
We have thus provided a protocol, to obtain all four effective linear response coefficients.
Importantly, both measurements, given by Eqs.~\eqref{eq:step_vs1} and \eqref{eq:step_ts1}, do not require fine-tuning of either the system bias or temperature.
This protocol, being fine-tuning free, is thus experimentally much more accessible, than tuning either $I_1$ or $S_1$ individually.

\end{widetext}

\end{document}